\documentclass[onecolumn]{utility/autart}

\usepackage{graphicx}
\graphicspath{{figures/}}
\usepackage{subfigure}

\usepackage[usenames,dvipsnames]{color}
\usepackage{epstopdf}
\usepackage{tabularx}
\usepackage{colortbl}
\usepackage{booktabs}
\usepackage{wrapfig}
\usepackage{stfloats}
\usepackage{enumerate}

\usepackage{hyperref} 

\usepackage{enumitem}

\usepackage{natbib}

\definecolor{steelblue}{RGB}{70,130,180}

\usepackage{amsmath,amssymb,mathrsfs,dsfont}

\usepackage{savesym}
\savesymbol{AND}
\savesymbol{OR}
\savesymbol{NOT}
\savesymbol{TO}
\savesymbol{COMMENT}
\savesymbol{BODY}
\savesymbol{IF}
\savesymbol{ELSE}
\savesymbol{ELSIF}
\savesymbol{FOR}
\savesymbol{WHILE}

\usepackage{algorithm}
\usepackage{algorithmicx, algpseudocode}
\usepackage{todonotes}
\newtheorem{theorem}{Theorem}
\newtheorem{lemma}{Lemma}
\newtheorem{corollary}{Corollary}

\newtheorem{example}{Example}
\newtheorem{assumption}{Assumption}
\newtheorem{remark}{Remark}

\newcommand{\PP}{\mathds{P}}
\newcommand{\EE}{\mathds{E}}
\newcommand{\RR}{\mathds{R}}

\newcommand{\NN}{\mathds{N}}

\newcommand{\Let}{: =}
\newcommand{\teL}{= :}

\newcommand{\tp}{\intercal}

\newcommand{\bigO}{\mathcal{O}}

\newcommand{\eps}{\varepsilon}

\DeclareMathOperator{\vect}{vec}
\DeclareMathOperator{\svect}{svec}
\DeclareMathOperator{\mat}{mat}
\DeclareMathOperator{\smat}{smat}

\newcommand{\bfe}{\mathbf{e}}
\newcommand{\bfZ}{\mathbf{Z}}
\newcommand{\bfY}{\mathbf{Y}}
\newcommand{\bfC}{\mathbf{C}}
\newcommand{\bfD}{\mathbf{D}}

\allowdisplaybreaks[4]

\begin{document}

\begin{frontmatter}

\title{Identification of Linear Systems with Multiplicative Noise from Multiple Trajectory Data\thanksref{footnoteinfo}} 

\thanks[footnoteinfo]{Y. Xing and B. Gravell contributed equally to this paper.\\ \emph{Email addresses:} \texttt{yuxing2@kth.se} (Y.~Xing), \texttt{Benjamin.Gravell@utdallas.edu} (B. Gravell), \texttt{xhe9@nd.edu} (X. He), \texttt{kallej@kth.se} (K. H. Johansson), \texttt{Tyler.Summers@utdallas.edu} (T. Summers).
}

\author[kth]{Yu Xing},    
\author[utd]{Benjamin Gravell},               
\author[nd]{Xingkang He},  
\author[kth]{Karl Henrik Johansson},
\author[utd]{Tyler Summers}

\address[kth]{Division of Decision and Control Systems, School of Electrical Engineering and Computer Science,\\ KTH Royal Institute of Technology, and Digital Futures, Stockholm, Sweden.}        
\address[utd]{Department of Mechanical Engineering, The University of Texas at Dallas, Richardson, TX, USA.}             
 \address[nd]{Department of Electrical Engineering, University
of Notre Dame, South Bend, IN, USA.} 
          

\begin{abstract}                          
The paper studies identification of linear systems with multiplicative noise from multiple-trajectory data. An algorithm based on the least-squares method and multiple-trajectory data is proposed for joint estimation of the nominal system matrices and the covariance matrix of the multiplicative noise. The algorithm does not need prior knowledge of the noise or stability of the system, 
but requires only independent inputs with pre-designed first and second moments and relatively small trajectory length. The study of identifiability of the noise covariance matrix shows that there exists an equivalent class of matrices that generate the same second-moment dynamic of system states. It is demonstrated how to obtain the equivalent class based on estimates of the noise covariance. Asymptotic consistency of the algorithm is verified under sufficiently exciting inputs and system controllability conditions. Non-asymptotic performance of the algorithm is also analyzed under the assumption that the system is bounded. The analysis provides high-probability bounds vanishing as the number of trajectories grows to infinity. The results are illustrated by numerical simulations.
\end{abstract}

\end{frontmatter}

\section{Introduction}
The study of stochastic systems with multiplicative noise (i.e., system states and inputs multiplied by noise) has a long history in control theory \citep{wonham1967optimal}, 
and is re-emerging in the context of complex networked systems and learning-based control. 
In contrast to the additive-noise setting, the multiplicative-noise modeling framework has the ability to capture the coupling between noise and system states. This situation occurs in modern control systems as diverse as robotics with distance-dependent sensor errors \citep{dutoit2011robot}, networked systems with noisy communication channels \citep{antsaklis2007special,hespanha2007survey}, modern power networks with high penetration of intermittent renewables \citep{guo2019a}, turbulent fluid flow \citep{lumley2007stochastic}, and neuronal brain networks \citep{breakspear2017dynamic}. 
Linear systems with multiplicative noise are particularly attractive as a stochastic modeling framework because they remain simple enough to admit closed-form expressions for stabilization \citep{boyd1994linear} and optimal control \citep{wonham1967optimal,kleinman1969optimal,gravell2020learning}. 

It is important to study identification of linear systems with multiplicative noise, because, when solving problems such as control design of multiplicative-noise linear quadratic regulator (LQR), system parameters including the nominal system matrices and the noise covariance matrix, especially the latter, generally need be known \citep{gravell2020learning}. In contrast, for the design problem of additive-noise LQR, the covariance matrix of additive noise needs not be known \citep{dean2019sample}. Moreover, the identification problem requires further investigation; for instance, it is unclear how to formally quantify identifiability issues resulting from coupling between system states and multiplicative noise, and how to design identification algorithms to efficiently tackle the influence of multiplicative noise. 

Another issue that must be addressed is how to perform system identification based on multiple-trajectory data, rather than on single-trajectory data.
Multiple-trajectory data arises in two broad situations: (1) episodic tasks where a system is reset to an initial state after a finite run time, as encountered in iterative learning control and reinforcement learning \citep{matni2019from}; and (2) data collected from multiple identical systems in parallel, for example, robotic-grasping dataset collected by Google running several robot arms concurrently \citep{gu2017deep,levine2018learning}. 
For multiple-trajectory data, the length of each trajectory may be small, but the number of trajectories can be large.
However, the classic literature of system identification mainly focuses on studying online estimation over a single trajectory, so there is a need to study how to identify systems based on multiple-trajectory data. 
In addition, system identification based on multiple trajectories can be a pre-step of conducting other tasks such as control design of LQR \citep{dean2019sample}. Thus, studying the performance of identification algorithms based on multiple trajectories is necessary for obtaining performance guarantees of later tasks. 

\subsection{Related Work}
For identification of a nominal linear system, recursive algorithms, such as the recursive least-squares algorithm, have been developed in the control literature \citep{lai1982least,ljung1986,chen2012identification}. These algorithms can be applied to identification of linear systems with multiplicative noise, provided that certain conditions of system stability and noise hold. Non-asymptotic performance analysis of identification methods can be found in \cite{weyer2002non,campi2002finite,campi2005guaranteed}. It has once again attracted attention from different domains and been investigated more extensively, because of recent development of random matrix theories, self-normalized martingales, and so on (see \cite{dean2019sample,matni2019tutorial,zheng2020non} and references therein). 

For estimation of noise covariance, both recursive and batch methods have been proposed over the last few decades~\citep{dunik2017noise}, but most of these methods focus on the additive-noise case. 
In order to estimate multiplicative noise covariance,~\cite{schon2011system} introduces a maximum-likelihood approach, and~\cite{kitagawa1998self,kantas2015particle} utilize Bayesian frameworks. These methods, however, require prior assumptions on the noise distributions, whose incorrectness may worsen algorithm performance. \cite{coppens2020sample,coppens2020data} study stochastic LQR design for a special case of linear systems with multiplicative noise. It is assumed that the multiplicative noise is observed directly so that a concentration inequality can be obtained for estimates of the noise covariance.
The most relevant work to our paper is \cite{diconfidence}, which studies simultaneously estimating the nominal system parameters and noise covariance matrix based on single-trajectory data. In that paper, a self-normalizing (ellipsoidal) bound and a Euclidean (box) bound are provided for least-squares estimates, but it is not clear whether the bounds converge to zero under the setting of linear systems with multiplicative noise.

There is a growing interest in system identification based on multiple-trajectory data, along with their applications in data-driven control \citep{dean2019sample,matni2019tutorial}, due to the powerful and convenient estimator schemes facilitated by resetting the system. This framework can be applied to both stable and unstable systems, because of the finite duration of each trajectory. The authors in \cite{tu2017least,sun2020finite} introduce the procedure of collecting multiple trajectories, to identify finite impulse response systems. In \cite{dean2019sample}, the authors develop a framework called coarse-ID control to solve the problem of LQR with unknown linear dynamics. The first step of this framework is to learn a coarse model of the unknown  linear system, by observing multiple independent trajectories with finite length. However, only the last input-state pairs of the trajectories are used in the theoretical analysis of the learning algorithm. The performance of a least-squares algorithm,
using all samples of every trajectory, is studied in \cite{zheng2020non}, for partially observed, possibly open-loop unstable, linear systems.

\subsection{Contributions}
This paper considers identification of linear systems with multiplicative noise from multiple-trajectory data. The contributions are three-fold:
\begin{itemize}
\item[1.] An algorithm (Algorithm~\ref{alg:A}) based on the least-squares method and multiple-trajectory data is proposed for joint identification of the nominal system matrices and the multiplicative noise covariance from multiple-trajectory data. 
The algorithm does not need prior knowledge of the noise or stability of the system, but requires only independent inputs with pre-designed first and second moments, relatively small length for each trajectory, and the assumption of independent and identically distributed (i.i.d.) noise with finite first and second moments. It is theoretically shown that, under the preceding conditions, the algorithm solves the identification problem.
\item[2.] Identifiability of the noise covariance matrix is investigated (Propositions~\ref{thm:identifiability} and~\ref{prop:unique_of_equiv_class}). It is shown that there exists an equivalent class of covariance matrices that generate the same second-moment dynamic of system states. In addition, it is studied when such equivalent class has a unique element, meaning that the covariance matrix can be uniquely determined. An explicit expression of the equivalent class is provided for the recovery of the noise covariance based on estimates given by the proposed algorithm.
\item[3.] Asymptotic consistency of the proposed algorithm is verified (Theorem~\ref{thm:consistency}), under sufficiently exciting inputs and system controllability conditions. Non-asymptotic estimation performance is also analyzed under the assumption that the system is bounded. This analysis provides high-probability error bounds, which vanish as the number of trajectories grows to infinity (Theorems~\ref{thm:hatAB_bounded} and~\ref{thm:hatSigmaAB_bounded}).
\end{itemize}

Compared with \cite{diconfidence}, the current paper provides high-probability error bounds, for the proposed algorithm, that converge to zero as the number of trajectories increases. In addition, identifiability of the noise covariance matrix is thoroughly studied, and conditions, under which the covariance matrix is uniquely determined, are provided. 
In our problem, because of the complicated structure of the second-moment dynamic of system states, both analysis of the error bounds and study of the identifiability require more elaborate use of tools from linear algebra and high-dimensional probability theory. The differences between this paper and its conference version~\citep{xing2020linear} are as follows. This paper studies identifiability of the noise covariance matrix in detail, demonstrating a framework to recover the equivalent class of  covariance matrices. Moreover, sharper bounds for the required length of each trajectory are obtained. Finally, finite sample analysis of the proposed algorithm is provided.

\subsection{Outline}
The remainder of the paper is organized as follows. The problem is formulated in Section~\ref{problemFormulation}.
In Section~\ref{sec:parameterEstimation} the algorithm is introduced and theoretical results are given.
Numerical simulation results are presented in Section~\ref{sec:numericalSimulations}. Section~\ref{conclusions} concludes the paper.  Some proofs are postponed to Appendix.

\textbf{Notation.}\\
Denote the $n$-dimensional Euclidean space by $\RR^n$, and the set of $n\times m$ real matrices by $\RR^{n\times m}$. 
Let $\NN$ stand for the set of nonnegative integers, and $\NN^+ \Let \NN \setminus \{0\}$. Let $[k] \Let \{1, 2, \dots, k\}$, $k \in \NN^+$. 
We use $\|\cdot\|$ to denote the Euclidean norm for vectors, and use $\|\cdot\|_F$ and $\|\cdot\|_2$ to denote the Frobenius and spectral norm for matrices. 
The probability of an event $E$ is denoted by $\PP\{E\}$, and the expectation of a random vector $x$ is represented by $\EE \{x\}$. An event happening almost surely (a.s.) means that it happens with probability one.
Let $A \times B$ be the Cartesian product of sets $A$ and $B$, namely, $A \times B = \{(a,b): a\in A, b \in B\}$.
For two sequences of real numbers $a_k$ and $b_k \not= 0$, $k \in \NN^+$, denote $a_k = \bigO(b_k)$, if there exists a positive constant $C$ such that $|a_k/b_k| \le C$ for all $k \in \NN^+$. 

Let $a_{ij}$ or $[A]_{ij}$ represent the $(i,j)$-th entry of $A\in\RR^{n\times m}$. Denote the $n$-dimensional all-one vector and all-zero vector by $\mathbf{1}_n$ and $\mathbf{0}_n$, respectively. The $n$-dimensional unit vector with $i$-th component being one is represented by $\bfe_i^n$. 
$I_n$ is the $n$-dimensional identity matrix.
For two symmetric matrices $A, B\in \RR^{n\times n}$, $A \succeq 0$ ($A \succ 0$) means that $A$ is positive semidefinite (positive definite), and $A \succeq B$ ($A \succ B$) means that $A - B \succeq 0$ ($A-B \succ 0$).
For a matrix $A \in \RR^{n\times n}$, $\rho(A)$ represents the spectral radius of $A$. For a symmetric matrix $A \in \RR^n$, denote its smallest and largest eigenvalue by $\lambda_{\min}(A)$ and $\lambda_{\max}(A)$ respectively. 
A block diagonal matrix $A$ with $A_1$, $\dots$, $A_k$ on its diagonal is denoted by $\text{blockdiag}(A_1, \dots, A_k)$. 

The Kronecker product of two matrices $A \in \RR^{m\times n}$ and $B \in \RR^{p\times q}$ is represented by $A \otimes B$.
The full vectorization of $A = [a_{ij}] \in \RR^{m\times n}$ is found by stacking the columns of $A$ (i.e., ${\vect(A) = [a_{11}~a_{21}~\cdots~a_{m1}~a_{12}~a_{22}~\cdots~a_{mn}]^\tp}$).
The symmetric vectorization (also called half-vectorization) of a symmetric matrix $A \in \RR^{n\times n}$ is found by stacking the upper triangular part of the columns of $A$ (i.e., ${\svect(A) = [a_{11}~a_{12}~a_{22}~\cdots~a_{1n}~a_{2n}~\cdots~a_{nn}]^\tp}$).
The inverse operations of $\vect(\cdot)$ and $\svect(\cdot)$, given $p,q \in \NN$, are the full matricization $\mat_{p \times q}(x) \Let (\vect(I_q)^\tp \otimes I_p)(I_q \otimes x)$ for a vector $x \in \RR^{pq}$ and symmetric matricization $\smat_p(y)$ for a vector $y \in \RR^{p(p+1)/2}$, respectively.
To generalize the vectorization and matricization operations to a block matrix
\begin{align*}
    B =
    \begin{bmatrix}
    B_{11} & B_{12} & \cdots & B_{1n}\\
    \vdots & \vdots &        & \vdots\\
    B_{m1} & B_{m2} & \cdots & B_{mn}
    \end{bmatrix}
    \in \RR^{mp \times nq},
\end{align*}
where $B_{ij} \in \RR^{p\times q}$, define the following matrix reshaping operator $F: \RR^{mp \times nq} \rightarrow \RR^{mn \times pq}$,
\begin{align*}
    &F(B, m, n, p, q) \Let 
    [\vect(B_{11}) ~ \vect(B_{21}) ~ \cdots ~ \vect(B_{m1}) \cdots ~\vect(B_{12}) ~ \vect(B_{22}) ~ \cdots ~ \vect(B_{mn})]^\tp . 
\end{align*}
Then it holds that $F(A \otimes A, m, n, m, n) = \vect(A)\vect(A)^\tp$ for $A \in \RR^{m \times n}$, which demonstrates the correspondence between the entries of $A \otimes A$ and those of $\vect(A) \vect(A)^\tp$. Note when $p=q=1$, $F(\cdot)$ degenerates to $\vect(\cdot)$. 
Define the inverse reshaping operator $G: \RR^{mn \times pq} \to \RR^{mp \times nq}$ as
\begin{align*}
    G(B, m, n, p, q)  \Let \begin{bmatrix}
    \mat_{p\times q}(B_1) & \cdots & \mat_{p\times q}(B_{(n-1)m+1}) \\
    \mat_{p\times q}(B_2) & \cdots & \mat_{p\times q}(B_{(n-1)m+2}) \\
    \vdots & & \vdots \\
    \mat_{p\times q}(B_m) & \cdots & \mat_{p\times q}(B_{mn})
    \end{bmatrix},
\end{align*}
where $B \in \RR^{mn\times pq}$, $B_i^\tp$ is the $i$-th row of $B$.
Thus $F$ and $G$ are inverses of each other in the sense that 
\begin{align*}
    F(G(A, m, n, p, q), m, n, p, q) &= A, \\
    G(F(B, m, n, p, q), m, n, p, q) &= B,
\end{align*}
for any $A \in \RR^{mn \times pq}$ and $B \in \RR^{mp \times nq}$. In this way, $G(\vect(A) \vect(A)^\tp, m, n, m, n) = A \otimes A$ for $A \in \RR^{m\times n}$. Note that both $F$ and $G$ are linear: $F(A+B,m,n,p,q) = F(A,m,n,p,q) + F(B,m,n,p,q)$ for $A, B \in \RR^{mp\times nq}$, and $G(A+B,m,n,p,q) = G(A,m,n,p,q) + G(B,m,n,p,q)$ for $A, B \in \RR^{mn \times pq}$.

\section{Problem Formulation}\label{problemFormulation}
Consider the linear system with multiplicative noise
\begin{equation}
\begin{aligned}\label{theSystem}
 x_{t+1} &= (A  + \bar{A}_t) x_t +  (B + \bar{B}_t) u_t,~t\in \NN,
\end{aligned}
\end{equation}
where  $x_t \in \RR^n$ is the system state, and $u_t \in \RR^m$ is the control input, $m \le n$. 
The system is described by the nominal dynamic matrix $A \in \RR^{n \times n}$ and the nominal input matrix $B \in \RR^{n \times m}$, and incorporates multiplicative noise terms modeled by i.i.d. and mutually independent random matrices $\bar{A}_t$ and $\bar{B}_t$, which have zero mean and covariance matrices $\Sigma_A \Let \EE \{\vect(\bar{A}_t)\vect(\bar{A}_t)^T\} \in \RR^{n^2 \times n^2}$ and  $\Sigma_B \Let \EE \{\vect(\bar{B}_t)\vect(\bar{B}_t)^T\} \in \RR^{nm \times nm}$, respectively. 
The multiplicative noise is assumed to be independent of the inputs. Note that if $\bar{A}_t$ and $\bar{B}_t$ have non-zero means $\bar{A}$ and $\bar{B}$, respectively, then we can consider a system with nominal matrix $[A+\bar{A}~B+\bar{B}]$, as well as noise terms $\bar{A}_t - \bar{A}$ and $\bar{B}_t - \bar{B}$, which satisfies the preceding zero-mean assumption. The term multiplicative noise refers to that noise, $\bar{A}_t$ and $\bar{B}_t$, enters the system as multipliers of $x_t$ and $u_t$, rather than as additions. The independence of $\bar{A}_t$ and $\bar{B}_t$ is assumed for simplicity, and under this assumption the covariance matrix of the entire multiplicative noise is a block diagonal matrix $\EE \{\vect([\bar{A}_t~ \bar{B}_t])\vect([\bar{A}_t~ \bar{B}_t])^\tp\} = \text{blockdiag}(\Sigma_A,\Sigma_B)$. Throughout the paper, we use $(\Sigma_A,\Sigma_B) \in \RR^{n^2\times n^2} \times \RR^{nm\times nm}$ to represent this matrix. If $\bar{A}_t$ and $\bar{B}_t$ are dependent, there is an extra but amenable term on their correlations, $\EE \{\vect(\bar{A}_t)\vect(\bar{B}_t)^\tp\}$.

An example of System~\eqref{theSystem} is the following system studied in the optimal control literature \citep{boyd1994linear,gravell2020learning},
\begin{align} \label{eq:system_eigen_noises}
 x_{t+1} &= \Big(A  + \sum_{i = 1}^r A_i p_{i,t}\Big) x_t + \Big(B + \sum_{j = 1}^s B_j q_{j, t}\Big) u_t,
\end{align}
where $\{p_{i,t}\}$ and $\{q_{i,t}\}$ are mutually independent scalar random variables, with $\EE\{ p_{i,t}\} = \EE \{q_{j,t}\} = 0$, $\EE \{p_{i,t}^2\} = \sigma^2_i$, and $\EE\{ q_{j,t}^2\} = \delta^2_j$, $\forall i \in [r], j \in [s], t \in \NN$. It can be seen that $\bar{A}_t = \sum_{i = 1}^r A_i p_{i,t}$ and $\bar{B}_t = \sum_{j = 1}^s B_j q_{j, t}$, where $\sigma_i$ and $\delta_j$ are the eigenvalues of $\Sigma_A$ and $\Sigma_B$, and $A_i$ and $B_j$ are the reshaped eigenvectors of $\Sigma_A$ and $\Sigma_B$. These parameters are necessary for optimal controller design \citep{gravell2020learning}. 
It is also possible to use System~\eqref{eq:system_eigen_noises} to model cyber-physical systems in which fault signals appear as multiplicative noise \citep{wang2020optimal}. For new systems with unknown parameters, the key problem is to identify the parameters in the first place. Another example of System~\eqref{theSystem} is interconnected systems, where the nominal part captures relationships between different subsystems, and multiplicative noise characterizes randomly varying topologies \citep{haber2014subspace}.

In the rest of the paper, a trajectory sample is referred to as a \emph{rollout}. Suppose that multiple rollouts consisting of system states and inputs (i.e., $\{[x_0^{(k)}, u_0^{(k)}, \dots, x_{\ell-1}^{(k)}, u_{\ell-1}^{(k)}, x_{\ell}^{(k)}], k \in [n_r]\}$) are available, where $[x_0^{(k)}, u_0^{(k)}, \dots, x_{\ell-1}^{(k)}, u_{\ell-1}^{(k)}, x_{\ell}^{(k)}]$ is the $k$-th trajectory, $\ell$ is the length (index of the final time-step) of every rollout, and $n_r$ is the number of rollouts. The problem considered in this paper is as follows.

\textbf{Problem.} Given multiple-trajectory data $\{[x_0^{(k)}, u_0^{(k)},$ $\dots,$ $x_{\ell-1}^{(k)}, u_{\ell-1}^{(k)}, x_{\ell}^{(k)}], k \in [n_r]\}$, estimate the nominal system matrix $[A~B]$ and the noise covariance matrix $(\Sigma_{A}, \Sigma_B)$. 

\section{Identification Algorithm Based on Least-Squares and Multiple-Trajectory Data}\label{sec:parameterEstimation}

In this section, we propose and study an identification algorithm solving the considered problem. Section~\ref{sec:algorithm_design} studies identifiability of the noise covariance matrix, paving the way to algorithm design. Consistency of the algorithm is given by Theorem~\ref{thm:consistency} in Section~\ref{subsec:consistency}. Finally, sample complexity of the algorithm is studied in Section~\ref{subsec:finite}, and the results are provided in Theorems~\ref{thm:hatAB_bounded} and~\ref{thm:hatSigmaAB_bounded}.

\subsection{Moment Dynamics and Algorithm Design} \label{sec:algorithm_design}
In this subsection, we propose an algorithm based on multiple trajectories collected independently to estimate system parameters.
Before algorithm design, the effect of multiplicative noise on moment dynamics is studied, and identifiability of the noise covariance matrix is clarified.

Taking the expectation of both sides of System~\eqref{theSystem} and denoting $\mu_t \Let \EE\{x_t\}$ and $\nu_t \Let \EE \{u_t\}$ yield the first-moment dynamic of system states (i.e., the dynamic of $\EE\{x_t\}$) as follows,
\begin{align}\label{eq:expectationDynamics}
    \mu_{t+1} = A \mu_t + B \nu_t,~t\in \NN.
\end{align}
Denote the vectorization of the second-moment matrices of state, state-input, and input at time $t$ by $X_t \Let \vect(\EE \{x_t x_t^\tp\})$, $W_t \Let \vect(\EE \{x_t u_t^\tp\})$, $W_t' \Let \vect(\EE \{u_t x_t^\tp\})$, and $U_t \Let \vect(\EE \{u_t u_t^\tp\})$.  
From the independence of $\bar{A}_t$ and $\bar{B}_t$, as well as vectorization, the second-moment dynamic of system states is
\begin{align}\nonumber
    X_{t+1} &= (A \otimes A) X_t + (B \otimes A) W_t + (A \otimes B) W_t'  + (B \otimes B) U_t + \EE \{(\bar{A}_t \otimes \bar{A}_t) \vect(x_t x_t^\tp) \} + \EE  \{(\bar{B}_t \otimes \bar{B}_t) \vect(u_t u_t^\tp)  \} \\\label{eq:vecCorrelationDynamics}
    &= (A \otimes A + \Sigma_A') X_t + (B \otimes B + \Sigma_B') U_t + (B \otimes A) W_t + (A \otimes B) W_t',~t \in \NN,
\end{align}
where $\Sigma_A' = \EE\{\bar{A}_t \otimes \bar{A}_t\} \in \RR^{n^2 \times n^2}$ and $\Sigma_B' = \EE\{\bar{B}_t \otimes \bar{B}_t\} \in \RR^{n^2 \times m^2}$. The relation between $(\Sigma_A, \Sigma_B)$ and $(\Sigma_A', \Sigma_B')$ can be illustrated by $F(\Sigma_A', n, n, n, n) = \Sigma_A$ and $F(\Sigma_B', n, m, n, m) = \Sigma_B$, where the reshaping operator $F(\cdot)$ is defined in the notation section. 

An intrinsic identifiability issue arises in the second-moment dynamic \eqref{eq:vecCorrelationDynamics}.
Since $\EE\{x_tx_t^\tp\}$ is symmetric, $X_t$ has $n(n-1)/2$ pairs of identical entries corresponding to the off-diagonal entries of $\EE\{x_tx_t^\tp\}$ (i.e., $\EE\{x_{t,i} x_{t,j}\} = \EE\{x_{t,j}x_{t,i}\}$ for all $i,j \in [n]$). 
To remove the redundant terms, introduce binary row- and column-selection matrices, which are also called elimination and duplication matrices \citep{magnus1980}. 

To begin, notice that the redundant entries of $X_t$ are associated with the index set $\{ (j-1)n+i: i,j \in [n], i<j \}$.
Define matrix $T_1 \in \RR^{n^2 \times n^2}$ by replacing the $[(j-1)n+i]$-th row of $I_{n^2}$ by $(\bfe_{(i-1)n+j}^{n^2})^\tp$ for all $i,j \in [n]$ with $i<j$. Note that $\EE\{x_{t,i}x_{t,j}\}$ is the $[(j-1)n+i]$-th entry of $X_t$, so $X_t$ is invariant under $T_1$ (i.e., $X_t = T_1 X_t$). Furthermore, define a binary elimination matrix $P_1$ that picks out only the unique entries of $X_t$, and a complementary binary duplication matrix $Q_1$ which in turn reconstructs $X_t$ from the unique representation, by repeating the redundant entries in the proper order. These matrices are defined explicitly as $P_1 \in \RR^{[n(n+1)/2] \times n^2}$ by removing the $[(j-1)n+i]$-th row of $I_{n^2}$, $i,j \in [n]$ with $i<j$, and $Q_1 \in \RR^{n^2 \times [n(n+1)/2]}$ by removing the $[(j-1)n+i]$-th column of $T_1$, $i,j \in [n]$ with $i<j$.
Then one is able to freely convert between the full vectorization (with redundant entries) $X_t$ and the symmetric vectorization (without redundant entries) $\tilde{X}_t := \svect(X_t)$, by employing the linear transformations defined by the matrices $P_1$ and $Q_1$:
\begin{align*}
    \tilde{X}_t = P_1 X_t, \qquad
    X_t = Q_1 \tilde{X}_t.
\end{align*}


Now apply the same arguments to the second moment of input $U_t$: $U_t$ has $m(m-1)/2$ pairs of identical entries corresponding to the off-diagonal entries of $\EE\{u_tu_t^\tp\}$, so define $T_2 \in \RR^{m^2 \times m^2}$, $P_2 \in \RR^{[m(m+1)/2] \times m^2}$, and $Q_2 \in \RR^{m^2 \times [m(m+1)/2]}$ by replacing $n$ by $m$ in the definitions of $T_1, P_1$, and $Q_1$, respectively.

Applying the symmetric vectorization transformations $\tilde{X}_t = P_1 X_t$ and $\tilde{U}_t = P_2 U_t$ yields the second-moment dynamic with unique entries,
\begin{align}\nonumber
    \tilde{X}_{t+1} &= P_1 X_{t+1} \\\nonumber
    &= P_1 (A \otimes A + \Sigma_A') X_t + P_1 (B \otimes B + \Sigma_B') U_t  + P_1 (B \otimes A) W_t + P_1 (A \otimes B) W_t'\\\nonumber
    &= P_1 (A \otimes A + \Sigma_A') Q_1 P_1 X_t + P_1 (B \otimes B + \Sigma_B') Q_2 P_2 U_t + P_1 (B \otimes A) W_t + P_1 (A \otimes B) W_t'\\\label{eq:vecCorrelationDynamics_simp}
    &= (\tilde{A} + \tilde{\Sigma}_A') \tilde{X}_t + (\tilde{B} + \tilde{\Sigma}_B') \tilde{U}_t + K_{BA} W_t + K_{AB} W_t',
\end{align}
where the penultimate equation follows from $T_1 = Q_1P_1$ and $T_2 = Q_2P_2$. In the last equation the following notations are introduced: 
\begin{align*}
    \tilde{A} &\Let P_1 (A \otimes A) Q_1 \in \RR^{[n(n+1)/2] \times [n(n+1)/2]},\\
    \tilde{\Sigma}_A' &\Let P_1 \Sigma_A' Q_1 \in \RR^{[n(n+1)/2] \times [n(n+1)/2]},\\
    \tilde{B} &\Let P_1 (B \otimes B) Q_2 \in \RR^{[n(n+1)/2] \times [m(m+1)/2]},\\
    \tilde{\Sigma}_B' &\Let P_1 \Sigma_B' Q_2 \in \RR^{[n(n+1)/2]\times [m(m+1)/2]},\\
    K_{BA} &\Let P_1 (B \otimes A), ~K_{AB} \Let P_1 (A \otimes B).
\end{align*} 
Note that $\tilde{X}_t$ and $\tilde{U}_t$ have no redundant entries but are able to capture the second-moment dynamic of system states.
By the definition of Kronecker product, $\Sigma_A'$ and $\Sigma_B'$ have the following structures.
\begin{align}\label{eq_structure_sigmaABp}
\begin{array}{lc}
\mbox{}&
\begin{array}{ccccc} &(k-1)n+l& &(l-1)n+k \end{array}\\
\begin{array}{c} ~\\ (i-1)n+j \\ ~\\ (j-1)n+i\\~ \end{array}&
\left[\begin{array}{ccccc}
 & \vdots & & \vdots & \\
\cdots & \EE\{[\bar{A}_{t}]_{ik}[\bar{A}_{t}]_{jl}\} & \cdots & \EE\{[\bar{A}_{t}]_{il}[\bar{A}_{t}]_{jk}\} &\cdots\\
 & \vdots & & \vdots & \\
 \cdots & \EE\{[\bar{A}_{t}]_{jk}[\bar{A}_{t}]_{il}\} & \cdots & \EE\{[\bar{A}_{t}]_{jl}[\bar{A}_{t}]_{ik}\} & \cdots\\
  & \vdots & & \vdots & 
\end{array}\right]
\end{array}
\begin{array}{lc}
\mbox{}&
\begin{array}{ccccc} &(p-1)m+q& &(q-1)m+p \end{array}\\
\begin{array}{c} ~\\ ~ \\ ~\\ ~ \\~ \end{array}&
\left[\begin{array}{ccccc}
 & \vdots & & \vdots & \\
\cdots & \EE\{[\bar{B}_{t}]_{ip}[\bar{B}_{t}]_{jq}\} & \cdots & \EE\{[\bar{B}_{t}]_{iq}[\bar{B}_{t}]_{jp}\} &\cdots\\
 & \vdots & & \vdots & \\
 \cdots & \EE\{[\bar{B}_{t}]_{jp}[\bar{B}_{t}]_{iq}\} & \cdots & \EE\{[\bar{B}_{t}]_{jq}[\bar{B}_{t}]_{ip}\} & \cdots\\
  & \vdots & & \vdots & 
\end{array}\right]
\end{array},
\end{align}
where $i,j,k,l \in [n]$, $p,q\in [m]$, and $[\bar{A}_{t}]_{ij}$ ($[\bar{B}_{t}]_{ip}$) is the $(i,j)$-th entry of $\bar{A}_t$ ($(i,p)$-th entry of $\bar{B}_t$). If $i=j$ ($k=l$), the corresponding two rows (two columns) coincide. The following proposition demonstrates the correspondences between the entries of $\tilde{\Sigma}_A^\prime$ and $\tilde{\Sigma}_B^\prime$ and those of $\Sigma_A^\prime$ and $\Sigma_B^\prime$, respectively.

\begin{prop}\label{thm:identifiability}
Denote the $(i,j)$-th entry of $\tilde{\Sigma}_A'$ by $[\tilde{\Sigma}_A']_{ij}$. It holds for $i,j,k,l \in [n]$ with $i < j$ and $k < l$ that
\begin{align*}
    &[\tilde{\Sigma}_A']_{(i-1)(n-i/2)+i,(k-1)(n-k/2)+k}= \EE\{[\bar{A}_{t}]_{ik}[\bar{A}_{t}]_{ik}\},\\
    &[\tilde{\Sigma}_A']_{(i-1)(n-i/2)+i,(k-1)(n-k/2)+l} = 2\EE\{[\bar{A}_{t}]_{ik}[\bar{A}_{t}]_{il}\},\\
    &[\tilde{\Sigma}_A']_{(i-1)(n-i/2)+j,(k-1)(n-k/2)+k}= \EE\{[\bar{A}_{t}]_{ik}[\bar{A}_{t}]_{jk}\},\\
    &[\tilde{\Sigma}_A']_{(i-1)(n-i/2)+j,(k-1)(n-k/2)+l}  = \EE\{[\bar{A}_{t}]_{ik}[\bar{A}_{t}]_{jl}\} + \EE\{[\bar{A}_{t}]_{il}[\bar{A}_{t}]_{jk}\}.
\end{align*}
Denote the $(i,j)$-th entry of $\tilde{\Sigma}_B^\prime$ by $[\tilde{\Sigma}_B^\prime]_{ij}$. It holds for $i,j \in [n]$ with $i<j$ and $p,q\in[m]$ with $p<q$ that
\begin{align*}
    &[\tilde{\Sigma}_B^\prime]_{(i-1)(n-i/2)+i,(p-1)(m-p/2)+p} =  \EE\{[\bar{B}_{t}]_{ip}[\bar{B}_{t}]_{ip}\},\\
    &[\tilde{\Sigma}_B^\prime]_{(i-1)(n-i/2)+i,(p-1)(m-p/2)+q} = 2\EE\{[\bar{B}_{t}]_{ip}[\bar{B}_{t}]_{iq}\},\\
    &[\tilde{\Sigma}_B^\prime]_{(i-1)(n-i/2)+j,(p-1)(m-p/2)+p} = \EE\{[\bar{B}_{t}]_{ip}[\bar{B}_{t}]_{jp}\},\\
    &[\tilde{\Sigma}_B^\prime]_{(i-1)(n-i/2)+j,(p-1)(m-p/2)+q}  = \EE\{[\bar{B}_{t}]_{ip}[\bar{B}_{t}]_{jq}\} + \EE\{[\bar{B}_{t}]_{iq}[\bar{B}_{t}]_{jp}\}.
\end{align*}
\end{prop}

\begin{pf}
By observing the definitions of $P_i$ and $Q_i$, $i=1,2$, and the structures of $\Sigma_A^\prime$ and $\Sigma_B^\prime$ shown in~\eqref{eq_structure_sigmaABp}, we can get the expressions of the entries of $\tilde{\Sigma}_A^\prime$ and $\tilde{\Sigma}_B^\prime$ as in the proposition. To determine their positions, note from the definition of $P_1$ that all of the $[(j-1)n+i]$-th rows of $\Sigma_A^\prime$ are removed during the transformation $P_1 \Sigma_A^\prime$, where $j>i$, $i,j \in [n]$. This means that the following rows above the $[(i-1)n+j]$-th row of $\Sigma_A^\prime$, $i\le j$, $i,j\in[n]$, are removed: $(i-1)n+1$, $\dots$, $(i-1)n+i-1$, $(i-2)n+1$, $\dots$, $(i-2)n+i-2$, $\dots$, $n+1$, whose total number is $i(i-1)/2$. Thus, the $[(i-1)n+j]$-th rows of $\Sigma_A^\prime$ becomes the $[(i-1)n+j-i(i-1)/2]$-th row of $\tilde{\Sigma}_A^\prime$, i.e., the $[(i-1)(n-i/2)+j]$-th row, where $i\le j$, $i,j\in[n]$. Applying the same argument to the columns of $\Sigma_A^\prime$ and to $\Sigma_B^\prime$, we obtain the correspondence given in the proposition.
\end{pf}


\begin{remark}\label{rmk:equiv_class_covar}
The preceding discussion indicates that $X_t$ is determined by $[A~B]$ and $[\tilde{\Sigma}_A'~\tilde{\Sigma}_B']$, and the proposition shows that there exists a set of equivalent covariance matrices in the sense that they generate the same second-moment dynamic of system states, given the nominal matrix $[A~B]$. This fact results from that the dynamic of $X_t = Q_1 \tilde{X}_t$ only depends on $[A~B]$ and $[\tilde{\Sigma}_A'~\tilde{\Sigma}_B']$, and is the same under all $(\Sigma_1^\prime, \Sigma_2^\prime)$ satisfying $P_1\Sigma_1^\prime Q_1 = \tilde{\Sigma}_A'$ and $P_2\Sigma_2^\prime Q_2 = \tilde{\Sigma}_B'$.

From an entry-wise point of view, $\EE\{[\bar{A}_{t}]_{ik}[\bar{A}_{t}]_{jl}\}$ and $\EE\{[\bar{A}_{t}]_{il}[\bar{A}_{t}]_{jk}\}$, $i\not = j$ and $k \not = l$, have a coupled effect on the second-moment dynamic of system states. 
We may only estimate the sum of these two entries out of $X_t$, rather than their exact values, since realizations of $\bar{A}_t$ and $\bar{B}_t$ are not observed directly but indirectly through their effect on system states.
Fortunately, some entries of $\Sigma_A'$ and $\Sigma_B'$ are identifiable, such as $\EE\{[\bar{A}_{t}]_{ik}[\bar{A}_{t}]_{ik}\}$, the variance of $[\bar{A}_{t}]_{ik}$, and $\EE\{[\bar{A}_{t}]_{ik}[\bar{A}_{t}]_{jk}\}$, the covariance between entries in the same column. 
Similar issues also appear, when estimating covariance matrices, in topics such as Kalman filtering \citep{mehra1970identification,moghe2019adaptive}.
Critically, since these identifiable quantities uniquely generate the second-moment dynamic of system states, it suffices to estimate $\tilde{\Sigma}_A'$ and $\tilde{\Sigma}_B'$ for LQR design. This fact can be verified by expanding the Bellman equation; we omit the details to keep the paper concise.
\end{remark}
 
Given $(\Sigma_A,\Sigma_B)$ with $\Sigma_A\succeq 0$ and $\Sigma_B \succeq 0$ (then $\tilde{\Sigma}_A^\prime = P_1 \Sigma_A^\prime Q_1$ and $\tilde{\Sigma}_B^\prime = P_2 \Sigma_B^\prime Q_2$), the set of
equivalent matrices discussed in Remark~\ref{rmk:equiv_class_covar} can be written explicitly as follows, where positive semidefinite conditions are imposed because $\Sigma_A$ and $\Sigma_B$ are covariance matrices, 
\begin{align}\nonumber
    S^*(\tilde{\Sigma}_A^\prime) &:= \Big\{\Sigma_A(\alpha) \in \RR^{n^2\times n^2}: \Sigma_A(\alpha) \succeq 0, \alpha \in \RR^{n^2(n-1)^2/4} \Big\},\\\nonumber
    S^*(\tilde{\Sigma}_B^\prime) &:= \Big\{\Sigma_B(\beta) \in \RR^{nm\times nm}: \Sigma_B(\beta) \succeq 0, \beta \in \RR^{nm(n-1)(m-1)/4} \Big\}, \\\label{eq_equivalent_class}
    S^*_{\Sigma} &:= S^*(\tilde{\Sigma}_A^\prime) \times S^*(\tilde{\Sigma}_B^\prime),
\end{align}
with $\Sigma_A(\alpha) \Let F( Q_1 \tilde{\Sigma}_A' Q_1^\tp D_n + E_{\alpha}, n, n, n, n)$ and $\Sigma_B(\beta) \Let F( Q_1 \tilde{\Sigma}_B' Q_2^\tp D_m + E_{\beta}, n, m, n, m)$.
Here
\begin{align*}
    E_{\alpha} &= \underset{i<j,k<l}{\underset{i,j,k,l \in [n]}{\sum}} \bigg[ \alpha_{ij,kl} \Big(\bfe_{(i-1)n+j}^{n^2} - \bfe_{(j-1)n+i}^{n^2}\Big) \Big(\bfe_{(k-1)n+l}^{n^2} - \bfe_{(l-1)n+k}^{n^2}\Big)^\tp\bigg],\\
    E_{\beta} &= \underset{p,q \in [m],p<q}{\underset{i,j \in [n],i<j}{\sum}} \bigg[ \beta_{ij,pq} \Big(\bfe_{(i-1)n+j}^{n^2} - \bfe_{(j-1)n+i}^{n^2}\Big) \Big(\bfe_{(p-1)m+q}^{m^2} - \bfe_{(q-1)m+p}^{m^2}\Big)^\tp\bigg],
\end{align*}
where $\alpha = [\alpha_{ij,kl}] \in \RR^{n^2(n-1)^2/4}$, $\beta = [\beta_{ij,pq}] \in \RR^{nm(n-1)(m-1)/4}$, $i,j,k,l \in [n]$, $p,q \in [m]$, $i< j$, $k<l$, $p<q$, $Q_1$ and $Q_2$ are given before \eqref{eq:vecCorrelationDynamics_simp}, $D_n$ is an $n^2$-dimensional diagonal matrix with $[(i-1)n+i]$-th diagonal entry being $1$ and the rest being $1/2$, $i\in[n]$, and $D_m$ is an $m^2$-dimensional diagonal matrix with $[(p-1)m+p]$-th diagonal entry being $1$ and the rest being $1/2$, $p\in[m]$. Note that $S^*_{\Sigma}$ is given by two inequalities which respectively depend on $\alpha$ and $\beta$. These two inequalities are linear matrix inequalities \citep{boyd1994linear}, since the reshaping operator $F$ is linear. Obviously $S^*_{\Sigma}$ is not empty, because $(\Sigma_A, \Sigma_B)$ is one of its elements. The following example provides an intuitive idea of previous discussions.

\begin{example}\label{exam:identifiability}
Consider System~\eqref{theSystem} with $n=2$ and $m=1$, where
$X_t = [\EE\{ x_{t,1} x_{t,1}\} ~ \EE\{ x_{t,2} x_{t,1}\} ~ \EE\{ x_{t,1} x_{t,2}\}$ $\EE\{ x_{t,2} x_{t,2}\}]^T$. So $\EE\{ X_{t,2} X_{t,1}\}$ and $\EE\{ X_{t,1} X_{t,2}\}$ are identical and have the same dynamic from \eqref{eq:vecCorrelationDynamics}. Thus, 
\begin{align*}
&\tilde{X}_{t} = \left[\EE\{ x_{t,1} x_{t,1}\} ~~ \EE\{ x_{t,2} x_{t,1}\} ~~ \EE\{ x_{t,2} x_{t,2}\}\right]^T,\\
&P_1 = \begin{bmatrix}
1 & 0 & 0 & 0\\
0 & 1 & 0 & 0\\
0 & 0 & 0 & 1
\end{bmatrix}, ~
Q_1 = \begin{bmatrix}
1 & 0 & 0 \\
0 & 1 & 0 \\
0 & 1 & 0 \\
0 & 0 & 1
\end{bmatrix}, ~
T_1 = \begin{bmatrix}
1 & 0 & 0 & 0 \\
0 & 1 & 0 & 0 \\
0 & 1 & 0 & 0 \\
0 & 0 & 0 & 1 
\end{bmatrix}, ~ P_2 = Q_2 = T_2 = 1.
\end{align*}
According to the previously discussed simplification, from 
\begin{align*}
\Sigma_A' = \begin{bmatrix}
\sigma_{a, 11,11} & \sigma_{a, 11,12} & \sigma_{a, 12,11} & \sigma_{a, 12,12}\\
\sigma_{a, 11,21} & \sigma_{a, 11,22} & \sigma_{a, 12,21} & \sigma_{a, 12,22}\\
\sigma_{a, 21,11} & \sigma_{a, 21,12} & \sigma_{a, 22,11} & \sigma_{a, 22,12}\\
\sigma_{a, 21,21} & \sigma_{a, 21,22} & \sigma_{a, 22,21} & \sigma_{a, 22,22}
\end{bmatrix},~
\Sigma_B' = \begin{bmatrix}
\sigma_{b, 11} & \sigma_{b, 12} & \sigma_{b,21} & \sigma_{b,22}
\end{bmatrix}^T,
\end{align*}
we have that
\begin{align*}
\tilde{\Sigma}_A' = \begin{bmatrix}
\sigma_{a, 11,11} & 2 \sigma_{a, 11,12} & \sigma_{a, 12,12}\\
\sigma_{a, 11,21} & \sigma_{a, 11,22} + \sigma_{a, 12,21} & \sigma_{a, 12,22}\\
\sigma_{a, 21,21} & 2 \sigma_{a, 21,22} & \sigma_{a, 22,22}
\end{bmatrix},~
\tilde{\Sigma}_B' = \begin{bmatrix}
\sigma_{b, 11} & \sigma_{b, 12} & \sigma_{b,22}
\end{bmatrix}^T,
\end{align*}
where $\sigma_{a, ij,kl} = \EE\{[\bar{A}_{t}]_{ij} [\bar{A}_{t}]_{kl}\}$ and $\sigma_{b, ij} = \EE\{[\bar{B}_{t}]_{i}, [\bar{B}_{t}]_{j}\}$, and that \begin{align*}
&\tilde{A} = \begin{bmatrix}
a_{11}a_{11} & a_{11}a_{12} + a_{12}a_{11} & a_{12}a_{12}\\
a_{11}a_{21} & a_{11}a_{22} + a_{12}a_{21} & a_{12}a_{22}\\
a_{21}a_{21} & a_{21}a_{22} + a_{22}a_{21} & a_{22}a_{22}
\end{bmatrix}, ~
\tilde{B} = \begin{bmatrix}
b_1b_1 & b_1b_2  & b_2b_2
\end{bmatrix} ^T, \\
&K_{AB} = \begin{bmatrix}
a_{11}b_1 & a_{12}b_1\\
a_{21}b_1 & a_{22}b_2\\
a_{21}b_2 & a_{22}b_2
\end{bmatrix}, ~
K_{BA} = \begin{bmatrix}
a_{11}b_1 & a_{12}b_1\\
a_{11}b_2 & a_{12}b_2\\
a_{21}b_2 & a_{22}b_2
\end{bmatrix}.
\end{align*}
In this example, $\Sigma_B$ is unique, but based on~\eqref{eq_equivalent_class} the covariance matrix $\Sigma_A(\alpha)$, equivalent to $\Sigma_A$, is given by
\begin{align*}
    \Sigma_A(\alpha) = \begin{bmatrix}
    \sigma_{a,11,11} & \sigma_{a,11,21} & \sigma_{a,11,12} & \sigma_{a,11,22} + \alpha \\
    \sigma_{a,21,11} & \sigma_{a,21,21} & \sigma_{a,21,12} - \alpha & \sigma_{a,21,22} \\
    \sigma_{a,12,11} & \sigma_{a,12,21} - \alpha & \sigma_{a,12,12} & \sigma_{a,12,22} \\
    \sigma_{a,22,11} + \alpha & \sigma_{a,22,21} & \sigma_{a,22,12} & \sigma_{a,22,22}
    \end{bmatrix},
\end{align*}
where $\alpha \in \RR$ is such that $\Sigma_A(\alpha) \succeq 0$.
\end{example}

\begin{example}\label{exam:system_eigen_noises}
Consider System~\eqref{eq:system_eigen_noises} with $\bar{A}_t = \sum_{i=1}^r A_i p_{i,t}$, $\bar{B}_t = \sum_{j=1}^s B_j q_{j,t}$. Hence,
\begin{align*}
    &\Sigma_A = \sum_{i=1}^r \EE\{p_{i,t}^2\} \vect(A_i) \vect(A_i)^\tp,~ \Sigma_A^\prime = \sum_{i=1}^r \EE\{p_{i,t}^2\} A_i \otimes A_i,~ \tilde{\Sigma}_A^\prime = \sum_{i=1}^r \EE\{p_{i,t}^2\} P_1(A_i \otimes A_i) Q_1,\\
    &\Sigma_B = \sum_{j=1}^s \EE\{q_{j,t}^2\} \vect(B_j) \vect(B_j)^\tp,~ \Sigma_B^\prime = \sum_{j=1}^s \EE\{q_{j,t}^2\} B_j \otimes B_j,~ \tilde{\Sigma}_B^\prime = \sum_{j=1}^s \EE\{q_{j,t}^2\} P_1(B_j \otimes B_j) Q_2.
\end{align*}
Suppose that for $A_i$, $i\in [r]$, there exist $k_i$, $l_i \in [n]$ such that $[A_i]_{k_i,l_i} \not=0$ and $[A_j]_{k_i,l_i}$ for all $j \in [r]\setminus\{i\}$. That is, the $(k_i,l_i)$-th entry of $A_i$ is nonzero but the $(k_i,l_i)$-th entry of $A_j$ is zero for all $j\not= i$. Then $\EE\{[\bar{A}_t]_{k_i,l_i}[\bar{A}_t]_{k_i,l_i}\} = [A_i]_{k_i,l_i}^2 \EE\{p_{i,t}^2\} = [A_i]_{k_i,l_i}^2 \sigma_i^2$. From Proposition~\ref{thm:identifiability} we know that $\sigma_i^2 = \EE\{p_{i,t}^2\}$ can be uniquely determined if second-moment dynamic~\eqref{eq:vecCorrelationDynamics}, or $\tilde{\Sigma}_A^\prime$, is given. A similar conclusion holds for $\{q_{j,t}\}$. However there are also situations where $\sigma_i^2$ cannot be uniquely determined. For instance, assume that $r\ge 2$ and for all $i\in [r]$, $[A_i]_{11} \not= 0$ but all other entries of $A_i$ are zero. Then we only have a single equation $\sum_{i=1}^r [A_i]_{11}^2 \sigma_i^2 = [\tilde{\Sigma}_A^\prime]_{11}$ for $\{\sigma_i^2\}$.
\end{example}

As shown in Example~\ref{exam:identifiability}, given $(\Sigma_A,\Sigma_B)$ with $\Sigma_A\succeq 0$ and $\Sigma_B \succeq 0$, the set $S^*_{\Sigma}$ is not empty but may have infinitely many elements, resulting in unidentifiable entries $\EE\{[\bar{A}_{t}]_{ik}[\bar{A}_{t}]_{jl}\}$ and $\EE\{[\bar{B}_{t}]_{ip}[\bar{B}_{t}]_{jq}\}$, $i\not = j$, $k \not = l$, $p\not=q$, $i,j,k,l\in[n]$, $p,q\in[m]$. The following proposition gives several conditions under which the covariance matrix of the multiplicative noise can or cannot be uniquely determined from $[\tilde{\Sigma}_A^\prime~\tilde{\Sigma}_B^\prime]$.

\begin{prop}\label{prop:unique_of_equiv_class}
Given $(\Sigma_A,\Sigma_B)$ with $\Sigma_A\succeq 0$ and $\Sigma_B \succeq 0$,  $\tilde{\Sigma}_A^\prime = P_1 \Sigma_A^\prime Q_1$ and $\tilde{\Sigma}_B^\prime = P_2 \Sigma_B^\prime Q_2$, the following results hold.\\
(i) If $n=m=1$, then $S^*_{\Sigma}$ has a unique element. If $m=1$, then $S^*(\tilde{\Sigma}_B^\prime)$ has a unique element. If $n\ge 2$ and $\Sigma_A \succ 0$ (resp. $m\ge 2$ and $\Sigma_B \succ 0$), then $S^*(\tilde{\Sigma}_A^\prime)$ (resp. $S^*(\tilde{\Sigma}_B^\prime)$) has infinitely many elements. As a result, under either condition, $S^*_{\Sigma}$ has infinitely many elements.\\
(ii) If $S^*(\tilde{\Sigma}_A^\prime)$ has infinitely many elements, then $S^*(\tilde{\Sigma}_A^\prime) \cap T_A$ has a unique element, where 
\begin{align*}
    T_A &:= \Big\{\Sigma \in \RR^{n^2\times n^2}: \gamma_{ij,kl}  \Sigma_{(k-1)n+i,(l-1)n+j} + \delta_{ij,kl}  \Sigma_{(l-1)n+i,(k-1)n+j} = \tau_{ij,kl}, i<j, k<l, i,j,k,l \in[n]  \Big\},
\end{align*}
with constants $\gamma_{ij,kl}, \delta_{ij,kl}, \tau_{ij,kl} \in \RR$ and $\gamma_{ij,kl} \not= \delta_{ij,kl}$ for all $i<j$, $k<l$, $i,j,k,l \in[n]$. The same result holds for $S^*(\tilde{\Sigma}_B^\prime)$ by modifying the definition of $T_A$ according to the dimension of $\Sigma_B$.
\end{prop}

\begin{pf}
The first two conclusions of~(i) are trivial. If $n\ge 2$, then $\tilde{\Sigma}_A^\prime$ has entries of the form $\EE\{[\bar{A}_{t}]_{ik}[\bar{A}_{t}]_{jl}\} + \EE\{[\bar{A}_{t}]_{il}[\bar{A}_{t}]_{jk}\}$. Since $\Sigma_A \succ 0$, its minimum eigenvalue is larger than zero. Note that from the definition of $S^*(\tilde{\Sigma}_A^\prime)$ there exists $\alpha^*$ such that $\Sigma_A = \Sigma_A(\alpha^*)$. Because the eigenvalues of a matrix depend continuously on its entries (Theorem~2.4.9.2 of~\cite{horn2012matrix}), $\Sigma_A(\alpha^* + \eps)$ is still a positive definite matrix for small enough $\eps > 0$. This proves the last result in~(i). From~(i), we know that if $S^*(\tilde{\Sigma}_A^\prime)$ has infinitely many elements, then $n\ge 2$. To show~(ii), just note that if $\Sigma = \vect(\bar{A})\vect(\bar{A})^\tp$ for some $A \in \RR^{n\times n}$, then the $[(k-1)n+i,(l-1)n+j]$-th entry of $\Sigma$ is $[\bar{A}]_{ik}[\bar{A}]_{jl}$ and the  $[(l-1)n+i,(k-1)n+j]$-th entry is $[\bar{A}]_{il}[\bar{A}]_{jk}$, $i\not=j$, $k\not=l$. Hence if $\gamma_{ij,kl} \not= \delta_{ij,kl}$ then we have two linearly independent equations for $[\bar{A}]_{ik}[\bar{A}]_{jl}$ and $[\bar{A}]_{il}[\bar{A}]_{jk}$ (the other one from Proposition~\ref{thm:identifiability} is $[\bar{A}]_{ik}[\bar{A}]_{jl} + [\bar{A}]_{il}[\bar{A}]_{jk} = [\tilde{\Sigma}_A']_{(i-1)(n-i/2)+j,(k-1)(n-k/2)+l}$). So these entries can be uniquely determined, and the conclusion follows.
\end{pf}

\begin{remark}
The first part of the proposition shows that if $\Sigma_A \succ 0$ or $\Sigma_B \succ 0$ and $n\ge m\ge 2$, then it is impossible to uniquely determine $(\Sigma_A,\Sigma_B)$ only based on second-moment dynamic~\eqref{eq:vecCorrelationDynamics_simp}. However the second part indicates that more conditions imposed on the covariance matrix can make all entries of $\Sigma_A$ and $\Sigma_B$ identifiable. The set $T_A$ introduces additional constraints for $\EE\{[\bar{A}_{t}]_{ik}[\bar{A}_{t}]_{jl}\}$, $i\not = j$, $k \not = l$.
For example, if entries in $\bar{A}_t$ are mutually independent, then $\Sigma_A$ is diagonal. In this case, it holds that $\EE\{[\bar{A}_{t}]_{ik}[\bar{A}_{t}]_{jl}\} - \EE\{[\bar{A}_{t}]_{il}[\bar{A}_{t}]_{jk}\} = 0$, $i\not = j$, $k \not = l$, and hence the covariance matrix of $\bar{A}_t$ is uniquely determined.
\end{remark}

Now we are ready to propose our estimation algorithm.
Following the previous discussion, we introduce an algorithm based on the first- and second-moment dynamics \eqref{eq:expectationDynamics} and~\eqref{eq:vecCorrelationDynamics_simp}.
Since the exact moment dynamics are unavailable, we average over multiple independent rollouts to obtain their estimates.
To get persistently exciting inputs, it is necessary to design their first and second moments in advance, in either a deterministic or a stochastic way. For example, generate the two moments from standard Gaussian and Wishart 
distributions \citep{gupta2018matrix}, respectively, or set them periodically. 
The initial states of different rollouts are assumed to be i.i.d. subject to a same distribution $\mathcal{X}_0$ with finite second moment (see Section~\ref{subsubsec:consistency}). The overall algorithm is shown in Algorithm \ref{alg:A}, where the superscript $(k)$ represents the $k$-th rollout. Note that Algorithm~\ref{alg:A} is different from classic recursive identification algorithms. The recursive least-squares algorithm \citep{lai1982least,chen2012identification}, for example, uses only one trajectory of a system. In contrast, Algorithm~\ref{alg:A} is based on multiple trajectories with finite length. 

Based on the estimates $\hat{\tilde{\Sigma}}_A^\prime$ and $\hat{\tilde{\Sigma}}_B^\prime$, it is able to obtain an estimate $\hat{S}^*_{\Sigma}$ of the equivalent class~\eqref{eq_equivalent_class}, via replacing $\tilde{\Sigma}_A^\prime$ and $\tilde{\Sigma}_B^\prime$ in the definition~\eqref{eq_equivalent_class} by their estimates. If the linear matrix inequalities are infeasible (i.e., $\hat{S}^*_{\Sigma} = \emptyset$), then project the estimates onto the positive semidefinite cone. However this situation is unlikely to happen when $n_r$ is large, because of the consistency of Algorithm~\ref{alg:A} given in the next section.

\begin{algorithm}[t]
\caption{\newline Multiple-trajectory averaging least-squares (MALS)}
\label{alg:A}
\begin{algorithmic}[1]
\Statex{\textbf{Input:} Rollout length $\ell$ and the number of rollouts $n_r$.}
\Statex{\textbf{Output:} $[\hat{A}~\hat{B}]$, $[\hat{\tilde{\Sigma}}_A'~\hat{\tilde{\Sigma}}_B']$.}
\Statex{// \textbf{Control-input design} }
\For{$t$ from $0$ to $\ell-1$}
\State{Generate $\nu_t \in \RR^m$ and $\bar{U}_t \in \RR^{m\times m}$ with $\bar{U}_t \succeq 0$.}
\EndFor
\Statex{// \textbf{Multiple-trajectory collection}}
\For{$k$ from $1$ to $n_r$}
\State{Generate $x_{0}^{(k)}$ independently from the initial  multivariate distribution $\mathcal{X}_0$.}
\For{$t$ from $0$ to $\ell-1$}
\State{Generate $u_t^{(k)}$ independently from a multivariate distribution with first moment $\nu_t$ and second central}
\Statex{\qquad\quad moment $\bar{U}_t$,}
\State{$x_{t+1}^{(k)} = (A + \bar{A}_t^{(k)}) x_t^{(k)} + (B + \bar{B}_t^{(k)}) u_t^{(k)}$.}
\EndFor
\EndFor
\Statex{// \textbf{Least-squares estimation}}
\For{$t$ from $0$ to $\ell$}
\State{Compute 
\begin{align*}
    \hat{\mu}_t & \Let \frac1{n_r}  \sum_{k = 1}^{n_r} x_t^{(k)} , \\ 
    \hat{\tilde{X}}_t   & \Let \frac{1}{n_r} P_1 \vect \left( \sum_{k = 1}^{n_r} x_t^{(k)} (x_t^{(k)})^\tp \right) , \\
    \hat{W}_t   & \Let \frac{1}{n_r} \vect \left( \sum_{k = 1}^{n_r} x_t^{(k)} \nu_t^\tp \right) = \vect(\hat{\mu}_t \nu_t^\tp ), \\
    \hat{W}_t'   & \Let \frac{1}{n_r} \vect \left( \sum_{k = 1}^{n_r} \nu_t {x_t^{(k)}}^\tp \right) = \vect(\nu_t \hat{\mu}_t^\tp ), \\
    \tilde{U}_t    & \Let P_2 \vect(\bar{U}_t + \nu_t\nu_t^\tp).
\end{align*}
}
\EndFor
\State{$[\hat{A}~\hat{B}] = \underset{{[A~B]}}{\text{argmin}} \{\sum_{t=0}^{\ell-1}  \|\hat{\mu}_{t+1} - (A\hat{\mu}_{t}+B \nu_{t}) \|_2^2 \}$,}
\State{Compute $\hat{\tilde{A}} = P_1 (\hat{A} \otimes \hat{A}) Q_1$, $\hat{\tilde{B}} = P_1 (\hat{B} \otimes \hat{B}) Q_2$, $\hat{K}_{BA} = P_1 (\hat{B} \otimes \hat{A})$, and $\hat{K}_{AB} = P_1 (\hat{A} \otimes \hat{B})$, where $P_1, P_2, Q_1$, and $Q_2$ are given before \eqref{eq:vecCorrelationDynamics_simp},}
\State{$[\hat{\tilde{\Sigma}}_A'~\hat{\tilde{\Sigma}}_B'] =
\underset{[\tilde{\Sigma}_A'~\tilde{\Sigma}_B']}{\text{argmin}} \{ \sum_{t=0}^{\ell-1}  \|\hat{\tilde{X}}_{t+1}$~$- [\tilde{A} \hat{\tilde{X}}_t + K_{BA} \hat{W}_t$ $ + K_{AB} \hat{W}_t' + \tilde{B} \tilde{U}_t  + \tilde{\Sigma}_A' \hat{\tilde{X}}_t + \tilde{\Sigma}_B' \tilde{U}_t] \|_2^2 \}$.}
\end{algorithmic}
\end{algorithm}

\subsection{Performance of Algorithm~\ref{alg:A}}\label{subsec:consistency}
This section analyzes performance of Algorithm~\ref{alg:A} by investigating the moment dynamics \eqref{eq:expectationDynamics} and \eqref{eq:vecCorrelationDynamics_simp}. 

\subsubsection{Moment Dynamics and Input Design}\label{sec:input_design}
Provided that $\mu_t$ and $\tilde{X}_t$ are known, it is possible to recover the parameters via least-squares as in lines $14$-$16$ in Algorithm \ref{alg:A}. Denote
\begin{equation}
\begin{aligned}\label{defLSMatrices}
    \bfY
    \Let [
    \mu_{\ell} ~ \cdots ~ \mu_1
    ],~
    \bfZ
    \Let \begin{bmatrix}
    \mu_{\ell-1} & \cdots & \mu_0\\
    \nu_{\ell-1} & \cdots & \nu_0
    \end{bmatrix},~
    \bfC \Let
    [
    C_\ell ~ \cdots ~ C_1
    ], ~
    \bfD \Let
    \begin{bmatrix}
    \tilde{X}_{\ell-1} & \cdots & \tilde{X}_0\\
    \tilde{U}_{\ell-1} & \cdots & \tilde{U}_0
    \end{bmatrix}, 
\end{aligned}
\end{equation}
where $C_t = \tilde{X}_t - \big(\tilde{A} \tilde{X}_{t-1} + K_{BA} W_{t-1} + K_{AB} W_{t-1}' + \tilde{B} \tilde{U}_{t-1}\big)$, $1 \le t \le \ell$. Then closed-form solutions of the least-squares problems are
\begin{align*}
    \big[\hat{A}~ \hat{B}\big] = \bfY \bfZ^\tp (\bfZ \bfZ^\tp)^{\dagger},~
    \Big[\hat{\tilde{\Sigma}}_A' ~\hat{\tilde{\Sigma}}_B' \Big] = \bfC \bfD^\tp (\bfD \bfD^\tp)^{\dagger},
\end{align*}
where $\dagger$ represents the pseudoinverse. When the inverse matrices exist, the solutions are identical to true values; that is, $[\hat{A}~ \hat{B}] = [A~B]$ and $[\hat{\tilde{\Sigma}}_A'~ \hat{\tilde{\Sigma}}_B'] = [\tilde{\Sigma}_A'~ \tilde{\Sigma}_B']$.
Hence, the first question towards the consistency of Algorithm~\ref{alg:A} is whether the matrices $\bfZ \bfZ^\tp$ and $\bfD \bfD^\tp$ are invertible.
As to be shown, designing a proper input sequence ensures this invertibility, if systems $(A, B)$ and $(\tilde{A} + \tilde{\Sigma}_A', \tilde{B} + \tilde{\Sigma}_B')$ are controllable, and the rollout length $\ell$ is large enough.

\begin{prop}\label{thm:as_Z_fullrank}
Suppose that $\ell \ge n+m$ and $(A, B)$ is controllable. For fixed $\mu_0 \in \RR^n$, the matrix $\bfZ$ has full row rank, and consequently $\bfZ \bfZ^\tp$ is invertible, for almost all $[\nu_0^\tp~\cdots~\nu_{\ell-1}^\tp]^\tp \in \RR^{m\ell}$.
\end{prop}
\begin{pf}
See Appendix \ref{appen:pf_thm:as_Z_fullrank}.
\end{pf}

\begin{remark}
The proposition shows that for large enough rollout length, the full row rankness of $\bfZ$ can be guaranteed for almost all $[\nu_0^\tp~\cdots~\nu_{\ell-1}^\tp]^\tp \in \RR^{m\ell}$. The controllability of $(A, B)$ plays a key role in the proof, similar to classic results on identification of linear systems \citep{chen2012identification}. The condition $\ell \ge n+m$ is necessary for the invertibility of $\bfZ\bfZ^\tp$. This lower bound is much smaller than that given in \cite{xing2020linear}. According to the proposition, $\bfZ\bfZ^\tp$ is invertible with probability one if the first moments of inputs are generated i.i.d. from a distribution absolutely continuous with respect to Lebesgue measure (e.g., Gaussian distribution or uniform distribution). This proposition can be seen as a generalization of the single-input case studied in~\cite{schmidt2005identification}.
\end{remark}

\begin{prop}\label{thm:as_D_fullrank}
Suppose that $\ell \ge [n(n+1) + m(m+1)]/2$ and $(\tilde{A} + \tilde{\Sigma}_A', \tilde{B} + \tilde{\Sigma}_B')$ is controllable. For fixed $\mu_0 \in \RR^n$ and $\tilde{X}_0 \in \RR^{n(n+1)/2}$, 
the matrix $\bfD$ has full row rank, and consequently $\bfD \bfD^\tp$ is invertible, for almost all $[\nu_0^\tp~\cdots~\nu_{\ell-1}^\tp~\svect(\bar{U}_0)^\tp$ $\cdots$ $\svect(\bar{U}_{\ell-1}))^\tp]^\tp \in \RR^{\ell m (m+3)/2}$, where $\bar{U}_t$ is defined in line~2 of Algorithm~\ref{alg:A}.
\end{prop}

\begin{pf}
See Appendix \ref{appen:pf_thm:as_D_fullrank}.
\end{pf}

\begin{remark}
The controllability condition in Proposition~\ref{thm:as_D_fullrank} reflects the nature of the multiplicative noise (i.e., coupling between $\bar{A}_t$ and $x_t$, and that between $\bar{B}_t$ and $u_t$). The result indicates that a controllability condition on~\eqref{eq:vecCorrelationDynamics_simp} may be necessary to ensure successful identification. The lower bound for $\ell$ is necessary for the invertibility of $\bfD\bfD^\tp$, and is much smaller than that given in \cite{xing2020linear}. As in Algorithm~\ref{alg:A}, $\tilde{U}_t = \svect(\bar{U}_t + \nu_t\nu_t^\tp)$, so random generation of $\nu_t$ and $\bar{U}_t$ ensures $\bfD\bfD^\tp$ is invertible with probability one.
\end{remark}

We summarize the preceding two results in the following corollary.

\begin{corollary}\label{cor_fullrank}
Suppose that $\ell \ge [n(n+1) + m(m+1)]/2$, and both $(A, B)$ and $(\tilde{A} + \tilde{\Sigma}_A', \tilde{B} + \tilde{\Sigma}_B')$ are controllable. For fixed $\mu_0 \in \RR^n$ and $\tilde{X}_0 \in \RR^{n(n+1)/2}$, the matrices $\bfZ\bfZ^\tp$ and $\bfD\bfD^\tp$ are invertible, for almost all $[\nu_0^\tp~\cdots~\nu_{\ell-1}^\tp~\svect(\bar{U}_0)^\tp$ $\cdots$ $\svect(\bar{U}_{\ell-1}))^\tp]^\tp \in \RR^{\ell m (m+3)/2}$, where $\bar{U}_t$ is defined in line~2 of Algorithm~\ref{alg:A}.
\end{corollary}

\begin{remark}
The corollary implies that the existence of $(\bfZ\bfZ^\tp)^{-1}$ and $(\bfD\bfD^\tp)^{-1}$ can be guaranteed with probability one, as long as both $\nu_t$ and $\bar{U}_t$ are independently generated from distributions that is absolutely continuous with respect to Lebesgue measure. For example, the entries of $\nu_t$ are generated i.i.d. from a non-degenerate Gaussian distribution and then $\bar{U}_t$ is generated i.i.d. from a non-degenerate Wishart distribution, $0 \le t \le \ell-1$. 
\end{remark}

\subsubsection{Asymptotic Consistency} \label{subsubsec:consistency}
In this subsection, we assume that the expectations and covariance matrices of inputs have been generated as discussed in the previous section, and that both $\bfZ\bfZ^\tp$ and $\bfD\bfD^\tp$ have been designed to be invertible. The closed-form estimates generated by Algorithm \ref{alg:A} are
\begin{align}\label{LSEstimator1}
    \big[\hat{A}~ \hat{B}\big] &= \hat{\bfY} \hat{\bfZ}^\tp (\hat{\bfZ} \hat{\bfZ}^\tp)^{\dagger},\\\label{LSEstimator2}
    \Big[\hat{\tilde{\Sigma}}_A'~ \hat{\tilde{\Sigma}}_B'\Big] &= \hat{\bfC} \hat{\bfD}^\tp (\hat{\bfD} \hat{\bfD}^\tp)^{\dagger},
\end{align}
where
\begin{align}
    &\hat{\bfY}\label{def:yhat_zhat}
    \Let \big[
    \hat{\mu}_{\ell} ~ \cdots ~ \hat{\mu}_1
    \big],\quad
    \hat{\bfZ}
    \Let \begin{bmatrix}
    \hat{\mu}_{\ell-1} & \cdots & \hat{\mu}_0\\
    \nu_{\ell-1} & \cdots & \nu_0
    \end{bmatrix}, \\\label{def:chat_dhat}
    &\hat{\bfC} \Let
    \big[
    \hat{C}_\ell ~ \cdots ~ \hat{C}_1
    \big], \quad
    \hat{\bfD} \Let
    \begin{bmatrix}
    \hat{\tilde{X}}_{\ell-1} & \cdots & \hat{\tilde{X}}_0\\
    \tilde{U}_{\ell-1} & \cdots & \tilde{U}_0
    \end{bmatrix}, 
\end{align}
and $\hat{C}_t = \hat{\tilde{X}}_t - \big(\hat{\tilde{A}} \hat{\tilde{X}}_{t-1} + \hat{K}_{BA} \hat{W}_{t-1} + \hat{K}_{AB} \hat{W}_{t-1}' + \hat{\tilde{B}} \tilde{U}_{t-1}\big)$, $1 \le t \le \ell$. Here $\hat{\tilde{A}}$, $\hat{\tilde{B}}$, $\hat{K}_{AB}$, and $\hat{K}_{BA}$ are estimates of $\tilde{A}$, $\tilde{B}$, $K_{AB}$, and $K_{BA}$, obtained from $\hat{A}$ and $\hat{B}$ given by Algorithm \ref{alg:A}. The estimates depend on the number of rollouts $n_r$, which is omitted for convenience. For the convergence result, we present the following assumptions.

\begin{assumption}\label{asmp1}
For all rollouts indexed by $k \in [n_r]$, the below conditions hold. \\
(i) The rollout length is $\ell \geq [n(n+1) + m(m+1)]/2$.\\
(ii) The initial states $x_0^{(k)}, k \in [n_r]$, are i.i.d. subject to the same distribution $\mathcal{X}_0$ with finite second moment, and are independent of the multiplicative noise and inputs. \\
(iii) $\{\bar{A}_t^{(k)}, 0 \le t \le \ell, k \in [n_r]\}$ and $\{\bar{B}_t^{(k)}, 0 \le t \le \ell, k \in [n_r]\}$, are i.i.d. sequences respectively and are mutually independent, both with zero mean and finite second moments (i.e., $\EE\{\bar{A}_t^{(k)}\}$ and $\EE\{\bar{B}_t^{(k)}\}$ are zero matrices, and $\|\Sigma_A\|_2, \|\Sigma_B\|_2 < \infty$). \\
(iv) The parameters of inputs are given by lines~$1$-$3$ of Algorithm~\ref{alg:A}, and the inputs are generated, according to line~$7$ of~Algorithm~\ref{alg:A}. The inputs and noise are independent.\\
(v) Both $\bfZ\bfZ^\tp$ and $\bfD\bfD^\tp$ are invertible.
\end{assumption}

\begin{remark}
    From Corollary~\ref{cor_fullrank}, the lower bound of the rollout length in Assumption~\ref{asmp1}~(i) is necessary for estimating the noise covariance matrix, whereas, from Proposition~\ref{thm:as_Z_fullrank}, trajectories with length $\ell \ge n+m$ may be enough for estimating the nominal system matrix. The initial states of different trajectories need not start with the same value, but it is required that they have the same first and second moments (Assumption~\ref{asmp1}~(ii)). The mutual independence of noise at different time steps in one trajectory is a standard assumption (Assumption~\ref{asmp1}~(iii)), but the results in this paper still hold, if the noise sequence in the same trajectory is dependent, but the noise sequences in different trajectories are mutually independent and the noise has zero mean and the same second moment. The physical meaning of the independence between the noise and the inputs in Assumption~\ref{asmp1}~(iv) is that the former is an intrinsic part of the system and cannot be influenced by inputs. To keep the analysis concise, we separately discuss the input design (Section~\ref{sec:input_design}) and the performance of Algorithm~\ref{alg:A}. Assumption~\ref{asmp1}~(v) indicates that the input design yields invertible $\bfZ\bfZ^\tp$ and $\bfD\bfD^\tp$, but note that it implicitly assumes the controllability of the first- and second-moment dynamics of system states.
\end{remark}

Under Assumption \ref{asmp1} the rollouts $[x_0^{(k)}, \dots, x_l^{(k)}]$, $k \in [n_r]$, are i.i.d., so the following consistency result can be obtained from strong law of large numbers.

\begin{theorem}{(Consistency)}\label{thm:consistency}
Suppose that Assumption \ref{asmp1} holds, then the estimators \eqref{LSEstimator1}-\eqref{LSEstimator2} are asymptotically consistent, namely,
\begin{align*}
    \big[\hat{A}~ \hat{B}\big] \to [A~ B] \ \ \text{ and } \ \
    \Big[ \hat{\tilde{\Sigma}}_A'~ \hat{\tilde{\Sigma}}_B' \Big] \to \big[\tilde{\Sigma}_A'~ \tilde{\Sigma}_B'\big], 
\end{align*}
with probability one as the number of rollouts $n_r \to \infty$.
\end{theorem}

\begin{pf}
See Appendix \ref{appen:pf_thm:consistency}.
\end{pf}

\begin{remark}
This theorem indicates that consistency of Algorithm~\ref{alg:A} may hold even when the rollout length is relatively small. In~\cite{diconfidence}, the estimation of the first and second moments of multiplicative noise is decoupled, whereas here the estimate of $ [\tilde{\Sigma}_A'~ \tilde{\Sigma}_B']$ relies on $[\hat{A}~ \hat{B}]$. The coupling exists because here the noise covariance matrix, which from definition depends on the mean of the noise, is estimated. Note that $\ell$ is assumed to be fixed and we do not consider the case where $\ell \to \infty$, since an averaging step is used in Algorithm~\ref{alg:A}. Study of the case with increasing rollout length is left to future work. 
\end{remark}

\subsection{Finite-Sample Analysis}\label{subsec:finite}

This subsection studies finite-sample performance of Algorithm~\ref{alg:A}, demonstrating its non-asymptotic behavior. 
The existence of multiplicative noise complicates the analysis, so the following assumptions, ensuring that the system is bounded a.s., are introduced.

\begin{assumption}\label{asmp:bounded_system}
For all rollouts indexed by $k \in [n_r]$, the following conditions hold.\\
(i) The initial state is bounded a.s. for all $k \in [n_r]$ as
\begin{align*}
    \| x_0^{(k)} \| \leq c_X < \infty.
\end{align*}
(ii) The inputs are bounded a.s. for all $0 \le t \le \ell-1$ and $k \in [n_r]$ as
\begin{align*}
    \| u_t^{(k)} \| \leq c_U < \infty.
\end{align*}
(iii) The multiplicative noise, $\bar{A}_t^{(k)}$ and $\bar{B}_t^{(k)}$, is bounded a.s. for all $0 \le t \le \ell-1$ and $k \in [n_r]$ as
\begin{align*}
    \| \bar{A}_t^{(k)} \|_2 \le c_{\bar{A}} < \infty,~
    \| \bar{B}_t^{(k)} \|_2 \le c_{\bar{B}} < \infty.
\end{align*}
\end{assumption}
\begin{remark}
The assumption of bounded multiplicative noise is reasonable for physical systems, which cannot have infinite variations. For example, in interconnected systems, the noise represents randomly varying topologies of subsystems, and is naturally bounded. 
\end{remark}

Introduce the state- and input-deviation quantities
\begin{align*}
    e_t^{(k)} &\Let x_t^{(k)} - \EE[x_t^{(k)} ] = x_t^{(k)} - \mu_t, \\
    d_t^{(k)} &\Let u_t^{(k)} - \EE[u_t^{(k)} ] = u_t^{(k)} - \nu_t.
\end{align*}
The next proposition is a natural consequence of Assumption~\ref{asmp:bounded_system}.

\begin{prop}
Under Assumption~\ref{asmp:bounded_system}, the following results hold.\\
(i) The initial state-deviation is bounded a.s. for all rollouts $k \in [n_r]$ as
\begin{align*}
    \| e_0^{(k)} \| \leq c_\mu < \infty.
\end{align*}
(ii) The outer product initial state deviation is bounded a.s. for all rollouts $k \in [n_r]$ as
\begin{align*}
    \left\| \vect \left( x_{0}^{(k)} (x_{0}^{(k)})^\tp - \EE \left[ x_{0}^{(k)} (x_{0}^{(k)})^\tp \right] \right) \right\| \leq c_{\Delta X}.
\end{align*}
(iii) The input-deviations are bounded a.s. for all $0\le t \le \ell - 1$ and $k \in [n_r]$ as
\begin{align*}
    \| d_t^{(k)} \| \leq c_\nu < \infty.
\end{align*}
(iv) The Kronecker products of $\bar{A}_t^{(k)}$ and $\bar{B}_t^{(k)}$ are bounded a.s. for all $0 \le t \le \ell-1$ and $k \in [n_r]$ as
\begin{align*}
    \| \bar{A}_t^{(k)} \otimes \bar{A}_t^{(k)} - \Sigma_A^\prime \|_2 \leq c_{\Sigma_A^\prime}, \\
    \| \bar{B}_t^{(k)} \otimes \bar{B}_t^{(k)} - \Sigma_B^\prime \|_2 \leq c_{\Sigma_B^\prime}.
\end{align*}
\end{prop}

\begin{remark}
This proposition captures the deviations of random components of System~\eqref{theSystem} from their expectations. Using the bounds in Assumption~\ref{asmp:bounded_system} one could upper-bound these deviations, for instance,
\begin{align*}
    \left\| \vect \left( x_{0}^{(k)} (x_{0}^{(k)})^\tp - \EE \left[ x_{0}^{(k)} (x_{0}^{(k)})^\tp \right] \right) \right\|
    \leq \left\| x_{0}^{(k)} ({x_{0}^{(k)}})^\tp \right\|_F + \left\| \EE \big\{ x_{0}^{(k)} ({x_{0}^{(k)}})^\tp \big\} \right\|_F
    \leq 2 c_X^2,
\end{align*}
and
\begin{align*}
    \| \bar{A}_t^{(k)} \otimes \bar{A}_t^{(k)} - \Sigma_A^\prime \|_2 &\leq \| \bar{A}_t^{(k)} \otimes \bar{A}_t^{(k)} \|_2 + \|\Sigma_A^\prime \|_2 = \|\bar{A}_t^{(k)} \|_2^2 + \|\Sigma_A^\prime \|_2 \leq c_{\bar{A}}^2 + \|\Sigma_A^\prime \|_2, \\
    \| \bar{B}_t^{(k)} \otimes \bar{B}_t^{(k)} - \Sigma_B^\prime \|_2 &\leq \| \bar{B}_t^{(k)} \otimes \bar{B}_t^{(k)} \|_2 + \|\Sigma_B^\prime \|_2 = \|\bar{B}_t^{(k)} \|_2^2 + \|\Sigma_B^\prime \|_2 \leq c_{\bar{B}}^2 + \|\Sigma_B^\prime \|_2.
\end{align*}
However, these bounds may not depend on those in Assumption~\ref{asmp:bounded_system}. For example, when $x_0^{(k)}$ is a nonzero constant, $c_{\mu} = 0$ but $c_X$ is positive.
\end{remark}



The boundedness of the states and state-deviations follows from Assumptions~\ref{asmp1} and~\ref{asmp:bounded_system} according to the following statement.
\begin{lemma}\label{lem:bounded_system}
Suppose that Assumptions~\ref{asmp1} and~\ref{asmp:bounded_system} hold, then for all $k \in [n_r]$ and $0 \le t \le \ell$ we have that
\begin{align*}
    \big\| x_{t}^{(k)} \big\| \le c_M, \quad
    \big\| e_{t}^{(k)} \big\| \le c_N, \quad
    \big\| x_{t}^{(k)} (x_{t}^{(k)})^\tp \big\|_2 \le c_M^2, \quad
    \big\| e_{t}^{(k)} (e_{t}^{(k)})^\tp \big\|_2 \le c_N^2,
\end{align*}
and
\begin{align*}
    \left\| \vect \left( x_{t}^{(k)} (x_{t}^{(k)})^\tp - \EE \left\{ x_{t}^{(k)} (x_{t}^{(k)})^\tp \right\} \right) \right\| \le c_F, \quad 
    \left\| \vect \left( x_{t}^{(k)} (u_{t}^{(k)})^\tp - \EE \left\{ x_{t}^{(k)} (u_{t}^{(k)})^\tp \right\} \right) \right\| \le c_W,
\end{align*}
where
\begin{align*}
    c_M &\Let \max_{0 \leq t \leq \ell} \left\{ c_A^t c_X + \sum_{i=0}^{t-1} c_A^i c_B c_U \right\}, \\
    c_N &\Let \max_{0 \leq t \leq \ell} \left\{ \|A\|_2^t c_\mu + \sum_{i=0}^{t-1} \|A\|_2^i (\|B\|_2 c_\nu + c_{\bar{A}} c_M + c_{\bar{B}} c_U) \right\}, \\
    c_F &\Let \max_{0 \leq t \leq \ell} \left\{  (\| A \|_2^2 + \| \Sigma_A^\prime \|_2)^t c_{\Delta X}  +  \sum_{i=0}^{t-1} (\| A \|_2^2 + \| \Sigma_A^\prime \|_2)^i (c_{FX} + c_{FU} + c_{FXU}) \right\}, \\
    c_W &\Let c_N c_U + c_M c_\nu,
\end{align*}
and
\begin{align*}
    c_A &\Let \| A \|_2 + c_{\bar{A}}, \\
    c_B &\Let \| B \|_2 + c_{\bar{B}}, \\
    c_{FX} &\Let (2 \| A \|_2 c_{\bar{A}} + c_{\Sigma_A^\prime} ) c_M^2, \\
    c_{FU} &\Let 3 (\| B \|_2^2 + \| \Sigma_B^\prime \|_2) c_U c_{\nu} + (2 \| B \|_2 c_{\bar{B}} + c_{\Sigma_B^\prime}) c_U^2 , \\
    c_{FXU} &\Let 2 \| A \|_2 \| B \|_2 c_W  + ( 2 \|A\|_2 c_{\bar{B}} + 2 \|B\|_2 c_{\bar{A}} + 2 c_{\bar{A}} c_{\bar{B}}) c_M c_U.
\end{align*}
\end{lemma}

\begin{pf}
See Appendix~\ref{appen:pf_lem:bounded_system}.
\end{pf}

\begin{remark}
The quantity $c_M$ can be interpreted as a bound on the radius from the origin to the outer boundary of the set of reachable states from any valid $x_0$ over $\ell$ time steps.
If the system is not robustly stable in the sense that $c_A > 1$, then the limit as $\ell \to \infty$ of $c_M$ could be infinite. However, since we consider only finite-length rollouts, $c_M$ is finite regardless of the stability properties of the system. 

Analogous interpretations follow for the quantity $c_N$ and the reachable state-deviations. Notice that the constants $c_N$ grows with increasing maximum initial state and input deviations $c_\mu$ and $c_\nu$, and maximum noise magnitudes $c_{\bar{A}}$ and $c_{\bar{B}}$. Conversely, $c_N$ vanishes as those quantities become smaller, i.e. in the case that the initial state $x_0$ is a fixed deterministic value, the inputs $u_t$ follow a deterministic sequence, and there is no multiplicative noise.
Likewise, $c_F$ vanishes in such a scenario, so that $c_{\Delta X} = c_{FX} = c_{FU} = c_{FXU} = 0$.
\end{remark}

The following theorems state finite-sample results for the estimates of $[A~B]$ and $[\tilde{\Sigma}_A'~\tilde{\Sigma}_B']$, whose proofs are given in Appendices~\ref{append:pf_thm:hatAB_bounded} and~\ref{append:pf_thm:hatSigmaAB_bounded}, respectively.

\begin{theorem}\label{thm:hatAB_bounded}
Suppose that Assumptions~\ref{asmp1} and~\ref{asmp:bounded_system} hold. Fix a failure probability $\delta\in(0,1)$. It holds with probability at least $1 - \delta$ that
\begin{align*}
    \big\|
    \big[\hat{A} ~\hat{B}\big] - [A ~ B]
    \big\|_2 
    \leq
    \bigO \left( \sqrt{\frac{\ell \log(\ell/\delta)}{n_r}} \right).
\end{align*}
\end{theorem}

\begin{theorem}\label{thm:hatSigmaAB_bounded}
Under the same condition of Theorem~\ref{thm:hatAB_bounded}, with probability at least~$1 - \delta$, it holds that
\begin{align*}
    \left\|\Big[\hat{\tilde{\Sigma}}'_A~ \hat{\tilde{\Sigma}}'_B\Big] - \big[\tilde{\Sigma}'_A~ \tilde{\Sigma}'_B \big]\right\|_2
    \leq
    \bigO \left( \sqrt{\frac{\ell\log(\ell/\delta)}{n_r}} \right).
\end{align*}
\end{theorem}

\begin{remark}
In Theorems \ref{thm:hatAB_bounded} and \ref{thm:hatSigmaAB_bounded}, high-probability upper bounds are given for the estimates of $[A~B]$ and $[\tilde{\Sigma}'_A~ \tilde{\Sigma}'_B]$. It can be observed that these bounds shrink as $\bigO \left( 1/\sqrt{n_r} \right)$ with the number of rollouts, and converge to zero as the number of rollouts grows to infinity, indicating the consistency of the estimators. 
Note that the bounds are deterministic, although they depend on the failure probability $\delta$. 
The theorems also indicate that the probability of the estimation error exceeding an arbitrary positive constant decays exponentially fast with the number of rollouts, which is illustrated in Section~\ref{subsec:simul_consist}.

The $\bigO(\cdot)$ notation hides the coefficients of the error bounds, and the polynomial and exponential factors of $n$ and $m$ in the logarithm term. Their explicit forms are given in Appendices~\ref{append:pf_thm:hatAB_bounded} and~\ref{append:pf_thm:hatSigmaAB_bounded}, respectively. The coefficient of the estimation error of $[A~B]$ increases with $\|\bfY\|_2$, $\|\bfZ\|_2$, and the bound of the system, but decreases with the minimum eigenvalue of $\bfZ\bfZ^\tp$. Similarly, the coefficient of the estimation error of $[\tilde{\Sigma}'_A~ \tilde{\Sigma}'_B]$ decreases with the minimum eigenvalue of $\bfD\bfD^\tp$, but increases with $\|\bfC\|_2$, $\|\bfD\|_2$, and the bound of the system. 
It also increases with $\|A\|_2$, $\|B\|_2$, and quantities related to the second-moment dynamic of system states, because of the dependence of $[\hat{\tilde{\Sigma}}'_A~ \hat{\tilde{\Sigma}}'_B]$ on $[\hat{A} ~\hat{B}]$.
From definition, $\bfY$, $\bfZ$, $\bfC$, and $\bfD$ depend on system parameters and inputs, so proper input design could reduce the estimation error. It remains for future study how to design the moments of inputs so that the coefficients of the bounds can achieve their smallest values, and how to obtain data-dependent bounds, because the nominal system matrix is unknown. 

In \cite{diconfidence}, the authors study identification of System~\eqref{theSystem} from single-trajectory data, by developing error bounds for a least-squares algorithm, but it is unclear under what conditions of System~\eqref{theSystem} these error bounds converge to zero. In contrast, our analysis provides sufficient conditions under which the error bounds for estimates given by Algorithm~\ref{alg:A} vanish.
The results show that a relatively small rollout length is enough to guarantee consistency, but the current bounds imply that longer rollout length $\ell$ may lead to worse performance, which seems to be contrary to the intuition that longer trajectory provides more information. This could result from the averaging step which eliminates some excitation. Future work will consider how to use the data more efficiently.
\end{remark}


\section{Numerical Simulations}\label{sec:numericalSimulations}


In this section we empirically validate the theoretical results for Algorithm~\ref{alg:A}, and compare its performance with the recursive least-squares algorithm based on single-trajectory data \citep{chen2012identification,lai1982least}.

\subsection{Consistency and Finite-Sample Result}\label{subsec:simul_consist}

\begin{figure*}
\centering
\subfigure{
\includegraphics[width=0.4\linewidth]{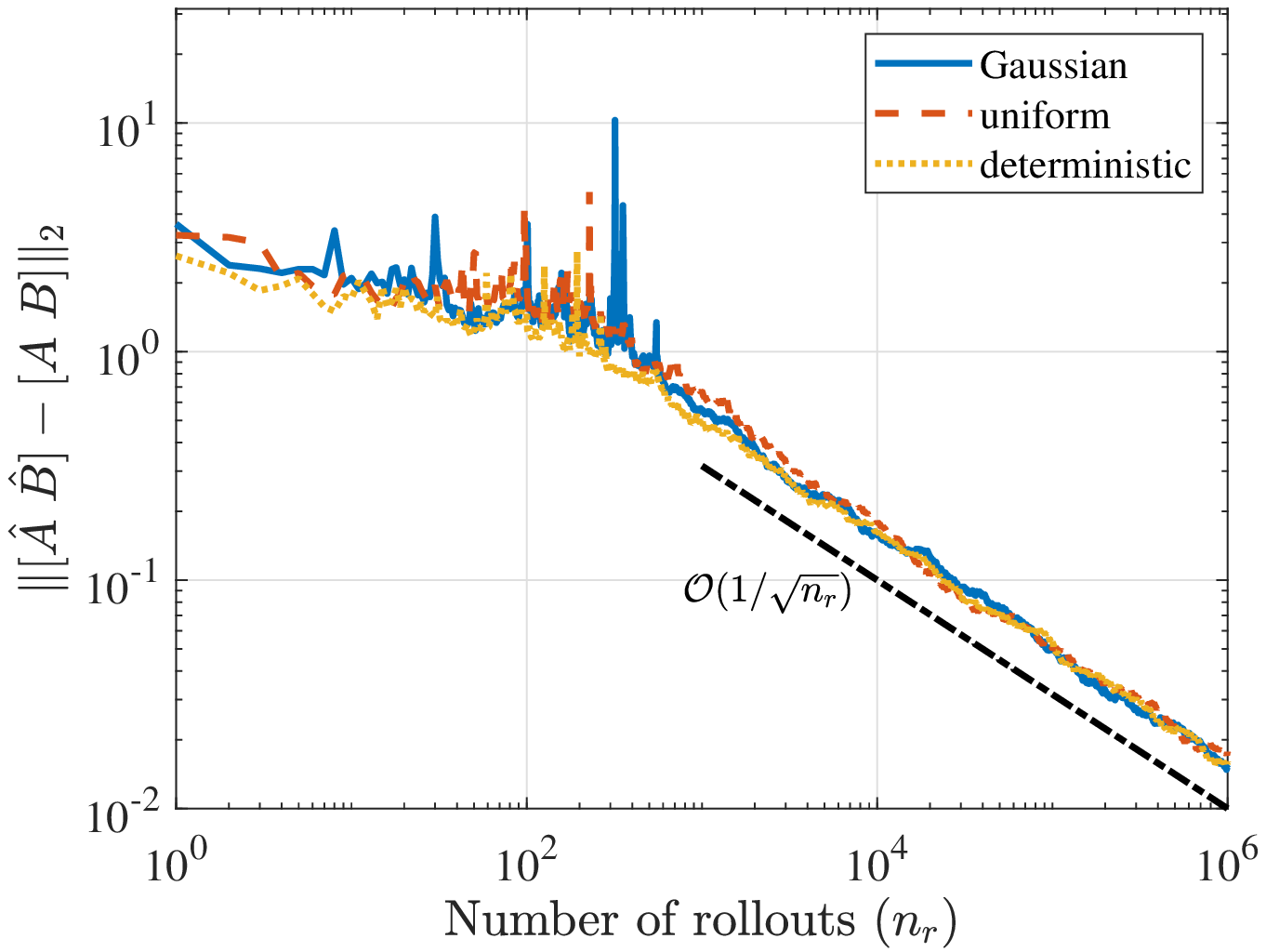}}\qquad
\subfigure{
\includegraphics[width=0.4\linewidth]{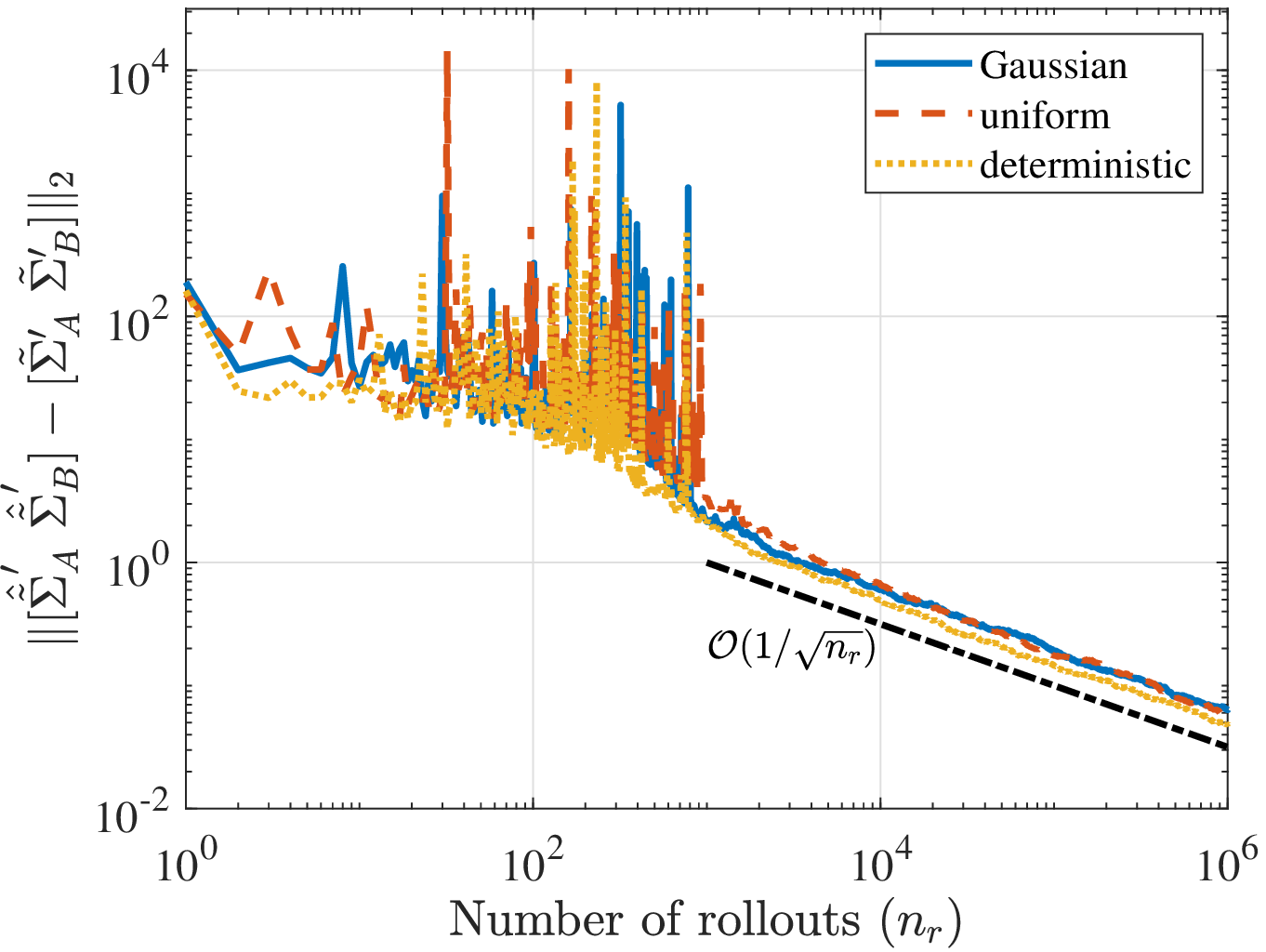}}
\caption{\label{fig:error_convergence}Consistency of Algorithm~\ref{alg:A}.}
\end{figure*}

This subsection considers identification of the $2$-dimensional system discussed in Example~\ref{exam:identifiability} with parameters 
\begin{align*}
    A  =
    \begin{bmatrix}
        1 & 0.2 \\
        0 & 1
    \end{bmatrix}, ~
    B = 
    \begin{bmatrix} 
        0.8 \\
        1
    \end{bmatrix}, ~ \Sigma_A = \frac{1}{40}
    \begin{bmatrix}
         8 & -2 & 0 &  0     \\
        -2 & 16 & 2 &  0     \\
         0 & 2  & 2 &  0     \\
         0 & 0  & 0 &  8   
    \end{bmatrix} ,~
    \Sigma_B = \frac{1}{40}
    \begin{bmatrix}
         5 &  -2 \\
        -2 &  20
    \end{bmatrix}.
\end{align*}
According to the reshaping operator $G$ defined in the notation section and the discussion in Example~\ref{exam:identifiability}, it holds that
\begin{align}\label{eq_simulation_example1_tildeSAB}
    \tilde{\Sigma}_A' = \frac{1}{40}
    \begin{bmatrix}
         8 & 0 & 2      \\
        -2 & 2 & 0     \\
         16 & 0 & 8  
    \end{bmatrix} ,~
    \tilde{\Sigma}_B' = \frac{1}{40}
    \begin{bmatrix}
         5 &  -2 & 20
    \end{bmatrix}^\tp.
\end{align}
A simulated experiment is conducted with rollout data of length $\ell = 4$. 
For $0\le t\le 3$, $\nu_t$ is generated independently from uniform distribution $\mathcal{U}([0,1])$ and then fixed. Three types of inputs are considered: Gaussian, uniform, and deterministic inputs. An identical sequence of input covariances, independently generated from $1$-dimensional Wishart distribution $W_p(0.1,1)$ and then fixed, is used in the former two cases. For the case of deterministic inputs, the covariances are set to be zero (i.e., $\bar{U}_t = 0$). In this setting $\bfD\bfD^\tp$ can be invertible because the second moment of the input at time $t$ satisfies that $U_t=\bar{U}_t+\nu_t\nu_t^\tp$, and the generation of $\nu_t$ provides randomness. For each case, Algorithm~\ref{alg:A} is run for $50$ times. The mean of estimation error in each case is shown in Fig.~\ref{fig:error_convergence}. It can be seen that Algorithm~\ref{alg:A} converges with convergence rate $\bigO(1/\sqrt{n_r})$, and performs similarly under all three types of inputs. The algorithm fluctuates when the number of rollouts is small, which may result from the averaging step.

\begin{figure*}
\centering
\includegraphics[width=0.35\linewidth]{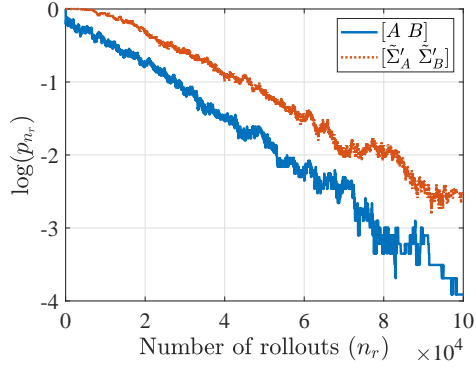}
\caption{\label{fig:expo_decay}Finite-sample result of Algorithm~\ref{alg:A}.}
\end{figure*}

Fig.~\ref{fig:expo_decay} provides the relative frequency of the normalized estimation errors, $\|[\hat{A}~\hat{B}] - [A~B]\|_2/\| [A~B]\|_2$ and $\|[\hat{\tilde{\Sigma}}_A^\prime~\hat{\tilde{\Sigma}}_B^\prime ]$ $- [\tilde{\Sigma}_A^\prime~\tilde{\Sigma}_B^\prime] \|_2/$ $\|[\tilde{\Sigma}_A^\prime~\tilde{\Sigma}_B^\prime] \|_2$, exceeding a given constant, under the uniform-input case. This result shows an exponential decay of the frequency and validates the finite-sample results. The relative frequency of $n_r$ rollouts is denoted by~$p_{n_r}$.

\begin{figure}
    \centering
    \includegraphics[width=0.45\linewidth]{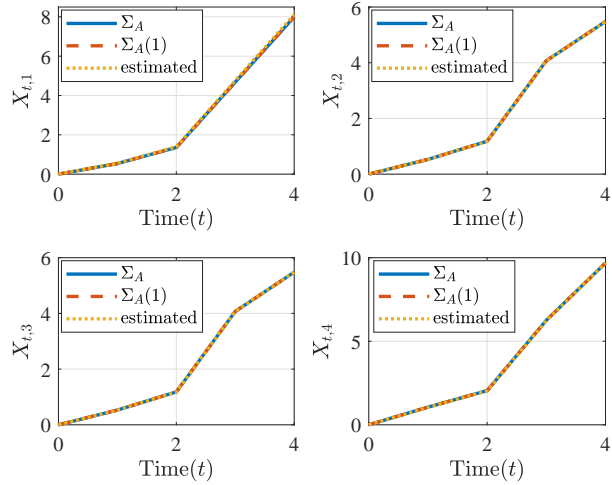}
    \caption{\label{fig:X_dynamics} The second-moment dynamic of system states defined by several noise covariance matrices.}
\end{figure}

From Remark~\ref{rmk:equiv_class_covar} and~\eqref{eq_equivalent_class}, it follows that~\eqref{eq_simulation_example1_tildeSAB} defines an equivalent class of covariance matrices that generates the same second-moment dynamic of system states. In the current example, $\Sigma_B$ is unique, but the following covariance matrix is equivalent to $\Sigma_A$,
\begin{align*}
    \Sigma_A(\alpha) = \frac{1}{40}
    \begin{bmatrix}
         8 & -2 & 0 &  1+\alpha     \\
        -2 & 16 & 1-\alpha &  0     \\
         0 & 1-\alpha  & 2 &  0     \\
         1+\alpha & 0  & 0 &  8   
    \end{bmatrix},
\end{align*}
with $\alpha \in \RR$ such that $\Sigma_A(\alpha) \succeq 0$. Fig.~\ref{fig:X_dynamics} illustrates the dynamic~\eqref{eq:vecCorrelationDynamics}, starting with the same initial condition $\mu_0 = \mathbf{0}_2$ and $X_0 = \mathbf{0}_4$, and with the noise covariance matrix given by $(\Sigma_A,\Sigma_B)$, $(\Sigma_A(1),\Sigma_B)$, and estimates from Algorithm~\ref{alg:A}, respectively. The parameters of inputs ($\nu_t$ and $\bar{U}_t$) are the same as the uniform-input case. Note that $\Sigma_A(-1) = \Sigma_A$, and $\Sigma_A(1) \succ 0$. It can be observed that the dynamics defined by $(\Sigma_A,\Sigma_B)$ and $(\Sigma_A(1),\Sigma_B)$ are identical, and the dynamic defined by the estimates from Algorithm~\ref{alg:A} is close to the former.

\begin{figure}
    \centering
    \includegraphics[width=0.4\linewidth]{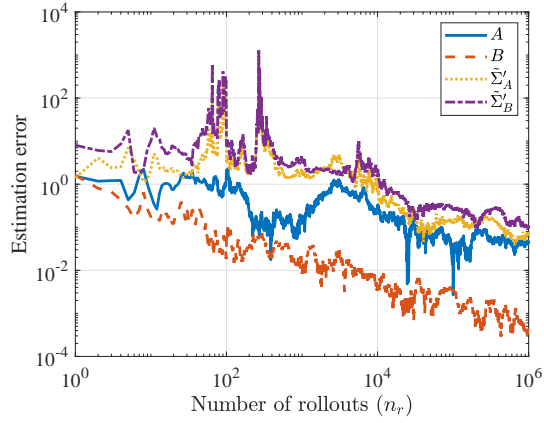}
    \caption{\label{fig:additive_noise}Consistency of Algorithm~\ref{alg:A} under both multiplicative and additive noise.}
\end{figure}

It is assumed that there is no additive noise in System~\eqref{theSystem}, but Algorithm~\ref{alg:A} can also be applied to identifying linear systems with both multiplicative and additive noise. If additive noise $w_t$, independent of the inputs and the multiplicative noise, exists, then write the system as
\begin{align}\nonumber
    x_{t+1} &= (A + \bar{A}_t) x_t + (B + \bar{B}_t) u_t + w_t \\\label{eq_additive_noise}
    &= (A + \bar{A}_t) x_t + \big[B+\bar{B}_t~w_t \big] 
    \begin{bmatrix}
    u_t \\ 1
    \end{bmatrix}.
\end{align}
In other words, $w_t$ can be considered as a part of multiplicative noise corresponding to a constant input equal to~one. Consider the above $2$-dimensional system with Gaussian noise $w_t \sim \mathcal{N}(\mathbf{0}_2, \sigma^2 I_2)$ and previously designed Gaussian inputs. Note that in this case $\ell = 6$ is needed because the dimension of inputs increases by one in~\eqref{eq_additive_noise}, compared with the original system. Fig.~\ref{fig:additive_noise} shows the consistency of Algorithm~\ref{alg:A} under the presence of additive noise.

\subsection{Performance Comparison}
The recursive form of the ordinary least-squares (OLS), namely, the recursive least-squares (RLS), is widely used in identification of dynamic systems \citep{chen2012identification,lai1982least}. It is possible to apply RLS to identify System~\eqref{theSystem} if certain conditions hold. Note that from System~\eqref{theSystem}, we have that
\begin{align*}
    x_{t+1} &= A x_t + B u_t + (\bar{A}_t x_t + \bar{B}_t u_t)\\
    &= A x_t + B u_t + w^{(1)}_t,
\end{align*}
where $w^{(1)}_t := \bar{A}_t x_t + \bar{B}_t u_t$ is considered to be noise. Under Assumption~\ref{asmp1}, $\{w^{(1)}_t, \mathcal{F}_t\}$ is a martingale difference sequence, i.e., $\EE\{w^{(1)}_t|\mathcal{F}_{t-1}\} = 0$, where $\mathcal{F}_t := \sigma(\bar{A}_k, \bar{B}_k, u_k, 0 \le k \le t)$. A mild condition for $w^{(1)}_t$ to ensure convergence of RLS in literature \citep{chen2012identification,lai1982least} is that $\sup_t \EE\{\|w^{(1)}_t\|^\beta|\mathcal{F}_{t-1}\} < \infty$ holds a.s. for some $\beta > 2$. However in our case $w^{(1)}_t$ is state-dependent, so certain stability assumption is needed to ensure this boundedness condition. This fact means that RLS could fail if the nominal part of System~\eqref{theSystem} is marginally stable ($\rho(A) = 1$) or unstable ($\rho(A) > 1$). In contrast, Algorithm~\ref{alg:A} can handle this situation with the help of multiple-trajectory data. Similarly, the noise covariance matrix of System~\eqref{theSystem} may be estimated using the following dynamic
\begin{align*}
    &P_1 \vect(x_{t+1} x_{t+1}^\tp) \\
    &=
    P_1 \big( (A+\bar{A}_t) \otimes (A + \bar{A}_t) \big) Q_1 P_1 \vect(x_t x_t^\tp) + P_1 \big( (B+\bar{B}_t) \otimes (B+\bar{B}_t) \big) Q_2 P_2 \vect(u_t u_t^\tp) \\
    &\quad + P_1 \big( (B+\bar{B}_t) \otimes (A+\bar{A}_t) \big) \vect(x_t u_t^\tp) + P_1 \big( (A + \bar{A}_t) \otimes (B+\bar{B}_t) \big) \vect(u_t x_t^\tp)\\
    &=
    P_1 \EE\big\{ (A+\bar{A}_t) \otimes (A + \bar{A}_t) \big\} Q_1 P_1 \vect(x_t x_t^\tp) + P_1 \EE\big\{ (B+\bar{B}_t) \otimes (B+\bar{B}_t) \big\} Q_2 P_2 \vect(u_t u_t^\tp) \\
    &\quad + P_1 (B\otimes A) \vect(x_t u_t^\tp)+ P_1 (A\otimes B) \vect(u_t x_t^\tp) +  w^{(2)}_t,
\end{align*}
where
\begin{align*}
    w^{(2)}_t &:= P_1 \big( (A+\bar{A}_t) \otimes (A + \bar{A}_t) - \EE \{ (A+\bar{A}_t) \otimes (A + \bar{A}_t) \} \big) Q_1 P_1 \vect(x_t x_t^\tp) \\
    &\quad + P_1 \big( (B+\bar{B}_t) \otimes (B+\bar{B}_t) - \EE \{ (B+\bar{B}_t) \otimes (B+\bar{B}_t) \}\big) Q_2 P_2 \vect(u_t u_t^\tp)\\
    &\quad + P_1 ( B \otimes \bar{A}_t + \bar{B}_t \otimes A + \bar{B}_t \otimes \bar{A}_t) \vect(x_t u_t^\tp) \\
    &\quad + P_1 ( A \otimes \bar{B}_t + \bar{A}_t \otimes B + \bar{A}_t \otimes \bar{B}_t) \vect(u_t x_t^\tp).
\end{align*}
It can be verified that, under Assumption~\ref{alg:A}, despite state-dependent, $\{w^{(2)}_t, \mathcal{F}_t\}$ is also a martingale difference sequence. To estimate the covariance matrix of the multiplicative noise, \cite{diconfidence} apply OLS, which is equivalent to RLS. Note that, when using OLS or RLS, one estimates the second moments of $A+\bar{A}_t$ and $B+\bar{B}_t$, rather than their covariance matrices, which are $\Sigma_A$ and $\Sigma_B$ in our context. The estimation of noise covariance is still coupled with the estimation of the nominal system, since $\Sigma_A^\prime = \EE \{ (A+\bar{A}_t) \otimes (A + \bar{A}_t) \} - A\otimes A$ and $\Sigma_B^\prime = \EE \{ (B+\bar{B}_t) \otimes (B + \bar{B}_t) \} - B\otimes B$.

\begin{figure}
    \centering
    \includegraphics[width=0.8\linewidth]{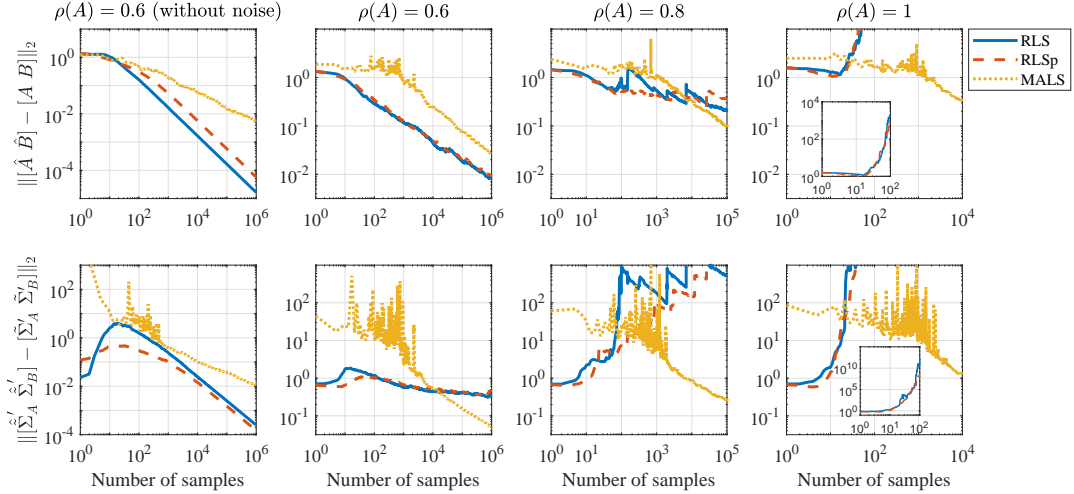}
    \caption{\label{fig:alg_comp}Performance comparison of RLS, RLSp, and Algorithm~\ref{alg:A}.}
\end{figure}

To compare the performance of RLS and Algorithm~\ref{alg:A}, we consider four systems. In the first case,  the nominal system matrices are
\begin{align*}
    A = \begin{bmatrix}
    0.6 & 0.2\\
    0 & 0.6
    \end{bmatrix},~B = \begin{bmatrix}
    0.8 \\ 1
    \end{bmatrix},
\end{align*}
and both $\Sigma_A$ and $\Sigma_B$ are zero matrices. That is, a linear system without noise and $\rho(A) = 0.6$, where $\rho(A)$ is the spectral radius of $A$. We use this case to show the consistency of RLS. In the other three cases, the matrix $A$ is set to be \[\begin{bmatrix}
0.6 & 0.2\\ 0 & 0.6
\end{bmatrix}, ~\begin{bmatrix}
0.8 & 0.2\\ 0 & 0.8
\end{bmatrix}, \text{ and } \begin{bmatrix}
1 & 0.2\\ 0 & 1
\end{bmatrix},\]
respectively. $B$ is the same as the first case, while $\Sigma_A$ and $\Sigma_B$ in Section~\ref{subsec:simul_consist} are adopted to be the covariance matrices. The implementation of Algorithm~\ref{alg:A} is the same as in Section~\ref{subsec:simul_consist}. That is, $\nu_t$ and $\bar{U}_t$ are randomly generated, and then fixed in all runs of the entire numerical experiment. The input $u_t$ at time $0\le t\le \ell-1$ in each rollout is generated from Gaussian distribution $\mathcal{N}(\nu_t, \bar{U}_t)$, and $\ell=4$. Since RLS is based on single-trajectory data, the length of the trajectory is set to be $\ell n_r$, so that the number of samples that RLS uses is the same as that of Algorithm~\ref{alg:A}. RLS with independent standard Gaussian inputs is considered as a baseline. In order to rule out the effect of different input design, we also run RLS with periodic inputs (RLSp) satisfying that, in each period, the inputs are generated in the same way as those in a rollout of Algorithm~\ref{alg:A}. 

For each system, the three algorithms, RLS, RLSp, and Algorithm~\ref{alg:A} are run for $50$ times, respectively. The mean of estimation error in each case is presented in Fig.~\ref{fig:alg_comp}. It can be observed that RLS and RLSp perform similarly in all cases. When multiplicative noise is absent, they converge slightly faster than Algorithm~\ref{alg:A}. They are also a little better than Algorithm~\ref{alg:A}, in the case $\rho(A) = 0.6$ with noise, for the estimation of $[A~B]$, indicating OLS could be applied to Algorithm~\ref{alg:A} as a way to estimate $[A~B]$. However, Algorithm~\ref{alg:A} surpasses RLS and RLSp when identifying the noise covariance matrix. Moreover, the performance of RLS gets worse as $\rho(A)$ grows. Interestingly, in the case of $\rho(A) = 0.8$, although the nominal system is stable, the second-moment dynamic of system states is not. This instability leads to degraded performance of RLS estimating $[A~B]$ and divergence of RLS estimating the covariance matrix. In the marginally stable case, namely $\rho(A) = 1$, RLS and RLSp explode in finite time. In contrast, Algorithm~\ref{alg:A} behaves almost identically for all cases (the consistency of Algorithm~\ref{alg:A} in the marginally stable case is shown in Fig.~\ref{fig:error_convergence}). To sum up, Algorithm~\ref{alg:A} can deal with the estimation of noise covariance matrix better and relies less on the stability of both the nominal system and the second-moment dynamic of system states.

\section{Conclusion and Future Work}\label{conclusions}
In this paper an identification algorithm based on multiple-trajectory data was proposed for linear systems with multiplicative noise. 
With appropriately designed exciting inputs, the proposed algorithm is able to jointly estimate the nominal system and the multiplicative noise covariance.
The asymptotic and non-asymptotic performance of the algorithm was analyzed theoretically, and illustrated by numerical experiments.
Future work include studying more efficient algorithms that can be used in online settings, optimal and adaptive input design, sparsity-promoting regularization for identification of networked systems, end-to-end finite-sample performance guarantees for identification-based optimal control, and applications to identification of cyber-physical systems with coupling between noise and inputs. 

\clearpage
\appendix

\textbf{Appendix}

\section{Proof of Proposition~\ref{thm:as_Z_fullrank}}\label{appen:pf_thm:as_Z_fullrank}    

We begin with a standard result from real analysis  \citep{caron2005zero,federer2014geometric} regarding the zero set of a polynomial.
\begin{lemma}\label{lem:polynomial}
A polynomial function $\RR^n$ to $\RR$ is either identically $0$ or non-zero almost everywhere.
\end{lemma}

It suffices to consider the case with $\ell = n+m$, implying $\bfZ$ is a square matrix. Note that when $(A,B)$ and $\mu_0$ are fixed, $|\bfZ|$ is a polynomial of $\nu \Let [\nu_0^\tp~\cdots~\nu_{\ell-1}^\tp]^\tp \in \RR^{m\ell}$. Hence the existence of a vector in $\RR^{m\ell}$ such that $|\bfZ| \not= 0$ implies that $|\bfZ| \not= 0$ almost everywhere from Lemma~\ref{lem:polynomial}.

First we verify the conclusion for $\mu_0 = \mathbf{0}_n$. It follows from the definition of controllability and the assumption that $[A^{n-1}B~A^{n-2}B~\cdots~B]$ has full row rank. Without loss of generality, let $B_1$, $AB_1$, $\dots$, $A^{r_1}B_1$, $B_2$, $\dots$, $A^{r_2}B_2$, $\dots$, $B_p$, $\dots$, $A^{r_p}B_p$ be a basis of $\RR^n$, where $B_i$ is the $i$-th column of $B$, $1\le p \le m$, $0\le r_i \le n-1$, $1\le i\le p$, and $p + \sum_{i=1}^p r_i = n$. Moreover, $A^k B_i$, $k > r_i$ and $1\le i \le p$, can be written as a linear combination of $B_1$, $\dots$, $A^{r_1}B_1$, $\cdots$, $A^{r_i} B_i$. 

Let $\tilde{\nu} \Let [\tilde{\nu}_0^\tp~\cdots~\tilde{\nu}_{n-1}^\tp]^\tp$ be such that $\tilde{\nu}_{0} = \bfe_1^m$, $\tilde{\nu}_{q_1} = \bfe_2^m$, $\dots$, $\tilde{\nu}_{q_k} = \bfe_{k+1}^m$, $\dots$, $\tilde{\nu}_{q_{p-1}} = \bfe_p^m$ (or any nonzero multiplier of respective unit vectors), where $q_k = k + \sum_{i=1}^k r_i$, $1\le k \le p$, and $\tilde{\nu}_{i} = \mathbf{0}_m$ for other $1\le i \le n-1$. Then for $\tilde{\mu}_{t+1} = A \tilde{\mu}_t + B \tilde{\nu}_t$, $0\le t \le n-1$, and $\tilde{\mu}_0 = \mathbf{0}_n$, it holds that
\begin{align*}
    \tilde{\mu}_t &= A^{t-1} B_1, &~1\le t \le q_1,\\
    \tilde{\mu}_t &= A^{t-q_1-1} B_2 + A^{t-1} B_1, &~q_1+1 \le t \le q_1+q_2,\\
    &\vdots  \\
    \tilde{\mu}_t &= A^{t-q_{p-1}-1} B_p + \cdots + A^{t-q_1-1} B_2 + A^{t-1} B_1, &~q_{p-1}+1 \le t \le n.
\end{align*}
The definitions of $B_1$, $\dots$, $A^{r_p}B_p$, and $q_k$ imply that $\tilde{\mu}_1$, $\dots$, $\tilde{\mu}_n$ are linearly independent. Hence it is shown that there exists a vector $[\nu_0^\tp~\cdots~\nu_{n-1}^\tp]^\tp$ such that $[\mu_1~\cdots~\mu_n]$ has full rank. If $m=1$, let $\nu = [\tilde{\nu}^\tp~0]^\tp$ and
\begin{align*}
    \bfZ = \begin{bmatrix}
    \tilde{\mu}_n & \cdots & \tilde{\mu}_1 & \mathbf{0}_n \\
    0 & \cdots & 0 & 1
    \end{bmatrix}
\end{align*}
has full rank. In the case of $m\ge 2$, fix $\nu_n$ to be zero. Set $\nu_{n+1} = c_1 \bfe_2^m$, $c_1 \in \RR$, and
\begin{align*}
    \left| 
    \begin{array}{cccccccc}
    \mu_{n+1} & \tilde{\mu}_n & \cdots & \cdots & \tilde{\mu}_{q_1} & \cdots & \cdots & \mathbf{0}_n\\
    0 & 0 & \cdots & \cdots & 0 & \cdots & 0 & 1 \\
    c_1 & 0 & \cdots & 0 & d_1 & 0 & \cdots & 0
    \end{array}
    \right| = c_1 \left|
    \begin{array}{cccc}
    \tilde{\mu}_n & \cdots & \tilde{\mu}_1 & \mathbf{0}_n\\
    0 & \cdots & 0 & 1 
    \end{array}
    \right| + (-1)^{2n+4-q_1} d_1 \left |
    \begin{array}{cccccccc}
    \mu_{n+1} & \tilde{\mu}_n & \cdots & \tilde{\mu}_{q_1+1} & \tilde{\mu}_{q_1-1} & \cdots & \mathbf{0}_n\\
    0 & 0 & \cdots & \cdots & \cdots & 0 & 1 
    \end{array}
    \right|,
\end{align*}
where $d_1 = 1$ if $p>1$ and $d_1=0$ if $p=1$. Hence there must exist $\tilde{c}_1 \in \RR$ such that the above determinant is nonzero. Inductively, set $\nu_{n+k} = c_k \bfe_{k+1}^m$, $2\le k \le m-1$, and it is able to find $\tilde{c}_k$, $2\le k \le m-1$, (consequently $\tilde{\nu}_0$, $\dots$, $\tilde{\nu}_{n+m-1}$) such that $|\bfZ| \not= 0$.
Now suppose that the system starts with $\mu_0 \not = \mathbf{0}_n$. Since $[A^{n-1}B~A^{n-2}B~\cdots~B]$ has full row rank, so does the matrix $[A^{n-1}B$ $A^{n-2}B$ $~\cdots~B~A^{n-1}\mu_0~\cdots~\mu_0]$. Thus, without loss of generality, assume that $\mu_0$, $A\mu_0$, $\dots$, $A^{r_0}\mu_0$, $B_1$, $\dots$, $A^{r_1}B_1$, $\dots$, $B_p$, $\dots$, $A^{r_p}B_p$ is a basis of $\RR^n$, where $0\le p\le m$ ($p=0$ means that there is no $B_i$), $0\le r_i\le n-1$, $0\le i\le p$, and $p + 1 + \sum_{i=0}^p r_i = n$. Moreover, $r_i \le n-2$ for all $1 \le i \le p$. It can be verified that there exists a vector $[\nu_0^\tp~\cdots~\nu_{n-1}^\tp]^\tp$ such that $[\mu_0~\cdots~\mu_{n-1}]$ has full rank. Therefore in a similar way the conclusion can be obtained.

\begin{remark}
From the proof we know that even if $(A,B)$ is not controllable, $\bfZ$ can still have full row rank as long as $[A^{n-1}B$ $A^{n-2}B$ $~\cdots~B~A^{n-1}\mu_0~\cdots~\mu_0]$ has full row rank for some $\mu_0 \in \RR^n$.
\end{remark}

\section{Proof of Proposition~\ref{thm:as_D_fullrank}}\label{appen:pf_thm:as_D_fullrank}
Write \eqref{eq:vecCorrelationDynamics_simp} as 
\begin{align*}
    \tilde{X}_{t+1} &= (\tilde{A} + \tilde{\Sigma}_A') \tilde{X}_t + (\tilde{B} + \tilde{\Sigma}_B') \tilde{U}_t  + (K_{BA} W_t + K_{AB} W_t')\\
    &\Let \breve{A} \tilde{X}_t + \breve{B} \tilde{U}_t + \eta_t. 
\end{align*}
Note that after setting $\nu_t = 0$ for all $0\le t \le \ell-1$, $\tilde{U}_t = \svect(\bar{U}_t)$ and the above dynamic becomes
\begin{align*}
    \tilde{X}_{t+1} = \breve{A} \tilde{X}_t + \breve{B} \tilde{U}_t.
\end{align*}
Similar to the proof of Proposition~\ref{thm:as_Z_fullrank}, first examine if $\tilde{X}_0$, $\breve{A}\tilde{X}_0$, $\dots$, $\breve{A}^{n(n+1)/2 - 1}\tilde{X}_0$ are linearly independent. If not, select some columns of $[\breve{A}^{n(n+1)/2 - 1}\breve{B}, \dots, \breve{B}]$ to together form a basis of $\RR^{n(n+1)/2}$. Considering $\tilde{U}_t$ as an input makes the rest of the proof essentially the same as the proof of Proposition~\ref{thm:as_Z_fullrank}.

\section{Proof of Theorem \ref{thm:consistency}}\label{appen:pf_thm:consistency}
Consider each rollout $[(x_0^{(k)})^\tp, \dots, (x_\ell^{(k)})^\tp]^\tp$ as an independent sample of the random vector $\mathbf{x}_{\ell} \Let [x_0^\tp, \dots, x_\ell^\tp]^\tp$, and from Assumption \ref{asmp1} (ii) and (iii) we know that the random vector $\mathbf{x}_{\ell}$ has finite first and second moments.
So it follows from the Kolmogorov's strong law of large numbers that $\hat{\bfY} \to \bfY$ a.s., and similarly $\hat{\bfZ} \to \bfZ$ a.s., as $n_r \to \infty$. Hence $\hat{\bfY}\hat{\bfZ}^\tp \to \bfY\bfZ^\tp$ and $\hat{\bfZ}\hat{\bfZ}^\tp \to \bfZ\bfZ^\tp$ a.s. From the assumption that $\bfZ \bfZ^\tp$ is invertible and the continuous mapping theorem (Theorem 2.3 of \citep{van2000asymptotic}), it can be obtained that as $n_r \to \infty$
\begin{align*}
    (\hat{\bfZ} \hat{\bfZ}^\tp)^{-1} \to (\bfZ \bfZ^\tp)^{-1}, \text{ a.s.}
\end{align*}
When $(\hat{\bfZ} \hat{\bfZ}^\tp)^{-1}$ does not exist, in Algorithm~\ref{alg:A} we replace it by $(\hat{\bfZ} \hat{\bfZ}^\tp)^\dagger$. Thus $(\hat{A},\hat{B}) \to (A, B)$.
Combining the above convergence with the Kolmogorov's strong law of large numbers, the convergence of $\hat{\bfC}$ and $\hat{\bfD}$ follows. Therefore, applying the continuous mapping theorem again, we obtain the consistency of the estimator $(\hat{\tilde{\Sigma}}_A', \hat{\tilde{\Sigma}}_B')$.

\clearpage
\section{Proof of Lemma~\ref{lem:bounded_system}}\label{appen:pf_lem:bounded_system}
For the first claim, regarding states, taking the norm of both sides of System~\eqref{theSystem}, at time step $t$ we have
\begin{align}
    \big\| x_{t+1}^{(k)} \big\| &= \big\| (A  + \bar{A}_t^{(k)}) x_t^{(k)} +  (B + \bar{B}_t^{(k)}) u_t^{(k)} \big\|.
\end{align}
Using the triangle inequality and the fact that the spectral norm is compatible with the Euclidean norm, we have
\begin{align*}
    \big\| x_{t+1}^{(k)} \big\| 
    &\leq \big\| (A  + \bar{A}_t^{(k)}) x_t^{(k)} \big\| + \big\| (B + \bar{B}_t^{(k)}) u_t^{(k)} \big\| \\
    &\leq \big\| A  + \bar{A}_t^{(k)} \big\|_2 \big\| x_t^{(k)} \big\| + \big\| B + \bar{B}_t^{(k)} \big\|_2 \big\| u_t^{(k)} \big\| .
\end{align*}
Using the triangle inequality and Assumption~\ref{asmp:bounded_system}~(iii) we have 
\begin{align*}
    \big\| A + \bar{A}_t^{(k)} \big\|_2 \leq \max_{0 \le t \le \ell} \big\| A + \bar{A}_t^{(k)} \big\|_2 \leq \| A \|_2 + \max_{0 \le t \le \ell} \big\| \bar{A}_t^{(k)} \big\|_2 &\teL c_A, \\
    \big\| B + \bar{B}_t^{(k)} \big\|_2 \leq \max_{0 \le t \le \ell} \big\| B + \bar{B}_t^{(k)} \big\|_2 \leq \| B \|_2 + \max_{0 \le t \le \ell} \big\| \bar{B}_t^{(k)} \big\|_2 &\teL c_B,
\end{align*}
so from Assumption~\ref{asmp:bounded_system}~(ii)
\begin{align} \label{eq:state_norm_bound}
    \big\| x_{t+1}^{(k)} \big\| 
    &\leq c_A \big\| x_t^{(k)} \big\| + c_B c_U .
\end{align}
For the base case when $t=0$, we have by Assumption \ref{asmp:bounded_system}~(i) that $\big\| x_{0}^{(k)} \big\| \leq c_X$.
Applying \eqref{eq:state_norm_bound} inductively with the base case proves the first claim.

For the second claim, regarding the state-deviations, we have
\begin{align*}
    e_{t+1}^{(k)} &=
    x_{t+1}^{(k)} - \EE \big\{ x_{t+1}^{(k)} \big\}  \\
    &=
    \big(A + \bar{A}_{t}^{(k)}\big) x_{t}^{(k)} + \big(B + \bar{B}_{t}^{(k)}\big) u_{t}^{(k)} - \EE\big\{ \big(A + \bar{A}_{t}^{(k)}\big) x_{t}^{(k)} + \big(B + \bar{B}_{t}^{(k)}\big) u_{t}^{(k)} \big\} \\
    &=
    \big(A + \bar{A}_{t}^{(k)}\big) x_{t}^{(k)} + \big(B + \bar{B}_{t}^{(k)}\big) u_{t}^{(k)} - A \EE \big\{ x_{t}^{(k)} \big\} - B \EE\big\{ u_{t}^{(k)} \big\} \\
    &=
    A \big(x_{t}^{(k)} - \EE \big\{ x_{t}^{(k)} \big\} \big) + B \big( u_{t}^{(k)} - \EE\big\{ u_{t}^{(k)} \big\} \big) + \bar{A}_{t}^{(k)} x_{t}^{(k)} + \bar{B}_{t}^{(k)} u_{t}^{(k)} \\
    &=
    A e_t^{(k)} + B d_t^{(k)} + \bar{A}_{t}^{(k)} x_{t}^{(k)} + \bar{B}_{t}^{(k)} u_{t}^{(k)}.
\end{align*}
Taking the norm of both sides, using submultiplicativity and triangle inequality, we have
\begin{align}
    \| e_{t+1}^{(k)} \| 
    &=
    \| A e_t^{(k)} + B d_t^{(k)} + \bar{A}_{t}^{(k)} x_{t}^{(k)} + \bar{B}_{t}^{(k)} u_{t}^{(k)} \| \nonumber \\
    &\leq
    \| A \|_2 \| e_t^{(k)} \| + \| B \|_2 \| d_t^{(k)} \| + \| \bar{A}_{t}^{(k)} \|_2 \| x_{t}^{(k)} \| + \| \bar{B}_{t}^{(k)} \|_2 \| u_{t}^{(k)} \| \nonumber \\
    &\leq
    \| A \|_2 \| e_t^{(k)} \| + \| B \|_2 c_\nu + c_{\bar{A}} c_M + c_{\bar{B}} c_U, \label{eq:state_deviation_norm_bound}
\end{align}
where the final inequality follows from the first part of Lemma~\ref{lem:bounded_system} and Assumptions \ref{asmp:bounded_system}~(ii)~and~(iii).
For the base case when $t=0$, we have by Assumption \ref{asmp:bounded_system}~(i) that $\big\| e_{0}^{(k)} \big\| \leq c_\mu$.
Applying \eqref{eq:state_deviation_norm_bound} inductively with the base case proves the second claim.

By the definition of the spectral norm and the first and second claims we have
\begin{align*}
    \big\| x_{t}^{(k)} {x_{t}^{(k)}}^\tp \big\|_2 &= \big\| x_{t}^{(k)} \big\|_2^2 \leq c_M^2, \\
    \big\| e_{t}^{(k)} {e_{t}^{(k)}}^\tp \big\|_2 &= \big\| e_{t}^{(k)} \big\|_2^2 \leq c_N^2,
\end{align*}
proving the third and fourth claims.

For the fifth and sixth claims, define the quantities
\begin{align*}
    \Delta X_t \Let  \vect\left( x_{t}^{(k)} ({x_{t}^{(k)}})^\tp - \EE \big\{ x_{t}^{(k)} ({x_{t}^{(k)}})^\tp \big\} \right) = \vect\left( x_{t}^{(k)} ({x_{t}^{(k)}})^\tp \right) - X_t, \\
    \Delta U_t \Let  \vect\left( u_{t}^{(k)} ({u_{t}^{(k)}})^\tp - \EE \big\{ u_{t}^{(k)} ({u_{t}^{(k)}})^\tp \big\} \right) = \vect\left( u_{t}^{(k)} ({u_{t}^{(k)}})^\tp \right) - U_t, \\
    \Delta W_t \Let  \vect\left( x_{t}^{(k)} ({u_{t}^{(k)}})^\tp - \EE \big\{ x_{t}^{(k)} ({u_{t}^{(k)}})^\tp \big\} \right) = \vect\left( x_{t}^{(k)} ({u_{t}^{(k)}})^\tp \right) - W_t.
\end{align*}
We can bound $\| \Delta U_t \|$ as
\begin{align*}
    \| \Delta U_t \|
    &= \left\| \vect\left( u_{t}^{(k)} ({u_{t}^{(k)}})^\tp - \EE \big\{ u_{t}^{(k)} ({u_{t}^{(k)}})^\tp \big\} \right) \right\| \\
    &= \left\| \vect\left( u_{t}^{(k)} \big({u_{t}^{(k)}} - \EE\{u_t^{(k)}\}\big)^\tp + \big(u_t^{(k)} - \EE\{u_t^{(k)}\}\big) \EE\{u_t^{(k)}\}^\tp - \EE \big\{ u_{t}^{(k)} \big({u_{t}^{(k)}} - \EE{u_t^{(k)}}\big)^\tp \big\} \right) \right\| \\
    &= \left\|  u_{t}^{(k)} \big({u_{t}^{(k)}} - \EE\{u_t^{(k)}\}\big)^\tp + \big(u_t^{(k)} - \EE\{u_t^{(k)}\}\big) \EE\{u_t^{(k)}\}^\tp - \EE \big\{ u_{t}^{(k)} \big({u_{t}^{(k)}} - \EE{u_t^{(k)}}\big)^\tp \big\} \right\|_F \\
    &\leq \left\| u_{t}^{(k)} \big({u_{t}^{(k)}} - \EE\{u_t^{(k)}\}\big)^\tp \right\|_F + \left\| \big(u_t^{(k)} - \EE\{u_t^{(k)}\}\big) \EE\{u_t^{(k)}\}^\tp \right\|_F + \EE \left\{ \left\| u_{t}^{(k)} \big({u_{t}^{(k)}} - \EE{u_t^{(k)}}\big)^\tp \right\|_F \right\}\\
    &\leq 3 c_U c_{\nu}.
\end{align*}
For the sixth claim, we can bound $\| \Delta W_t \|$ as
\begin{align*}
    \| \Delta W_t \|
    &=\left\| \vect\left( x_{t}^{(k)} ({u_{t}^{(k)}})^\tp - \EE \big\{ x_{t}^{(k)} ({u_{t}^{(k)}})^\tp \big\} \right) \right\| \\
    &=
    \left\| \vect\left( \big( x_{t}^{(k)} - \EE \big\{ x_{t}^{(k)} \big\} \big) ({u_{t}^{(k)}})^\tp + \EE \big\{x_{t}^{(k)} \big\} \big(u_t^{(k)} - \EE \big\{ u_t^{(k)} \big\} \big)^\tp \right) \right\| \\
    &=
    \Big\| \big( x_{t}^{(k)} - \EE \big\{ x_{t}^{(k)} \big\} \big) ({u_{t}^{(k)}})^\tp + \EE \big\{x_{t}^{(k)} \big\} \big(u_t^{(k)} - \EE \big\{ u_t^{(k)} \big\} \big)^\tp \Big\|_F\\
    &\le
    \big\|  x_{t}^{(k)} - \EE \big\{ x_{t}^{(k)} \big\} \big\| \big\| {u_{t}^{(k)}} \big\| + \big\| \EE \big\{x_{t}^{(k)} \big\} \big\| \big\| u_t^{(k)} - \EE \big\{ u_t^{(k)}\big\} \big\| \\
    &\leq
    c_N c_U + c_M c_\nu \teL c_W.
\end{align*}
For the fifth claim, substituting the dynamics and expanding the products we have
\begin{align}\nonumber
    \Delta X_{t+1} &= \vect\left( x_{t+1}^{(k)} ({x_{t+1}^{(k)}})^\tp - \EE \big\{ x_{t+1}^{(k)} ({x_{t+1}^{(k)}})^\tp \big\} \right)\\\nonumber
    &=
    \vect\left( \left[ (A + \bar{A}_t^{(k)}) x_t^{(k)} + (B + \bar{B}_t^{(k)}) u_t^{(k)} \right] \left[ (A + \bar{A}_t^{(k)}) x_t^{(k)} + (B + \bar{B}_t^{(k)}) u_t^{(k)} \right]^\tp \right. \\\nonumber
    & \quad \left. - \EE \left\{ \left[ (A + \bar{A}_t^{(k)}) x_t^{(k)} + (B + \bar{B}_t^{(k)}) u_t^{(k)} \right] \left[ (A + \bar{A}_t^{(k)}) x_t^{(k)} + (B + \bar{B}_t^{(k)}) u_t^{(k)} \right]^\tp \right\} \right)  \\\nonumber
    &= 
    \vect \bigg( (A + \bar{A}_t^{(k)}) x_t^{(k)} ({x_t^{(k)}})^\tp (A + \bar{A}_t^{(k)})^\tp 
    + (A + \bar{A}_t^{(k)}) x_t^{(k)} ({u_t^{(k)}})^\tp (B + \bar{B}_t^{(k)})^\tp \\\nonumber
    &\quad + (B + \bar{B}_t^{(k)}) u_t^{(k)} ({x_t^{(k)}})^\tp (A + \bar{A}_t^{(k)})^\tp 
    + (B + \bar{B}_t^{(k)}) u_t^{(k)} ({u_t^{(k)}})^\tp (B + \bar{B}_t^{(k)})^\tp \\\nonumber
    & - \EE \big\{(A + \bar{A}_t^{(k)}) x_t^{(k)} ({x_t^{(k)}})^\tp (A + \bar{A}_t^{(k)})^\tp 
    + (A + \bar{A}_t^{(k)}) x_t^{(k)} ({u_t^{(k)}})^\tp (B + \bar{B}_t^{(k)})^\tp \\\nonumber
    &\quad + (B + \bar{B}_t^{(k)}) u_t^{(k)} ({x_t^{(k)}})^\tp (A + \bar{A}_t^{(k)})^\tp 
    + (B + \bar{B}_t^{(k)}) u_t^{(k)} ({u_t^{(k)}})^\tp (B + \bar{B}_t^{(k)})^\tp\big\} \bigg) \\\nonumber
    &=
    \vect \bigg( (A + \bar{A}_t^{(k)}) x_t^{(k)} ({x_t^{(k)}})^\tp (A + \bar{A}_t^{(k)})^\tp - \EE \big\{ (A + \bar{A}_t^{(k)}) x_t^{(k)} ({x_t^{(k)}})^\tp (A + \bar{A}_t^{(k)})^\tp \big\} \\\nonumber
    &      \quad (A + \bar{A}_t^{(k)}) x_t^{(k)} ({u_t^{(k)}})^\tp (B + \bar{B}_t^{(k)})^\tp - \EE \big\{ (A + \bar{A}_t^{(k)}) x_t^{(k)} ({u_t^{(k)}})^\tp (B + \bar{B}_t^{(k)})^\tp \big\} \\\nonumber
    &      \quad (B + \bar{B}_t^{(k)}) u_t^{(k)} ({x_t^{(k)}})^\tp (A + \bar{A}_t^{(k)})^\tp - \EE \big\{ (B + \bar{B}_t^{(k)}) u_t^{(k)} ({x_t^{(k)}})^\tp (A + \bar{A}_t^{(k)})^\tp \big\} \\\label{eq:lemma1_fifth_claim}
    &      \quad (B + \bar{B}_t^{(k)}) u_t^{(k)} ({u_t^{(k)}})^\tp (B + \bar{B}_t^{(k)})^\tp - \EE \big\{ (B + \bar{B}_t^{(k)}) u_t^{(k)} ({u_t^{(k)}})^\tp (B + \bar{B}_t^{(k)})^\tp \big\} \bigg).
\end{align}
Considering the first of the four terms of~\eqref{eq:lemma1_fifth_claim}, we have
\begin{align*}
    \boxed{1} &\Let \vect \left( (A + \bar{A}_t^{(k)}) x_t^{(k)} ({x_t^{(k)}})^\tp (A + \bar{A}_t^{(k)})^\tp - \EE \big\{ (A + \bar{A}_t^{(k)}) x_t^{(k)} ({x_t^{(k)}})^\tp (A + \bar{A}_t^{(k)})^\tp \big\} \right) \\
    &= \vect \bigg( A x_t^{(k)} ({x_t^{(k)}})^\tp A^\tp + A x_t^{(k)} ({x_t^{(k)}})^\tp (\bar{A}_t^{(k)})^\tp + \bar{A}_t^{(k)} x_t^{(k)} ({x_t^{(k)}})^\tp A^\tp + \bar{A}_t^{(k)} x_t^{(k)} ({x_t^{(k)}})^\tp (\bar{A}_t^{(k)})^\tp \\
    & - A \EE \big\{ x_t^{(k)} ({x_t^{(k)}})^\tp \big\} A^\tp  - 0 - 0 - \EE \big\{ \bar{A}_t^{(k)} x_t^{(k)} ({x_t^{(k)}})^\tp (\bar{A}_t^{(k)})^\tp \big\} \bigg) \\
    &= ( A \otimes A ) \vect \left( x_t^{(k)} ({x_t^{(k)}})^\tp - \EE \big\{ x_t^{(k)} ({x_t^{(k)}})^\tp \big\} \right) \\
    & \quad + (A \otimes \bar{A}_t^{(k)}) \vect \left( x_t^{(k)} ({x_t^{(k)}})^\tp \right) + (\bar{A}_t^{(k)} \otimes A) \vect \left( x_t^{(k)} ({x_t^{(k)}})^\tp \right) \\
    & \quad + (\bar{A}_t^{(k)} \otimes \bar{A}_t^{(k)}) \vect \left( x_t^{(k)} ({x_t^{(k)}})^\tp \right) - \EE \big\{ (\bar{A}_t^{(k)} \otimes \bar{A}_t^{(k)})\big\} \EE \big\{ \vect \left( x_t^{(k)} ({x_t^{(k)}})^\tp \right) \big\} \\
    &= ( A \otimes A ) \vect \left( x_t^{(k)} ({x_t^{(k)}})^\tp - \EE \big\{ x_t^{(k)} ({x_t^{(k)}})^\tp \big\} \right) \\
    & \quad + (A \otimes \bar{A}_t^{(k)}) \vect \left( x_t^{(k)} ({x_t^{(k)}})^\tp \right) + (\bar{A}_t^{(k)} \otimes A) \vect \left( x_t^{(k)} ({x_t^{(k)}})^\tp \right) \\
    & \quad + \left[ (\bar{A}_t^{(k)} \otimes \bar{A}_t^{(k)}) - \EE \big\{ (\bar{A}_t^{(k)} \otimes \bar{A}_t^{(k)})\big\} \right] \vect \left( x_t^{(k)} ({x_t^{(k)}})^\tp \right) \\
    & \quad + \EE \big\{ (\bar{A}_t^{(k)} \otimes \bar{A}_t^{(k)})\big\} \vect \left( x_t^{(k)} ({x_t^{(k)}})^\tp - \EE \big\{  x_t^{(k)} ({x_t^{(k)}})^\tp  \big\} \right).
\end{align*}
Taking norms, and substituting notated quantities, we have
\begin{align*}
    \left\| \boxed{1} \right\|  &\leq
    \|  A \otimes A  \|_2 \| \Delta X_{t} \| 
    + \| A \otimes \bar{A}_t^{(k)} \|_2 \| \vect (x_t^{(k)} ({x_t^{(k)}})^\tp) \| + \| \bar{A}_t^{(k)} \otimes A \|_2 \| \vect (x_t^{(k)} ({x_t^{(k)}})^\tp) \| \\
    &\quad + \| (\bar{A}_t^{(k)} \otimes \bar{A}_t^{(k)}) - \Sigma_A^\prime \|_2 \| \vect (x_t^{(k)} ({x_t^{(k)}})^\tp)\| 
    + \| \Sigma_A^\prime \|_2 \| \Delta X_{t} \| \\
    &\leq
    (\| A \|_2^2 + \| \Sigma_A^\prime \|_2) \| \Delta X_{t} \| + \left(2 \| A \|_2 \| \bar{A}_t^{(k)} \|_2 + \| (\bar{A}_t^{(k)} \otimes \bar{A}_t^{(k)}) - \Sigma_A^\prime \|_2 \right) \| x_t^{(k)} \|^2 \\
    &\leq
    (\| A \|_2^2 + \| \Sigma_A^\prime \|_2) \| \Delta X_{t} \| + (2 \| A \|_2 c_{\bar{A}} + c_{\Sigma_A^\prime} ) c_M^2.
\end{align*}
Applying identical arguments to the fourth term of~\eqref{eq:lemma1_fifth_claim}
\begin{align*}
    \boxed{4} &\Let \vect \left( (B + \bar{B}_t^{(k)}) u_t^{(k)} ({u_t^{(k)}})^\tp (B + \bar{B}_t^{(k)})^\tp - \EE \big\{ (B + \bar{B}_t^{(k)}) u_t^{(k)} ({u_t^{(k)}})^\tp (B + \bar{B}_t^{(k)})^\tp \big\} \right),
\end{align*}
we obtain the norm bound
\begin{align*}
    \left\| \boxed{4} \right\| &\leq
    (\| B \|_2^2 + \| \Sigma_B^\prime \|_2) \| \Delta U_{t} \| + (2 \| B \|_2 c_{\bar{B}} + c_{\Sigma_B^\prime} ) c_U^2\\
    &\le 3 (\| B \|_2^2 + \| \Sigma_B^\prime \|_2) c_U c_{\nu} + (2 \| B \|_2 c_{\bar{B}} + c_{\Sigma_B^\prime} ) c_U^2.
\end{align*}
Likewise, for the second term of~\eqref{eq:lemma1_fifth_claim}, we have
\begin{align*}
    \boxed{2} 
    &\Let \vect \left( (A + \bar{A}_t^{(k)}) x_t^{(k)} ({u_t^{(k)}})^\tp (B + \bar{B}_t^{(k)})^\tp - \EE \big\{ (A + \bar{A}_t^{(k)}) x_t^{(k)} ({u_t^{(k)}})^\tp (B + \bar{B}_t^{(k)})^\tp \big\} \right) \\
    &= \vect \bigg( A x_t^{(k)} ({u_t^{(k)}})^\tp B^\tp + A x_t^{(k)} ({u_t^{(k)}})^\tp (\bar{B}_t^{(k)})^\tp + \bar{A}_t^{(k)} x_t^{(k)} ({u_t^{(k)}})^\tp B^\tp + \bar{A}_t^{(k)} x_t^{(k)} ({u_t^{(k)}})^\tp (\bar{B}_t^{(k)})^\tp \\
    & - A \EE \big\{ x_t^{(k)} ({u_t^{(k)}})^\tp \big\} B^\tp  - 0 - 0 - \EE \big\{ \bar{A}_t^{(k)} x_t^{(k)} ({u_t^{(k)}})^\tp (\bar{B}_t^{(k)})^\tp \big\} \bigg) \\
    &= ( B \otimes A ) \vect \left( x_t^{(k)} ({u_t^{(k)}})^\tp - \EE \big\{ x_t^{(k)} ({u_t^{(k)}})^\tp \big\} \right) \\
    & \quad + (B \otimes \bar{A}_t^{(k)}) \vect \left( x_t^{(k)} ({u_t^{(k)}})^\tp \right) + (\bar{B}_t^{(k)} \otimes A) \vect \left( x_t^{(k)} ({u_t^{(k)}})^\tp \right) \\
    & \quad + (\bar{B}_t^{(k)} \otimes \bar{A}_t^{(k)}) \vect \left( x_t^{(k)} ({u_t^{(k)}})^\tp \right) - \EE \big\{ (\bar{B}_t^{(k)} \otimes \bar{A}_t^{(k)})\big\} \EE \big\{ \vect \left( x_t^{(k)} ({u_t^{(k)}})^\tp \right) \big\} \\
    &= ( B \otimes A ) \vect \left( x_t^{(k)} ({u_t^{(k)}})^\tp - \EE \big\{ x_t^{(k)} ({u_t^{(k)}})^\tp \big\} \right) \\
    & \quad + (B \otimes \bar{A}_t^{(k)}) \vect \left( x_t^{(k)} ({u_t^{(k)}})^\tp \right) + (\bar{B}_t^{(k)} \otimes A) \vect \left( x_t^{(k)} ({u_t^{(k)}})^\tp \right) \\
    & \quad + (\bar{B}_t^{(k)} \otimes \bar{A}_t^{(k)}) \vect \left( x_t^{(k)} ({u_t^{(k)}})^\tp \right). 
\end{align*}
Taking norms, and substituting notated quantities, we have
\begin{align*}
    \left\| \boxed{2} \right\| &\leq
    \| A \|_2 \| B \|_2 \| \Delta W_t \| + (\|A\|_2 c_{\bar{B}} + \|B\|_2 c_{\bar{A}} + c_{\bar{A}} c_{\bar{B}}) c_M c_U.
\end{align*}
The third term of~\eqref{eq:lemma1_fifth_claim} is simply the transpose of the second term, so an identical norm bound holds.

Putting together the four terms of~\eqref{eq:lemma1_fifth_claim}, we obtain
\begin{align}
    \| \Delta X_{t+1} \|
    &\leq 
    (\| A \|_2^2 + \| \Sigma_A^\prime \|_2) \| \Delta X_{t} \| + (2 \| A \|_2 c_{\bar{A}} + c_{\Sigma_A^\prime} ) c_M^2 \nonumber \\
    & \ + 3 (\| B \|_2^2 + \| \Sigma_B^\prime \|_2) c_U c_{\nu} + (2 \| B \|_2 c_{\bar{B}} + c_{\Sigma_B^\prime} ) c_U^2 \nonumber \\
    & \ + 2 \| A \|_2 \| B \|_2 c_W  + ( 2 \|A\|_2 c_{\bar{B}} + 2 \|B\|_2 c_{\bar{A}} + 2 c_{\bar{A}} c_{\bar{B}}) c_M c_U. \label{eq:state_deviation_outer_norm_bound}
\end{align}
For the base case when $t=0$, we have by Assumption \ref{asmp:bounded_system}~(v) that $\| \Delta X_0 \| \leq c_{\Delta X}$.
Applying \eqref{eq:state_deviation_outer_norm_bound} inductively with the base case proves the fifth claim.

\clearpage
\section{Basic identities and inequalities}
In the proofs of Theorems~\ref{thm:hatAB_bounded} and \ref{thm:hatSigmaAB_bounded}, the following facts will be used.
\begin{description}
\item[Submultiplicativity of spectral norm] For $A\in \RR^{m\times n}$ and $B \in \RR^{n\times p}$, 
\begin{align*}
    \|AB\|_2 \le \|A\|_2 \|B\|_2 .
\end{align*}

\item[Norm of Kronecker product] For $A\in \RR^{m\times n}$ and $B \in \RR^{p\times q}$, 
\begin{align*}
    \|A\otimes B\|_2 = \|A\|_2 \|B\|_2 .
\end{align*}

\item[Inverse of spectral norm]
For any invertible matrix $A \in \RR^{n\times n}$ we have
\begin{align} 
    \|A^{-1}\|_2 = \frac{1}{\sqrt{\lambda_{\min}(AA^\tp)}} \label{eq:inverse_of_spectral_norm}.
\end{align}

\item[Difference of matrix inverses]
Suppose $A, E \in \RR^{n\times n}$ are invertible square matrices. Then
\begin{align} \label{eq:difference_of_matrix_inverses}
    A^{-1} - E^{-1} 
    = E^{-1}(((E-A)+A)A^{-1}) - E^{-1} 
    = E^{-1} (E-A) A^{-1}.
\end{align}

\item[Matrix inverse perturbation bound](Equation (5.8.1) of \citep{horn2012matrix})~\\
Suppose $A, A+F \in \RR^{n\times n}$ are invertible square matrices. Then
\begin{align} \label{eq:matrix_inverse_perturbation_bound}
    \|A^{-1} - (A+F)^{-1}\| \le \|A^{-1}\| \cdot \|F\| \cdot \|(A+F)^{-1}\| .
\end{align}
This follows from taking $E = A+F$ in \eqref{eq:difference_of_matrix_inverses} and using submultiplicativity of spectral norm.

\item[Probability bound on the sum of random variables] \ \\ \noindent
Consider $k$ random variables, $X_1$, $\dots$, $X_k$, and a positive number $\eps$. Note that $X_i < \eps/k$ for all $i \in [k]$, implies $\sum_{i=1}^k X_i < \eps$, so it follows from the union bound that
\begin{align}\label{eq:prob_addition_bound}
    \mathbb{P}\left\{\sum_{i=1}^k X_i \ge \eps\right\} \le \sum_{i=1}^k \mathbb{P}\{X_i \ge \eps/k\}.
\end{align}

\item[Probability bound on the product of nonnegative random variables] \ \\ \noindent
Consider two nonnegative random variables, $X_1$ and $X_2$, and a positive number $\eps$. Since $X_1 < \sqrt{\eps}$ and $X_2 < \sqrt{\eps}$ implies $X_1 X_2 < \eps$ we have
\begin{align}\label{eq:prob_times_bound}
    \mathbb{P}\{X_1 X_2 \ge \eps\} \le \mathbb{P}\{X_1 \ge \sqrt{\eps}\} + \mathbb{P}\{X_2 \ge \sqrt{\eps}\} .
\end{align}
\end{description}

We will need the following geometrical result later in the use of covering arguments.
\begin{lemma}[Covering numbers of the Euclidean Sphere]
Consider a minimal $\gamma$-net $\{w_k, k \in [M_\gamma]\}$ of the $n$-dimensional sphere surface $\mathcal{S}_{n-1} \Let \{w \in \RR^{n} : \|w\|=1\}$. That is, for all $w \in \mathcal{S}_{n-1}$ there exists $w_i \in \{w_k, k \in [M_\gamma]\}$ such that $\|w - w_i\| \le \gamma$, and $M_{\gamma}$ is the smallest number satisfies this condition. Then for any $\gamma > 0$, the covering number $M_{\gamma}$, i.e., the cardinality of the $\gamma$-net satisfies
\begin{align}\label{eq:covering_number_sphere}
    \left( \frac{1}{\gamma} \right)^n \leq M_{\gamma} \leq \left( \frac{2}{\gamma} + 1 \right)^n .
\end{align}
\end{lemma}
\begin{pf}
A standard volume comparison involving Euclidean balls, e.g. Corollary 4.2.13 of \citep{vershynin2018high}, yields the result.
\end{pf}

\clearpage
We also need the following matrix concentration inequality.
\begin{lemma}[Matrix Bernstein inequality \citep{tropp2015introduction}]\label{lem:bernstein}
Consider a finite sequence of independent random matrices $\{X_k,$ $k \in [N]\}$ with common dimension $m\times n$. Assume that $\EE\{X_k\} = 0$ and $\|X_k\|_2 \le L$, $k\in [N]$. 
Introduce $S \Let \sum_{k=1}^N X_k$ and let $v \geq \max\{\|\EE\{SS^\tp\}\|_2, \|\EE\{S^\tp S\}\|_2\}$. Then, for all $\eps \ge 0$, 
\begin{align}
    \PP\{\|S\|_2 \ge \eps\} \le (n+m) \exp\left\{ - \frac{3}{2} \cdot \frac{\eps^2}{3 v + L \eps}\right\}.
\end{align}
\end{lemma}
We obtain the following corollary from Lemma \ref{lem:bernstein}.
\begin{corollary} \label{cor:bernstein}
Consider a finite sequence of independent random matrices $\{Y_k, k \in [N]\}$ with common dimension $m\times n$. Assume that $\EE\{Y_k\} = 0$ and $\|Y_k\|_2 \le M$, $k\in[N]$. Then, for all $\eps \ge 0$, 
\begin{align*}
    \PP \left\{ \left\| \frac{1}{N} \sum_{k=1}^N Y_k \right\|_2 \ge \eps \right\} \le \delta(\eps),
\end{align*}
where
\begin{align*}
    \delta(\eps) \Let (n+m) \exp\left\{ - \frac{3}{2} \cdot \frac{N \eps^2}{3 M^2 + M \eps}\right\} .
\end{align*}
\end{corollary}
\begin{pf}
Towards application of Lemma \ref{lem:bernstein}, assign $X_k = Y_k / N$ and thus $L = M / N$.
Now we get a crude bound on $v$ as
\begin{align*}
    \|\EE\{SS^\tp\}\|_2
    &=
    \left\|\EE \left\{ \left( \sum_{k=1}^N X_k \right) \left( \sum_{j=1}^N X_j^\tp \right) \right\} \right\|_2 \\
    &=
    \left\| \sum_{k=1}^N \sum_{j=1}^N \EE \left\{X_k  X_j^\tp \right\} \right\|_2 \tag*{(linearity of $\EE \{ \cdot \}$)} \\
    &=
    \left\| \sum_{k=1}^N \EE \left\{X_k  X_k^\tp \right\} \right\|_2 \tag*{(mutual independence of $X_k$)} \\
    &\leq 
    \sum_{k=1}^N \left\| \EE \left\{X_k  X_k^\tp \right\} \right\|_2 \tag*{(triangle inequality)} \\
    &\leq
    \sum_{k=1}^N \EE \left\{ \| X_k  X_k^\tp \|_2 \right\} \tag*{(Jensen's inequality)} \\
    &= 
    \sum_{k=1}^N \EE \left\{ \| X_k \|_2^2 \right\} \tag*{(property of the spectral norm)} \\
    &\leq 
    \sum_{k=1}^N (M / N) ^2
    = M^2 / N .
\end{align*}
An identical argument shows $\|\EE\{SS^\tp\}\|_2 \leq M^2/N$ so we can take $v = M^2/N$. Applying Lemma \ref{lem:bernstein} with $X_k = Y_k / N$, $L = M / N$, and $v = M^2/N$ yields the claim.
\end{pf}

\clearpage
\section{Proof of Theorem \ref{thm:hatAB_bounded}}\label{append:pf_thm:hatAB_bounded}
In this section we obtain bounds for system parameter error matrix 
$[\hat{A} ~ \hat{B}] - [A ~ B]$ 
by decomposing the difference using their representation in the least-squares estimators as 
\begin{align*}
    \big[\hat{A} ~ \hat{B}\big] - [A ~ B]
    &= 
    \hat{\bfY}\hat{\bfZ}^\tp (\hat{\bfZ}\hat{\bfZ}^\tp)^{\dagger} - \bfY\bfZ^\tp (\bfZ\bfZ^\tp)^{-1}\\
    &= 
    \big[ \hat{\bfY}\hat{\bfZ}^\tp - \bfY\bfZ^\tp \big] (\bfZ\bfZ^\tp)^{-1} 
     + \bfY\bfZ^\tp  \big[ (\hat{\bfZ}\hat{\bfZ}^\tp)^{\dagger} - (\bfZ\bfZ^\tp)^{-1} \big] 
     + \big[ \hat{\bfY}\hat{\bfZ}^\tp - \bfY\bfZ^\tp \big] \big[ (\hat{\bfZ}\hat{\bfZ}^\tp)^{\dagger} - (\bfZ\bfZ^\tp)^{-1} \big].
\end{align*}
In this form it is obvious that there are four unique terms, which fall into two groups. 
The first group is $\hat{\bfY}\hat{\bfZ}^\tp - \bfY\bfZ^\tp$ and $(\hat{\bfZ}\hat{\bfZ}^\tp)^{\dagger} - (\bfZ\bfZ^\tp)^{-1}$, which represent error terms amenable to analysis. 
The second group is $(\bfZ\bfZ^\tp)^{-1}$ and $\bfY\bfZ^\tp$, which are inherent to the system and do not depend on the estimator quality.
The terms $\hat{\bfY}\hat{\bfZ}^\tp - \bfY\bfZ^\tp$ and $(\hat{\bfZ}\hat{\bfZ}^\tp)^{\dagger} - (\bfZ\bfZ^\tp)^{-1}$ are treated first, then the bound on $[\hat{A}~\hat{B}] - [A~B]$ is obtained.

Throughout this section, small probability bounds are denoted by $\delta_{[\cdot]}$, where $[\cdot]$ are various subscripts, and each of these bounds decreases monotonically towards $0$ with increasing number of rollouts $n_r$.

The following bounds for $\hat{\bfY} - \bfY$ and $\hat{\bfZ} - \bfZ$ follow from Corollary \ref{cor:bernstein}.


\begin{lemma}\label{lem:bounded_Y_and_Z}
Suppose that Assumptions~\ref{asmp1} and~\ref{asmp:bounded_system} hold. Then for all $\eps > 0$,
\begin{align*}
      \PP \big\{ \big\|\hat{\bfY} - \bfY \big\|_2 \ge \eps \big\} 
    = \PP \big\{ \big\|\hat{\bfZ} - \bfZ \big\|_2 \ge \eps \big\} 
    = \PP \Big\{ \big[ \big\|\hat{\bfY} - \bfY \big\|_2 \ge \eps \big] \bigcup \big[ \big\|\hat{\bfZ} - \bfZ\big\|_2 \ge \eps \big] \Big\} &\le \delta_{Y}(\eps), 
\end{align*}
where
\begin{align*}
    \delta_Y(\eps) 
    &\Let 
    (n+\ell) \exp\left\{- \frac{3}{2} \cdot \frac{n_r \eps^2}{3 \ell c_N^2 + \eps \sqrt{\ell c_N^2}} \right\}, \\
\end{align*}
\end{lemma}

\begin{pf}
Using the bound $\| e_{t}^{(k)} \| = \| x_{t}^{(k)} - \EE \big\{ x_{t}^{(k)} \big\} \| \leq c_N$ from Lemma \ref{lem:bounded_system}, and denoting $\hat{\bfY}_k \Let \big[ x_{\ell}^{(k)}~\cdots~x_{1}^{(k)} \big]$ so $\hat{\bfY} = (\sum\nolimits_{k=1}^{n_r} \hat{\bfY}_k)/n_r$ and $\EE \big\{\hat{\bfY}_k \big\} = \bfY$, we obtain
\begin{align*}
    \big\| \hat{\bfY}_k - \bfY \big\|_2 
    \le  \big\|\hat{\bfY}_k - \bfY \big\|_F 
    = \sqrt{ \sum_{t=1}^\ell \big\| x_{t}^{(k)} - \EE \big\{ x_{t}^{(k)} \big\} \big\|^2 } 
    \leq \sqrt{ \ell c_N^2}.
\end{align*}

Applying Corollary \ref{cor:bernstein} with $Y_k = \hat{\bfY}_k - \bfY$, $N = n_r$, and $M = \sqrt{ \ell c_N^2}$, we conclude
\begin{align*}
    \PP\{\|\hat{\bfY} - \bfY\|_2 \ge \eps \} &\le \delta_{Y}(\eps).
\end{align*}
Denote $\hat{\bfZ}_k = \begin{bmatrix} x_{\ell-1}^{(k)} & \cdots & x_0^{(k)}\\ \nu_{\ell-1} & \cdots & \nu_0 \end{bmatrix}$ so $\hat{\bfZ} = (\sum\nolimits_{k=1}^{n_r} \hat{\bfZ}_k)/n_r$ and $\EE \big\{\hat{\bfZ}_k\big\} = \bfZ$.
Noticing that the last $m$ rows of $\hat{\bfZ}_k - \bfZ$ all have zero entries, we have that
\begin{align*}
    \big\|\hat{\bfY} - \bfY \big\|_2 = \big\|\hat{\bfZ} - \bfZ \big\|_2.
\end{align*}
Hence the events $\big\{\big\|\hat{\bfY} - \bfY \big\|_2 \ge \eps \big\}$ and  $\big\{\big\|\hat{\bfZ} - \bfZ \big\|_2 \ge \eps\big\}$ are precisely the same, concluding the proof.
\end{pf}

\begin{remark} \label{rem:bounded_Y_and_Z}
The reason that the last $m$ rows of $\hat{\bfZ}_k - \bfZ$ all have zero entries is that the first moments of the inputs are known in Algorithm~\ref{alg:A}, and appear identically in both $\hat{\bfZ}_k$ and $\bfZ$. 
Hence the probability bounds are independent of the input dimension $m$. 
\end{remark}

\begin{lemma}\label{lem:deltaYZ}
Suppose that Assumptions~\ref{asmp1} and~\ref{asmp:bounded_system} hold. Then for all $ \eps > 0$,
\begin{align*}
    \PP \big\{\big\| \hat{\bfY}\hat{\bfZ}^\tp - \bfY\bfZ^\tp \big\|_2 \ge \eps \big\} \le \delta_{YZ}(\eps)
\end{align*}
where
\begin{align*}
    \delta_{YZ}(\eps)
    &\Let
    \delta_Y \left( \sqrt{\eps + \left( \frac{\| \bfY \|_2 + \| \bfZ \|_2}{2}\right)^2 } - \frac{\| \bfY \|_2 + \| \bfZ \|_2}{2}  \right).
\end{align*}
\end{lemma}


\begin{pf}
We begin with the decomposition
\begin{align*}
    \hat{\bfY}\hat{\bfZ}^\tp - \bfY\bfZ^\tp 
    =
    (\hat{\bfY} - \bfY) (\hat{\bfZ}-\bfZ)^\tp 
    + 
    (\hat{\bfY} - \bfY)\bfZ^\tp 
    + 
    \bfY(\hat{\bfZ}- \bfZ)^\tp .
\end{align*}
By the triangle inequality and submultiplicativity we have
\begin{align*}
    \big\| \hat{\bfY}\hat{\bfZ}^\tp - \bfY\bfZ^\tp \big\|_2
    &=
    \big\| (\hat{\bfY} - \bfY) (\hat{\bfZ}-\bfZ)^\tp 
    + (\hat{\bfY} - \bfY)\bfZ^\tp 
    + \bfY(\hat{\bfZ}- \bfZ)^\tp \big\|_2 \\ 
    &\leq
    \big\| (\hat{\bfY} - \bfY) (\hat{\bfZ}-\bfZ)^\tp \big\|_2
    + \big\| (\hat{\bfY} - \bfY)\bfZ^\tp \big\|_2
    + \big\| \bfY(\hat{\bfZ}- \bfZ)^\tp \big\|_2 \\
    &\leq
    \big\| \hat{\bfY} - \bfY \big\|_2 \big\| \hat{\bfZ}-\bfZ \big\|_2
    + \big\| \hat{\bfY} - \bfY \big\|_2 \| \bfZ \|_2
    + \| \bfY \|_2 \big\| \hat{\bfZ}- \bfZ \big\|_2 \\
    &=
    \big\| \hat{\bfY} - \bfY \big\|_2^2 
    + \big\| \hat{\bfY}- \bfY \big\|_2 \left(  \left\| \bfY \right\|_2 + \left\| \bfZ \right\|_2 \right)
    .
\end{align*}
Considering a probability bound, solving the quadratic inequality in $\big\| \hat{\bfY}- \bfY \big\|_2$, and applying Lemma \ref{lem:bounded_Y_and_Z} we have
\begin{align*}
    \PP \big\{ \big\| \hat{\bfY}\hat{\bfZ}^\tp - \bfY\bfZ^\tp \big\|_2 \ge \eps \big\} 
    &\leq
    \PP \big\{ \big\| \hat{\bfY} - \bfY \big\|_2^2 
    + \big\| \hat{\bfY}- \bfY \big\|_2 ( \| \bfY \|_2 + \| \bfZ \|_2 ) \ge \eps \big\} \\
    &=
    \PP\left\{ \big\| \hat{\bfY} - \bfY \big\|_2 \geq \frac{1}{2} (\| \bfY \|_2 + \| \bfZ \|_2 ) \left( \sqrt{1 + \frac{4 \eps}{(\| \bfY \|_2 + \| \bfZ \|_2)^2 }} - 1 \right) \right\} \\
    &\leq
    \delta_Y \left( \frac{\| \bfY \|_2 + \| \bfZ \|_2}{2} \left( \sqrt{1 + \frac{4 \eps}{(\| \bfY \|_2 + \| \bfZ \|_2)^2 }} - 1 \right) \right),
\end{align*}
which was the claimed inequality.
\end{pf}

\begin{lemma}\label{lem:deltaZZ}
Suppose Assumptions~\ref{asmp1} and~\ref{asmp:bounded_system} hold. Given a positive constant $\eps_{\max}$, then for all $0 < \eps < \eps_{\max}$,
\begin{align*}
    \PP \big\{\big\| (\hat{\bfZ}\hat{\bfZ}^\tp)^{\dagger} - (\bfZ\bfZ^\tp)^{-1} \big\|_2 \ge \eps \big\} \le \delta_{ZZ}(\eps, \eps_{\max}),
\end{align*}
where
\begin{align*}
    \delta_{ZZ}(\eps, \eps_{\max}) 
    &\Let
    \delta_0 \left( \frac12 \lambda_{\min}^2 (\bfZ\bfZ^\tp)  \left(1 - \frac{\eps}{\eps_{\max}}\right) \eps  \right)
    +
    \delta_m \left( \frac{\eps \lambda_{\min}(\bfZ\bfZ^\tp)}{ \eps_{\max} ( 2 + \lambda_{\min}(\bfZ\bfZ^\tp)/\lambda_{\max}(\bfZ\bfZ^\tp) ) } \right),\\
    \delta_0(\eps)
    &\Let
    \delta_{Y} \left(\sqrt{\lambda_{\max}(\bfZ\bfZ^\tp) + \eps} - \sqrt{\lambda_{\max}(\bfZ\bfZ^\tp)}  \right), \\
    \delta_m(\eps)
    &\Let
    \left(9^{n+m} + \left( \frac{ 16 \lambda_{\max} (\bfZ\bfZ^\tp)}{\lambda_{\min} (\bfZ\bfZ^\tp)} + 1 \right)^{n+m} \right) \delta_0(\eps).
\end{align*}
\end{lemma}

\begin{remark}
The additional parameter $\eps_{\max}$ arises when bounding $\lambda_{\min}^2 (\hat{\bfZ}\hat{\bfZ}^\tp)$, and is in fact independent of the estimation bound. 
\end{remark}

\begin{pf}
Later we will show that $\hat{\bfZ}\hat{\bfZ}^\tp$ is invertible with high probability, so we now assume the existence of $(\hat{\bfZ}\hat{\bfZ}^\tp)^{-1}$, i.e., $(\hat{\bfZ}\hat{\bfZ}^\tp)^{\dagger} = (\hat{\bfZ}\hat{\bfZ}^\tp)^{-1}$. Hence we may apply \eqref{eq:matrix_inverse_perturbation_bound} to obtain the decomposition
\begin{align*}
    \big\|(\hat{\bfZ}\hat{\bfZ}^\tp)^{-1} - (\bfZ\bfZ^\tp)^{-1}\big\|_2
    &=
    \big\|[ (\bfZ\bfZ^\tp)^{-1}  (\hat{\bfZ}\hat{\bfZ}^\tp)^{-1} \big( \hat{\bfZ}\hat{\bfZ}^\tp - \bfZ\bfZ^\tp \big) \big\|_2 \\
    &\le 
    \big\|(\bfZ\bfZ^\tp)^{-1} \big\|_2 \big\|(\hat{\bfZ}\hat{\bfZ}^\tp)^{-1} \big\|_2 \big\|\hat{\bfZ}\hat{\bfZ}^\tp - \bfZ\bfZ^\tp\big\|_2.
\end{align*}
Considering a probability bound, rearranging, and using \eqref{eq:inverse_of_spectral_norm} we obtain
\begin{align} 
    \PP \big\{\big\|(\hat{\bfZ}\hat{\bfZ}^\tp)^{-1} - (\bfZ\bfZ^\tp)^{-1}\big\|_2 \ge \eps \big\}
    &\le 
    \PP \big\{\big\|(\bfZ\bfZ^\tp)^{-1} \big\|_2 \big\|(\hat{\bfZ}\hat{\bfZ}^\tp)^{-1}\big\|_2 \big\|(\hat{\bfZ}\hat{\bfZ}^\tp) - (\bfZ\bfZ^\tp)\big\|_2 \ge \eps \big\} \nonumber \\
    &= 
    \PP\left\{\big\|\hat{\bfZ}\hat{\bfZ}^\tp - \bfZ\bfZ^\tp \big\|_2 \ge \frac{\eps}{\big\|(\bfZ\bfZ^\tp)^{-1} \big\|_2 \big\|(\hat{\bfZ}\hat{\bfZ}^\tp)^{-1}\big\|_2} \right\} \nonumber  \\
    &= 
    \PP \big\{\big\|\hat{\bfZ}\hat{\bfZ}^\tp - \bfZ\bfZ^\tp \big\|_2 \ge \eps \lambda_{\min}(\bfZ\bfZ^\tp) \lambda_{\min}(\hat{\bfZ}\hat{\bfZ}^\tp) \big\}. \label{eq:bound_of_ZZinverse_ZZhat}
\end{align}
We are now faced with providing bounds on both $\big\| \hat{\bfZ}\hat{\bfZ}^\tp - \bfZ\bfZ^\tp \big\|_2$ on the left side and $\lambda_{\min}(\hat{\bfZ}\hat{\bfZ}^\tp)$ on the right side of the inequality inside the probability.

First, we consider the bound of $\big\| \hat{\bfZ}\hat{\bfZ}^\tp - \bfZ\bfZ^\tp \big\|_2$ by beginning with the decomposition
\begin{align*}
    \hat{\bfZ}\hat{\bfZ}^\tp - \bfZ\bfZ^\tp
    = 
    (\hat{\bfZ}-\bfZ)(\hat{\bfZ}-\bfZ)^\tp + (\hat{\bfZ}-\bfZ)\bfZ^\tp + \bfZ(\hat{\bfZ}-\bfZ)^\tp .
\end{align*}
Using the triangle inequality, submultiplicativity, and solving the quadratic inequality in $\big\|\hat{\bfZ} - \bfZ \big\|_2$, we obtain
\begin{align*} 
    \PP \big\{\big\|\hat{\bfZ}\hat{\bfZ}^\tp - \bfZ\bfZ^\tp \big\|_2 \ge \eps \big\} 
    &=
    \PP\big\{\big\|(\hat{\bfZ}-\bfZ)(\hat{\bfZ}-\bfZ)^\tp + (\hat{\bfZ}-\bfZ)\bfZ^\tp 
    + \bfZ(\hat{\bfZ}-\bfZ)^\tp \big\|_2 \ge \eps \big\}\\
    &\le 
    \PP\big\{\big\|(\hat{\bfZ}-\bfZ)(\hat{\bfZ}-\bfZ)^\tp \big\|_2 + \big\| (\hat{\bfZ}-\bfZ)\bfZ^\tp \big\|_2 
    + \big\| \bfZ(\hat{\bfZ}-\bfZ)^\tp \big\|_2 \ge \eps \big\}\\
    &\le 
    \PP \big\{\big\|\hat{\bfZ}-\bfZ \big\|_2^2 + 2 \big\| \hat{\bfZ}-\bfZ \big\|_2 \big\|\bfZ  \big\|_2 \ge \eps \big\}\\
    &=
    \PP \left\{ \big\|\hat{\bfZ} - \bfZ \big\|_2 \ge \|\bfZ\|_2 \left(\sqrt{1 + \frac{\eps}{\|\bfZ\|_2^2}} - 1\right) \right\}.
\end{align*}
Applying Lemma \ref{lem:bounded_Y_and_Z} with the appropriate settings of $\eps$ yields
\begin{align}
    \PP \big\{\big\|\hat{\bfZ}\hat{\bfZ}^\tp - \bfZ\bfZ^\tp \big\|_2 \ge \eps \big\} 
    &\le 
    \delta_{Y} \left( \|\bfZ\|_2 \left(\sqrt{1 + \frac{\eps}{\|\bfZ\|_2^2}} - 1\right) \right) \label{eq:bound_of_ZZ} 
    \teL \delta_0(\eps).
\end{align}

Now we seek a lower bound of $\lambda_{\min}(\hat{\bfZ}\hat{\bfZ}^\tp)$. 
First an upper bound of $ \big\| \hat{\bfZ}\hat{\bfZ}^\tp \big\|_2$ is needed.
To obtain this, we put forward a covering argument. 
Begin by constructing a quadratic form of $\hat{\bfZ}\hat{\bfZ}^\tp$ with $ w \in \RR^{n+m}$, $\|w\| = 1$ and use the earlier result in \eqref{eq:bound_of_ZZ} to obtain
\begin{align}
    \PP \big\{ w^\tp \hat{\bfZ}\hat{\bfZ}^\tp w > \|\bfZ\bfZ^\tp\|_2 + \eps \big\} 
    &\le
    \PP \big\{ w^\tp \hat{\bfZ}\hat{\bfZ}^\tp w > w^\tp {\bfZ}{\bfZ}^\tp w + \eps \big\} \tag*{(definition of spectral norm)} \nonumber \\
    &=
    \PP \big\{ w^\tp \hat{\bfZ}\hat{\bfZ}^\tp w - w^\tp \bfZ\bfZ^\tp w \ge \eps \big\} \nonumber \\
    &=
    \PP \big\{ w^\tp \big( \hat{\bfZ}\hat{\bfZ}^\tp - \bfZ\bfZ^\tp \big) w \ge \eps \big\} \nonumber \\
    &\le
    \PP \big\{ \big\| \hat{\bfZ}\hat{\bfZ}^\tp - \bfZ\bfZ^\tp \big\|_2 \ge \eps \big\} \tag*{(definition of spectral norm, $\|w\| = 1$)} \nonumber \\
    &\leq
    \delta_0(\eps) \label{eq:Zhat_quadform_bound_upr}.
\end{align}
Consider a minimal $\gamma$-net $\{w_k, k \in [M_\gamma]\}$ of the $(n+m)$-sphere surface $\mathcal{S}_{n+m-1} \Let \{w \in \RR^{n+m} : \|w\|=1\}$. 
Hence for all $w \in \mathcal{S}_{n+m-1}$ there exists $k \in [M_{\gamma}]$ such that
\begin{align*}
    w^\tp \hat{\bfZ}\hat{\bfZ}^\tp w
    &= (w-w_k)^\tp \hat{\bfZ}\hat{\bfZ}^\tp w + w_k^\tp \hat{\bfZ}\hat{\bfZ}^\tp (w - w_k) + w_k^\tp \hat{\bfZ}\hat{\bfZ}^\tp w_k\\
    &\le 2\gamma \| \hat{\bfZ}\hat{\bfZ}^\tp \|_2 + \max_{k \in [M_{\gamma}]} w_k^\tp \hat{\bfZ}\hat{\bfZ}^\tp w_k,
\end{align*}
where the inequality follows by $\|A\|_2 = \max_{\|x\|=1,\|y\|=1} y^\tp A x$ for $A\in \RR^n$ and $\|w - w_k\| \le \gamma$.
Taking supremum of the left side of the inequality over $w$, using the definition of the spectral norm, and rearranging implies that
\begin{align} \label{eq:ZZhat_bound_net}
    \big\| \hat{\bfZ}\hat{\bfZ}^\tp \big\|_2 \le \frac{1}{1-2\gamma} \max_{k \in [M_{\gamma}]} w_k^\tp \hat{\bfZ}\hat{\bfZ}^\tp w_k.
\end{align}
By construction $\| w_k \| = 1$, so we may apply \eqref{eq:Zhat_quadform_bound_upr} to $w_k$ in place of $w$ from earlier in the proof:
\begin{align*}
    \PP\left\{ w_k^\tp \hat{\bfZ}\hat{\bfZ}^\tp w_k > \|\bfZ\bfZ^\tp\|_2 + \eps \right\} 
    \leq
    \delta_0(\eps).
\end{align*}

Now apply the union bound over the $M_{\gamma}$ terms to obtain
\begin{align*}
    \PP\left\{ \left[ \max_{k \in [M_{\gamma}]} w_k^\tp \hat{\bfZ}\hat{\bfZ}^\tp w_k \right] > \|\bfZ\bfZ^\tp\|_2 + \eps \right\} 
    &=
    \PP\left\{ \bigcup_{k=1}^{M_{\gamma}} \big[ w_k^\tp \hat{\bfZ}\hat{\bfZ}^\tp w_k  > \|\bfZ\bfZ^\tp\|_2 + \eps \big] \right\} \\
    &\le 
    \sum_{k=1}^{M_{\gamma}} \PP \big\{ \big[ w_k^\tp \hat{\bfZ}\hat{\bfZ}^\tp w_k  > \|\bfZ\bfZ^\tp\|_2 + \eps \big] \big\} \\
    &\le 
    \sum_{k=1}^{M_{\gamma}} \delta_0(\eps) 
    =
    M_{\gamma} \delta_0(\eps) .
\end{align*}
Note that tighter bounds can be obtained via more complicated arguments e.g. as in \citep{tropp2012user, wainwright2019high}.
For definiteness, choose $\gamma = 1/4$. By \eqref{eq:covering_number_sphere} we know $M_{\gamma} \le 9^{n+m}$. Thus \eqref{eq:ZZhat_bound_net} becomes
\begin{align*}
    \big\| \hat{\bfZ}\hat{\bfZ}^\tp \big\|_2 \le 2 \max_{k \in [9^{n+m}]} w_k^\tp \hat{\bfZ}\hat{\bfZ}^\tp w_k.
\end{align*}
Considering a probability bound we have
\begin{align}
    \PP \big\{ \| \hat{\bfZ}\hat{\bfZ}^\tp \|_2 \geq 2 \left( \|\bfZ\bfZ^\tp\|_2 + \eps \right) \big\}
    &\le
    \PP \left\{ 2 \max_{k \in [9^{n+m}]} w_k^\tp \hat{\bfZ}\hat{\bfZ}^\tp w_k \geq 2 \left( \|\bfZ\bfZ^\tp\|_2 + \eps \right) \right\} \nonumber \\
    &=
    \PP \left\{ \max_{k \in [9^{n+m}]} w_k^\tp \hat{\bfZ}\hat{\bfZ}^\tp w_k \geq \|\bfZ\bfZ^\tp\|_2 + \eps \right\} \nonumber \\
    &\le
    9^{n+m} \delta_0(\eps) 
    \teL \delta_1(\eps). \label{eq:sigma_max_Zhat_prob_bound}
\end{align}
We are now in a position to derive a lower bound for $\lambda_{\min}(\hat{\bfZ}\hat{\bfZ}^\tp)$.
Again we construct a quadratic form of $\hat{\bfZ}\hat{\bfZ}^\tp$ with $ w \in \RR^{n+m}$, $ \|w\| = 1$ and use the earlier result in \eqref{eq:bound_of_ZZ} to obtain
\begin{align}
    \PP \big\{ w^\tp \hat{\bfZ}\hat{\bfZ}^\tp w  < \lambda_{\min} (\bfZ\bfZ^\tp) - \eps \big\} 
    &=
    \PP \big\{ \lambda_{\min} (\bfZ\bfZ^\tp) > w^\tp \hat{\bfZ}\hat{\bfZ}^\tp w + \eps \big\} \nonumber \\ 
    &\le
    \PP \big\{ w^\tp {\bfZ}{\bfZ}^\tp w > w^\tp \hat{\bfZ}\hat{\bfZ}^\tp w + \eps \big\} \tag*{(property of minimum eigenvalue)} \nonumber \\
    &=
    \PP \big\{ w^\tp \bfZ\bfZ^\tp w - w^\tp \hat{\bfZ}\hat{\bfZ}^\tp w  \ge \eps \big\} \nonumber \\
    &=
    \PP \big\{ w^\tp \big( \bfZ\bfZ^\tp - \hat{\bfZ}\hat{\bfZ}^\tp \big) w \ge \eps \big\} \nonumber \\
    &\le
    \PP \big\{ \big\| \bfZ\bfZ^\tp - \hat{\bfZ}\hat{\bfZ}^\tp \big\|_2 \ge \eps \big\} \tag*{(definition of spectral norm, $\|w\| = 1$)} \nonumber \\
    &\leq
    \delta_0(\eps) . \label{eq:Zhat_quadform_bound_lwr}
\end{align}
Consider again a minimal $\gamma$-net $\{w_k , k\in [M_\gamma]\}$ of $\mathcal{S}_{n+m-1}$, and for all $ w \in \mathcal{S}_{n+m-1}$, there exists $k \in [M_{\gamma}]$ such that
\begin{align}
    w^\tp \hat{\bfZ}\hat{\bfZ}^\tp w
    &= (w-w_k)^\tp \hat{\bfZ}\hat{\bfZ}^\tp w + w_k^\tp \hat{\bfZ}\hat{\bfZ}^\tp (w - w_k) + w_k^\tp \hat{\bfZ}\hat{\bfZ}^\tp w_k \nonumber \\
    &\ge - 2\gamma \| \hat{\bfZ}\hat{\bfZ}^\tp \|_2 + \min_{k \in [M_{\gamma}]} w_k^\tp \hat{\bfZ}\hat{\bfZ}^\tp w_k, \label{eq:sigma_min_Zhat_quadform_bound}
\end{align}
where the inequality follows by $\|A\|_2 = \max_{\|x\|=1,\|y\|=1} y^\tp A x$ for $A\in \RR^n$ and $\|w_k - w\| \le \gamma$.
By construction $\| w_k \| = 1$, so we may apply \eqref{eq:Zhat_quadform_bound_lwr} to $w_k$ in place of $w$ from earlier in the proof:
\begin{align*}
    \PP \big\{ w_k^\tp \hat{\bfZ}\hat{\bfZ}^\tp w_k  < \lambda_{\min} (\bfZ\bfZ^\tp) - \eps \big\} 
    \leq
    \delta_0(\eps).
\end{align*}
As before, apply the union bound over the $M_{\gamma}$ terms to obtain
\begin{align*}
    \PP\left\{ \left[ \min_{k \in [M_{\gamma}]} w_k^\tp \hat{\bfZ}\hat{\bfZ}^\tp w_k \right] < \lambda_{\min} (\bfZ\bfZ^\tp) - \eps\right\} 
    &=
    \PP\left\{ \bigcup_{k=1}^{M_{\gamma}} \big[ w_k^\tp \hat{\bfZ}\hat{\bfZ}^\tp w_k  < \lambda_{\min} (\bfZ\bfZ^\tp) - \eps \big] \right\} \\
    &\le 
    \sum_{k=1}^{M_{\gamma}} \PP \big\{\big[ w_k^\tp \hat{\bfZ}\hat{\bfZ}^\tp w_k  < \lambda_{\min} (\bfZ\bfZ^\tp) - \eps \big] \big\} \\
    &\le 
    \sum_{k=1}^{M_{\gamma}} \delta_0(\eps) 
    =
    M_{\gamma} \delta_0(\eps) .
\end{align*}
For definiteness choose $\gamma = \lambda_{\min} (\bfZ\bfZ^\tp)/(8 \lambda_{\max} (\bfZ\bfZ^\tp)) = \lambda_{\min} (\bfZ\bfZ^\tp)/(8 \|\bfZ\bfZ^\tp\|_2)$.
By \eqref{eq:covering_number_sphere} we know \[M_{\gamma} \le \left( \frac{16 \|\bfZ\bfZ^\tp\|_2}{\lambda_{\min} (\bfZ\bfZ^\tp)} + 1 \right)^{n+m}.\]
Hence, 
\begin{align}
    \PP\left\{ \left[ \min_{k \in [M_{\gamma}]} w_k^\tp \hat{\bfZ}\hat{\bfZ}^\tp w_k \right] < \lambda_{\min} (\bfZ\bfZ^\tp) - \eps \right\} 
    &\le 
    \left( \frac{ 16 \|\bfZ\bfZ^\tp\|_2}{\lambda_{\min} (\bfZ\bfZ^\tp)} + 1 \right)^{n+m} \delta_0(\eps) 
    \teL \delta_2(\eps) \label{eq:sigma_min_Zhat_prob_bound1}.
\end{align}
Also, using our choice of $\gamma = \lambda_{\min} (\bfZ\bfZ^\tp)/(8 \|\bfZ\bfZ^\tp\|_2)$, we have
\begin{align*}
    - 4 \gamma (\|\bfZ\bfZ^\tp\|_2 + \eps) 
    &= 
    - 4 \cdot \frac{\lambda_{\min} (\bfZ\bfZ^\tp)}{8 \|\bfZ\bfZ^\tp\|_2}  (\|\bfZ\bfZ^\tp\|_2 + \eps) \\
    &=
    - \frac{1}{2} \lambda_{\min} (\bfZ\bfZ^\tp) \left(  1 + \frac{\eps}{\|\bfZ\bfZ^\tp\|_2} \right).
\end{align*}
Considering a probability bound and using the result on $ \big\| \hat{\bfZ}\hat{\bfZ}^\tp \big\|_2$ in \eqref{eq:sigma_max_Zhat_prob_bound}, we have
\begin{align}
    &\PP \left\{ - 2\gamma \big\| \hat{\bfZ}\hat{\bfZ}^\tp \big\|_2 \leq - \frac{1}{2} \lambda_{\min} (\bfZ\bfZ^\tp) \left(  1 + \frac{\eps}{\|\bfZ\bfZ^\tp\|_2} \right)  \right\} \nonumber\\
    &=  
    \PP \big\{ - 2\gamma \big\| \hat{\bfZ}\hat{\bfZ}^\tp \big\|_2 \leq - 4 \gamma  (\| \bfZ\bfZ^\tp \|_2 + \eps)  \big\} \nonumber \\
    &= 
    \PP \big\{ \big\| \hat{\bfZ}\hat{\bfZ}^\tp \big\|_2 \geq 2 ( \|\bfZ\bfZ^\tp\|_2 + \eps ) \big\} \nonumber \\
    &\le \delta_1(\eps). \label{eq:sigma_min_Zhat_prob_bound2}
\end{align}
Combining \eqref{eq:sigma_min_Zhat_prob_bound1} and \eqref{eq:sigma_min_Zhat_prob_bound2} and using the union bound we obtain that
\begin{align*}
    - 2\gamma \| \hat{\bfZ}\hat{\bfZ}^\tp \|_2 + \min_{k \in [M_{\gamma}]} w_k^\tp \hat{\bfZ}\hat{\bfZ}^\tp w_k
    &\ge
    - \frac{1}{2} \lambda_{\min} (\bfZ\bfZ^\tp) \left(  1 + \frac{\eps}{\|\bfZ\bfZ^\tp\|_2} \right)  +  \lambda_{\min} (\bfZ\bfZ^\tp) - \eps \\
    &= 
    \frac12 \lambda_{\min} (\bfZ\bfZ^\tp) - \left( 1 +  \frac{\lambda_{\min} (\bfZ\bfZ^\tp)}{2 \| \bfZ\bfZ^\tp \|_2} \right) \eps
\end{align*}
takes place with high probability at least $1 - \left[ \delta_1(\eps) + \delta_2(\eps) \right]$.
Recalling \eqref{eq:sigma_min_Zhat_quadform_bound} we have
\begin{align*}
    \lambda_{\min} (\hat{\bfZ}\hat{\bfZ}^\tp) 
    = 
    \min_{w \in \mathcal{S}_{n+m-1}} w^\tp \hat{\bfZ}\hat{\bfZ}^\tp w
    &\ge
    - 2\gamma \| \hat{\bfZ} \|_2^2 + \min_{k \in [M_{\gamma}]} w_k^\tp \hat{\bfZ}\hat{\bfZ}^\tp w_k,
\end{align*}
and thus the probability bound
\begin{align}
    \PP \left\{ \lambda_{\min} (\hat{\bfZ}\hat{\bfZ}^\tp) < \frac12 \lambda_{\min} (\bfZ\bfZ^\tp) - \left( 1 +  \frac{\lambda_{\min} (\bfZ\bfZ^\tp)}{2 \| \bfZ\bfZ^\tp \|_2} \right) \eps \right\}
    \leq 
    \delta_1(\eps) + \delta_2(\eps)
    \teL \delta_m(\eps)    . \label{eq:sigma_min_Zhat_prob_bound_total}
\end{align}
In fact, this means that $\hat{\bfZ}\hat{\bfZ}^\tp$ is invertible with probability at least $1 - \delta_m(\eps)$, if we take $\eps$ small enough, since $\lambda_{\min}(\bfZ\bfZ^\tp) > 0$ from Assumption~\ref{asmp1}~(iv).
We conclude the proof by returning to the overall bound in \eqref{eq:bound_of_ZZinverse_ZZhat} and using both the bound of $\hat{\bfZ}\hat{\bfZ}^\tp - \bfZ\bfZ^\tp$ in \eqref{eq:bound_of_ZZ} and the bound of $\lambda_{\min} (\hat{\bfZ}\hat{\bfZ}^\tp)$ in \eqref{eq:sigma_min_Zhat_prob_bound_total} to obtain
\begin{align*}
    &\PP \big\{ \big\|(\hat{\bfZ}\hat{\bfZ}^\tp)^{\dagger} - (\bfZ\bfZ^\tp)^{-1} \big\|_2 \ge \eps \big\} \\
    &\le 
    \PP \left\{ \big[\big\|(\hat{\bfZ}\hat{\bfZ}^\tp)^{\dagger} - (\bfZ\bfZ^\tp)^{-1} \big\|_2 \ge \eps \big] \bigcap \left[\lambda_{\min}(\hat{\bfZ}\hat{\bfZ}^\tp) \ge \frac{1-\tau}{2} \lambda_{\min}(\bfZ\bfZ^\tp) \right] \right\} + \PP\left\{\lambda_{\min}(\hat{\bfZ}\hat{\bfZ}^\tp) < \frac{1-\tau}{2} \lambda_{\min}(\bfZ\bfZ^\tp) \right\} \\
    &=
    \PP \left\{ \big[\big\|(\hat{\bfZ}\hat{\bfZ}^\tp)^{-1} - (\bfZ\bfZ^\tp)^{-1} \big\|_2 \ge \eps \big] \bigcap \left[\lambda_{\min}(\hat{\bfZ}\hat{\bfZ}^\tp) \ge \frac{1-\tau}{2} \lambda_{\min}(\bfZ\bfZ^\tp) \right] \right\} + \PP\left\{\lambda_{\min}(\hat{\bfZ}\hat{\bfZ}^\tp) < \frac{1-\tau}{2} \lambda_{\min}(\bfZ\bfZ^\tp) \right\} \\
    &\le 
    \PP \left\{ \big[ \big\|\hat{\bfZ}\hat{\bfZ}^\tp - \bfZ\bfZ^\tp \big\|_2 \ge \eps \lambda_{\min} (\bfZ\bfZ^\tp) \lambda_{\min} (\hat{\bfZ}\hat{\bfZ}^\tp) \big] \bigcap \left[\lambda_{\min}(\hat{\bfZ}\hat{\bfZ}^\tp) \ge \frac{1-\tau}{2} \lambda_{\min}(\bfZ\bfZ^\tp) \right] \right\} \\
    & \quad + \PP\left\{\lambda_{\min}(\hat{\bfZ}\hat{\bfZ}^\tp) < \frac{1-\tau}{2} \lambda_{\min}(\bfZ\bfZ^\tp) \right\} \\
    &\le 
    \PP\left\{ \big\|\hat{\bfZ}\hat{\bfZ}^\tp - \bfZ\bfZ^\tp \big\|_2 \ge \frac{1-\tau}{2} \eps \lambda_{\min}^2 (\bfZ\bfZ^\tp) \right\}
    + \PP\left\{\lambda_{\min}(\hat{\bfZ}\hat{\bfZ}^\tp) < \frac{1-\tau}{2} \lambda_{\min}(\bfZ\bfZ^\tp) \right\} \\
    &\le
    \delta_0 \left( \frac{1-\tau}{2}  \lambda_{\min}^2 (\bfZ\bfZ^\tp) \eps \right)
    +
    \delta_m \left( \frac{\tau \lambda_{\min}(\bfZ\bfZ^\tp)}{ 2 + \lambda_{\min}(\bfZ\bfZ^\tp)/\|\bfZ\bfZ^\tp \|_2} \right),
\end{align*}
where $0 < \tau < 1$.
Note that $\tau$ may be chosen arbitrarily small. In order to preserve useful dependence of the bound on $\eps$, fix a maximum $\eps_{\max} > \eps$ and set $\tau = \eps/\eps_{\max}$, so the bound becomes
\begin{align}
    &\PP \big\{ \big\|(\hat{\bfZ}\hat{\bfZ}^\tp)^{\dagger} - (\bfZ\bfZ^\tp)^{-1} \big\|_2 \ge \eps \big\} \nonumber\\
    &\le
    \delta_0 \left( \frac12 \lambda_{\min}^2 (\bfZ\bfZ^\tp)  \left(1 - \frac{\eps}{\eps_{\max}}\right) \eps  \right)
    +
    \delta_m \left( \frac{\eps \lambda_{\min}(\bfZ\bfZ^\tp)}{ \eps_{\max} ( 2 + \lambda_{\min}(\bfZ\bfZ^\tp)/\|\bfZ\bfZ^\tp \|_2 ) } \right)
    \teL \delta_{ZZ}(\eps, \eps_{\max}). \label{eq:bound_of_ZZinverse_2}
\end{align}
\end{pf}

\begin{theorem}[Theorem \ref{thm:hatAB_bounded} restated] \label{thm:hatAB_bounded_restate}
Suppose Assumptions~\ref{asmp1} and~\ref{asmp:bounded_system} hold. Given a positive value $\eps_{\max}$, then for all $0 < \eps < 3 \eps_{\max} \min\{\sqrt{\lambda_{\max}(\bfY\bfY^\tp)\lambda_{\max}(\bfZ\bfZ^\tp)}, \eps_{\max} \}$,
\begin{align*}
    \PP\big\{ \big\|
    \big[\hat{A} ~ \hat{B} \big] - [A ~ B]
    \big\|_2 \ge \eps \big\} \le \delta_{AB}(\eps),
\end{align*}
where
\begin{align*}
    \delta_{AB}(\eps) &= \delta_{AB}(\eps,\eps_{\max})\\
    &\Let
    \delta_{YZ} \left( \frac13 \lambda_{\min} (\bfZ\bfZ^\tp) \eps \right)
    +
    \delta_{YZ} \left( \sqrt{\frac{\eps}{3}} \right)
    +
    \delta_{ZZ} \left( \frac{\eps}{ 3 \sqrt{\lambda_{\max}(\bfY\bfY^\tp)\lambda_{\max}(\bfZ\bfZ^\tp)} }, \eps_{\max} \right)
    + \delta_{ZZ} \left( \sqrt{\frac{\eps}{3}}, \eps_{\max} \right) .
\end{align*}
\end{theorem}

\begin{pf}
Decompose the system parameter error matrix using the least-squares estimators, as discussed earlier, as
\begin{align*}
    \big[\hat{A} ~ \hat{B} \big] - [A ~ B]
    &= 
    \hat{\bfY}\hat{\bfZ}^\tp (\hat{\bfZ}\hat{\bfZ}^\tp)^{\dagger} - \bfY\bfZ^\tp (\bfZ\bfZ^\tp)^{-1}\\
    &= 
    \underbrace{ \big[ \hat{\bfY}\hat{\bfZ}^\tp - \bfY\bfZ^\tp \big] (\bfZ\bfZ^\tp)^{-1}}_{\teL \Pi_1}
    + \underbrace{ \bfY\bfZ^\tp \big[ (\hat{\bfZ}\hat{\bfZ}^\tp)^{\dagger} - (\bfZ\bfZ^\tp)^{-1} \big]}_{\teL \Pi_2}
    + \underbrace{\big[ \hat{\bfY}\hat{\bfZ}^\tp - \bfY\bfZ^\tp \big] \big[ (\hat{\bfZ}\hat{\bfZ}^\tp)^{\dagger} - (\bfZ\bfZ^\tp)^{-1} \big]}_{\teL \Pi_3}.
\end{align*}
Considering a probability bound and using \eqref{eq:prob_addition_bound} we obtain
\begin{align*}
    \PP \left\{ 
    \big[\hat{A} ~ \hat{B} \big] - [A ~ B]    \geq \eps \right\}
    &=
    \PP \left\{ \Pi_1 + \Pi_2 + \Pi_3 \geq \eps \right\} \\
    &\leq 
    \PP \left\{ \Pi_1 \geq \eps/3 \right\} + \PP \left\{ \Pi_2 \geq \eps/3 \right\} + \PP \left\{ \Pi_3 \geq \eps/3 \right\} .
\end{align*}
For the first term, use the submultiplicative property, rearrange, and use \eqref{eq:inverse_of_spectral_norm} to obtain
\begin{align*}
    \PP \left\{ \Pi_1 \geq \frac{\eps}{3} \right\}
    &=
    \PP\left\{\big\| \big[ \hat{\bfY}\hat{\bfZ}^\tp - \bfY\bfZ^\tp \big] (\bfZ\bfZ^\tp)^{-1} \big\|_2 \ge \frac{\eps}{3} \right\} \\
    &\leq 
    \PP\left\{\big\| \hat{\bfY}\hat{\bfZ}^\tp - \bfY\bfZ^\tp \big\|_2 \big\| (\bfZ\bfZ^\tp)^{-1} \big\|_2 \ge \frac{\eps}{3} \right\}  \\
    &= 
    \PP\left\{\big\| \hat{\bfY}\hat{\bfZ}^\tp - \bfY\bfZ^\tp \big\|_2 \ge \frac{\eps}{3 \| (\bfZ\bfZ^\tp)^{-1} \|_2 } \right\}   \\
    &= 
    \PP\left\{ \big\| \hat{\bfY}\hat{\bfZ}^\tp - \bfY\bfZ^\tp \big\|_2 \ge \frac13 \lambda_{\min} (\bfZ\bfZ^\tp) \eps   \right\}  \\
    &\leq
    \delta_{YZ} \left( \frac13 \lambda_{\min} (\bfZ\bfZ^\tp) \eps \right),
\end{align*}
where the last step follows by applying Lemma \ref{lem:deltaYZ} with the appropriate setting of $\eps$.

For the second term, use submultiplicativity and rearrange to obtain
\begin{align*}
    \PP \left\{ \Pi_2 \geq \frac{\eps}{3} \right\}
    &=
    \PP\left\{ \big\| \bfY\bfZ^\tp \big[ (\hat{\bfZ}\hat{\bfZ}^\tp)^{\dagger} - (\bfZ\bfZ^\tp)^{-1} \big] \big\|_2 \ge \frac{\eps}{3} \right\} \\
    &\le 
    \PP\left\{ \| \bfY \|_2 \| \bfZ \|_2 \big\|  (\hat{\bfZ}\hat{\bfZ}^\tp)^{\dagger} - (\bfZ\bfZ^\tp)^{-1} \big\|_2 \ge \frac{\eps}{3} \right\} \\
    &=
    \PP\left\{ \big\|(\hat{\bfZ}\hat{\bfZ}^\tp)^{\dagger} - (\bfZ\bfZ^\tp)^{-1} \big\|_2 \ge \frac{\eps}{ 3 \|\bfY\|_2\|\bfZ\|_2 } \right\} \\
    &\leq
    \delta_{ZZ} \left( \frac{\eps}{ 3 \|\bfY\|_2\|\bfZ\|_2 }, \eps_{\max} \right),
\end{align*}
where the last step follows by applying Lemma \ref{lem:deltaZZ} with the appropriate setting of $\eps$.

For the third term, use submultiplicativity and \eqref{eq:prob_times_bound} to obtain
\begin{align*}
    \PP \left\{ \Pi_3 \geq \frac{\eps}{3} \right\}
    &=
    \PP\left\{ \big\| \big[ \hat{\bfY}\hat{\bfZ}^\tp - \bfY\bfZ^\tp \big] \big[ (\hat{\bfZ}\hat{\bfZ}^\tp)^{\dagger} - (\bfZ\bfZ^\tp)^{-1} \big] \big\|_2 \ge \frac{\eps}{3} \right\} \\
    &\le 
    \PP\left\{\big\| \hat{\bfY}\hat{\bfZ}^\tp - \bfY\bfZ^\tp \big\|_2 \big\| (\hat{\bfZ}\hat{\bfZ}^\tp)^{\dagger} - (\bfZ\bfZ^\tp)^{-1} \big\|_2 \ge \frac{\eps}{3} \right\} \\
    &\le 
     \PP\left\{ \big\| \hat{\bfY}\hat{\bfZ}^\tp - \bfY\bfZ^\tp  \big\|_2 \ge \sqrt{\frac{\eps}{3}} \right\}
    +\PP\left\{ \big\| (\hat{\bfZ}\hat{\bfZ}^\tp)^{\dagger} - (\bfZ\bfZ^\tp)^{-1} \big\|_2 \ge \sqrt{\frac{\eps}{3}} \right\} \\
    &\le 
    \delta_{YZ} \left( \sqrt{\frac{\eps}{3}} \right) + \delta_{ZZ} \left( \sqrt{\frac{\eps}{3}}, \eps_{\max} \right),
\end{align*}
where the last step follows by applying Lemmas \ref{lem:deltaYZ} and \ref{lem:deltaZZ} with the appropriate settings of $\eps$.
The conclusion follows by combining the probability bounds for each term.
\end{pf}

\begin{pf*}{PROOF OF THEOREM \ref{thm:hatAB_bounded}.} \ \\
The qualitative claim in Theorem \ref{thm:hatAB_bounded} is found by inverting the bound of Theorem \ref{thm:hatAB_bounded_restate} and examining the behavior of the bound as $n_r \to \infty$. 
To be specific, from Lemma~\ref{lem:bounded_Y_and_Z}, given fixed $\delta \in (0,1)$, we can find $\eps_Y(\delta)$ such that 
\begin{align*}
    \PP\big\{\big\|\hat{\bfY} - \bfY \big\|_2 \ge \eps_Y(\delta) \big\} \le \delta,
\end{align*}
where $\eps_Y(\delta)$ satisfies
\begin{align*}
    \delta_Y(\eps_Y(\delta)) = (n+\ell) \exp\left\{- \frac{3}{2} \cdot \frac{n_r \eps_Y^2(\delta)}{3 \ell c_N^2 + \eps_Y(\delta) \sqrt{\ell c_N^2}} \right\} = \delta.
\end{align*}
Solving for $\eps_Y(\delta)$ in terms of $\delta$ using the quadratic formula, we obtain
\begin{align*}
    \eps_Y(\delta) = \frac{1}{2n_r} \bigg( \frac{2}{3} \sqrt{\ell c_N^2} \log \frac{n+\ell}{\delta} \pm \sqrt{\frac49 \ell c_N^2 \log^2 \frac{n+\ell}{\delta} + 8 n_r \ell c_N^2 \log \frac{n+\ell}{\delta}} \bigg).
\end{align*}
Since $\eps_Y(\delta) \ge 0$, we have that
\begin{align*}
    \eps_Y(\delta) &= \frac{1}{2n_r} \bigg( \frac{2}{3} \sqrt{\ell c_N^2} \log \frac{n+\ell}{\delta} + \sqrt{\frac49 \ell c_N^2 \log^2 \frac{n+\ell}{\delta} + 8 n_r \ell c_N^2 \log \frac{n+\ell}{\delta}} \bigg)\\
    &=
    \frac{1}{3n_r} \sqrt{\ell c_N^2} \log \frac{n+\ell}{\delta} + \sqrt{\frac{1}{9n_r^2} \ell c_N^2 \log^2 \frac{n+\ell}{\delta} + \frac{2}{n_r} \ell c_N^2 \log \frac{n+\ell}{\delta}} \\
    &=
    \bigO \left( \sqrt{\frac{\ell c_N^2 \log [ (n+\ell)/\delta]}{n_r}} \right).
\end{align*}
An identical argument holds for $\big\|\hat{\bfZ} - \bfZ\big\|_2$.
Therefore, for fixed $\delta \in (0,1)$, with probability at least $1-\delta$,
\begin{align*}
    \big\|\hat{\bfY} - \bfY\big\|_2 < \eps_Y(\delta) = \bigO \left( \sqrt{\frac{\ell c_N^2 \log [ (n+\ell)/\delta]}{n_r}} \right),\\
    \big\|\hat{\bfZ} - \bfZ\big\|_2 < \eps_Y(\delta) = \bigO \left( \sqrt{ \frac{\ell c_N^2\log [ (n+\ell)/\delta]}{n_r}} \right).
\end{align*}
We proceed with this argument. Now from Lemma~\ref{lem:deltaYZ} it follows that for fixed $\delta \in (0,1)$ there is $\eps_{YZ}(\delta)$ such that
\begin{align*}
    \PP\big\{ \big\| \hat{\bfY}\hat{\bfZ}^\tp - \bfY\bfZ^\tp \big\|_2 \ge \eps_{YZ}(\delta) \big\} \le \delta_{YZ}(\eps_{YZ}(\delta)) = \delta.
\end{align*}
This implies
\begin{align*}
    \delta_{YZ}(\eps_{YZ}(\delta)) = \delta_Y  \left( \sqrt{\eps_{YZ}(\delta) + \left( \frac{\| \bfY \|_2 + \| \bfZ \|_2}{2}\right)^2 } - \frac{\| \bfY \|_2 + \| \bfZ \|_2}{2}  \right) = \delta,
\end{align*}
and solving for $\eps_{YZ}(\delta)$ we obtain
\begin{align*}
    \eps_{YZ}(\delta) = \eps_Y^2(\delta) + (\| \bfY \|_2 + \| \bfZ \|_2) \eps_Y(\delta).
\end{align*}
Thus when $n_r$ is large enough it holds that
\begin{align*}
    \eps_{YZ}(\delta) = \bigO \big( (\| \bfY \|_2 + \| \bfZ \|_2) \eps_Y(\delta) \big) = \bigO \left( \big(\sqrt{\lambda_{\max}(\bfY\bfY^\tp)} + \sqrt{\lambda_{\max}(\bfZ\bfZ^\tp)}\big) \sqrt{\frac{ \ell c_N^2 \log [ (n+\ell)/\delta]}{n_r}} \right),
\end{align*}
which is the bound of $\|\hat{\bfY}\hat{\bfZ}^\tp - \bfY\bfZ^\tp \|_2$ with probability at least $1-\delta$.

Similarly, under the condition of Lemma~\ref{lem:deltaZZ}, we have that for fixed $\delta \in (0,1)$ and large enough $n_r$, there are $\eps_0(\delta), \eps_m(\delta) > 0$ with $\delta_0(\eps_0(\delta)) = \delta$ and $\delta_m(\eps_m(\delta)) = \delta$ such that
\begin{align*}
    \eps_0(\delta) &= \eps_Y^2(\delta) + 2 \eps_Y(\delta) \sqrt{\lambda_{\max}(\bfZ\bfZ^\tp)} = \bigO \left( \sqrt{\lambda_{\max}(\bfZ\bfZ^\tp)} \sqrt{\frac{\ell c_N^2 \log [ (n+\ell)/\delta]}{n_r}} \right),\\
    \eps_m(\delta) &= \bigO \left( \sqrt{\lambda_{\max}(\bfZ\bfZ^\tp)} \sqrt{\frac{\ell c_N^2 \log \{ (n+\ell) [9^{n+m} + \left( 16 \lambda_{\max} (\bfZ\bfZ^\tp)/\lambda_{\min} (\bfZ\bfZ^\tp) + 1 \right)^{n+m}]/\delta\}}{n_r}} \right)
\end{align*}
For fixed $\delta,\eps_{\max} \in (0,1)$ and $\eps_{Z1}(\delta) \in (0,\eps_{\max})$, suppose that
\begin{align*}
    \delta_0 \left( \frac12 \lambda_{\min}^2 (\bfZ\bfZ^\tp)  \left(1 - \frac{\eps_{Z1}(\delta)}{\eps_{\max}}\right) \eps_{Z1}(\delta)  \right) = \delta,
\end{align*}
and we know that
\begin{align*}
    \frac12 \lambda_{\min}^2 (\bfZ\bfZ^\tp)  \left(1 - \frac{\eps_{Z1}(\delta)}{\eps_{\max}}\right) \eps_{Z1}(\delta) = \eps_0(\delta),
\end{align*}
which implies
\begin{align*}
    \eps_{Z1}(\delta) = \frac{1}{2} \eps_{\max} \bigg(1 \pm \sqrt{1 - \frac{8 \eps_0(\delta)}{\lambda_{\min}^2 (\bfZ\bfZ^\tp)}} \bigg).
\end{align*}
Because for any nonnegative random variable $X$ and constants $a > b > 0$, $\PP\{X \ge a\} \le \PP\{X \ge b\}$, we choose the smaller root. Thus, 
\begin{align*}
    \eps_{Z1}(\delta) &= \frac{1}{2} \eps_{\max} \bigg(1 - \sqrt{1 - \frac{8 \eps_0(\delta)}{\lambda_{\min}^2 (\bfZ\bfZ^\tp)}} \bigg)\\
    &= 
    \frac{1}{2} \eps_{\max} \bigg(1 - \bigg( 1 - \bigO \bigg(\frac{\eps_0(\delta)}{\lambda_{\min}^2 (\bfZ\bfZ^\tp)} \bigg) \bigg) \bigg)\\
    &=
    \bigO \left( \frac{\sqrt{\lambda_{\max}(\bfZ\bfZ^\tp)}}{\lambda_{\min}^2 (\bfZ\bfZ^\tp)} \sqrt{\frac{\ell c_N^2 \log [ (n+\ell)/\delta]}{n_r}} \right).
\end{align*}
For $\delta, \eps_{\max} \in (0,1)$ and $\eps_{Z2}(\delta) \in (0,\eps_{\max})$, if
\begin{align*}
    \delta_m \left( \frac{\eps_{Z2}(\delta) \lambda_{\min}(\bfZ\bfZ^\tp)}{ \eps_{\max} ( 2 + \lambda_{\min}(\bfZ\bfZ^\tp)/\lambda_{\max}(\bfZ\bfZ^\tp) ) } \right) = \delta,
\end{align*}
then
\begin{align*}
    \eps_{Z2}(\delta) &= \frac{\eps_{\max}(2 + \lambda_{\min}(\bfZ\bfZ^\tp)/\lambda_{\max}(\bfZ\bfZ^\tp))}{\lambda_{\min}(\bfZ\bfZ^\tp)} \eps_m(\delta)\\
    &=
    \bigO \left( \bigg(2 + \frac{\lambda_{\min}(\bfZ\bfZ^\tp)}{\lambda_{\max}(\bfZ\bfZ^\tp)} \bigg) \frac{\sqrt{\lambda_{\max}(\bfZ\bfZ^\tp)}}{\lambda_{\min}(\bfZ\bfZ^\tp)} \sqrt{\frac{\ell c_N^2 \log \{ (n+\ell) [9^{n+m} + \left( 16 \lambda_{\max} (\bfZ\bfZ^\tp)/\lambda_{\min} (\bfZ\bfZ^\tp) + 1 \right)^{n+m}]/\delta\}}{n_r}} \right)
\end{align*}
Note that $c_1 \eps^2/(c_2 + c_3 \eps)$ is monotonically increasing on $(0,+\infty)$ for any positive constants $c_1$, $c_2$, and $c_3$, so from the monotonicity of composite functions, we know that $\delta_Y(\eps)$ is monotonically decreasing. So are $\delta_{YZ}(\eps)$, $\delta_0(\eps)$, and $\delta_m(\eps)$. In addition, $(1 - \eps/\eps_{\max})\eps$ is monotonically increasing on $(0,\eps_{\max}/2)$ for fixed $\eps_{\max} > 0$, implying $\delta_{ZZ}(\eps,\eps_{\max})$ is monotonically decreasing on $(0,\eps_{\max}/2)$. Let $\eps_{ZZ}(\delta) := \max\{\eps_{Z1}(\delta/2),\eps_{Z2}(\delta/2)\}$. Since for fixed $\delta, \eps_{\max} \in (0,1)$, when $n_r$ is large enough, $\eps_{Z1}(\delta/2),\eps_{Z2}(\delta/2) < \eps_{\max}/2$, it holds that
\begin{align*}
    &\PP \big\{\big\| (\hat{\bfZ}\hat{\bfZ}^\tp)^{\dagger} - (\bfZ\bfZ^\tp)^{-1} \big\|_2 \ge \eps_{ZZ}(\delta) \big\} \\
    &\le \delta_{ZZ}(\eps_{ZZ}(\delta), \eps_{\max})\\
    &= 
    \delta_0 \left( \frac12 \lambda_{\min}^2 (\bfZ\bfZ^\tp)  \left(1 - \frac{\eps_{ZZ}(\delta)}{\eps_{\max}}\right) \eps_{ZZ}(\delta)  \right)
    +
    \delta_m \left( \frac{\eps_{ZZ}(\delta) \lambda_{\min}(\bfZ\bfZ^\tp)}{ \eps_{\max} ( 2 + \lambda_{\min}(\bfZ\bfZ^\tp)/\lambda_{\max}(\bfZ\bfZ^\tp) ) } \right)\\
    &\le
    \delta_0 \left( \frac12 \lambda_{\min}^2 (\bfZ\bfZ^\tp)  \left(1 - \frac{\eps_{Z1}(\delta/2)}{\eps_{\max}}\right) \eps_{Z1}(\delta/2)  \right)
    +
    \delta_m \left( \frac{\eps_{Z2}(\delta/2) \lambda_{\min}(\bfZ\bfZ^\tp)}{ \eps_{\max} ( 2 + \lambda_{\min}(\bfZ\bfZ^\tp)/\lambda_{\max}(\bfZ\bfZ^\tp) ) } \right)
    =\delta.
\end{align*}
Therefore, we know that with probability at least $1-\delta$, 
\begin{align*}
    &\big\| (\hat{\bfZ}\hat{\bfZ}^\tp)^{\dagger} - (\bfZ\bfZ^\tp)^{-1} \big\|_2 \\
    &< \eps_{ZZ}(\delta)\\
    &=
    \bigO \Bigg( \max \Bigg\{\frac{\sqrt{\lambda_{\max}(\bfZ\bfZ^\tp)}}{\lambda_{\min}^2 (\bfZ\bfZ^\tp)} \sqrt{\frac{\ell c_N^2 \log [ 2(n+\ell)/\delta]}{n_r}}, \\
    & \bigg(2 + \frac{\lambda_{\min}(\bfZ\bfZ^\tp)}{\lambda_{\max}(\bfZ\bfZ^\tp)} \bigg) \frac{\sqrt{\lambda_{\max}(\bfZ\bfZ^\tp)}}{\lambda_{\min}(\bfZ\bfZ^\tp)} \sqrt{\frac{\ell c_N^2 \log \{2 (n+\ell) [9^{n+m} + \left( 16 \lambda_{\max} (\bfZ\bfZ^\tp)/\lambda_{\min} (\bfZ\bfZ^\tp) + 1 \right)^{n+m}]/\delta\}}{n_r}} \Bigg\}\Bigg).
\end{align*}
Similarly, let 
\begin{align}\label{eq_append_AB_epsAB}
    \eps_{AB}(\delta) := \max \bigg\{ \frac{3\eps_{YZ}(\delta/4)}{\lambda_{\min}(\bfZ\bfZ^\tp)}, 3 \eps^2_{YZ}(\delta/4), 3\sqrt{\lambda_{\max}(\bfY\bfY^\tp)\lambda_{\max}(\bfZ\bfZ^\tp)}\eps_{ZZ}(\delta/4), 3\eps^2_{ZZ}(\delta/4) \bigg\},
\end{align}
and from Theorem~\ref{thm:hatAB_bounded_restate} it holds for large enough $n_r$ that
\begin{align*}
    &\PP\big\{ \big\|
    \big[\hat{A} ~ \hat{B} \big] - [A ~ B]
    \big\|_2 \ge \eps_{AB}(\delta) \big\} \\
    &\le \delta_{AB}(\eps_{AB}(\delta))\\
    &= 
    \delta_{YZ} \left( \frac13 \lambda_{\min} (\bfZ\bfZ^\tp) \eps_{AB}(\delta) \right)
    +
    \delta_{YZ} \left( \sqrt{\frac{\eps_{AB}(\delta)}{3}} \right)+
    \delta_{ZZ} \left( \frac{\eps_{AB}(\delta)}{ 3 \sqrt{\lambda_{\max}(\bfY\bfY^\tp)\lambda_{\max}(\bfZ\bfZ^\tp)} }, \eps_{\max} \right)\\
    &\quad + \delta_{ZZ} \left( \sqrt{\frac{\eps_{AB}(\delta)}{3}}, \eps_{\max} \right)\\
    &\le
    \delta_{YZ} ( \eps_{YZ}(\delta/4) )
    +
    \delta_{YZ} ( \eps_{YZ}(\delta/4) )+
    \delta_{ZZ} (\eps_{ZZ}(\delta/4), \eps_{\max}) + \delta_{ZZ} (\eps_{ZZ}(\delta/4), \eps_{\max}) = \delta.
\end{align*}
Therefore, we have that with probability $1-\delta$, for large enough~$n_r$, 
\begin{align*}
    \big\|
    \big[\hat{A} ~ \hat{B} \big] - [A ~ B]
    \big\|_2 < \eps_{AB}(\delta),
\end{align*}
where
\begin{align*}
    &\eps_{AB}(\delta) \\
    &= \bigO \Bigg( \max\Bigg\{  \frac{\sqrt{\lambda_{\max}(\bfY\bfY^\tp)} + \sqrt{\lambda_{\max}(\bfZ\bfZ^\tp)}}{\lambda_{\min}(\bfZ\bfZ^\tp)} \sqrt{\frac{\ell c_N^2 \log [ 4(n+\ell)/\delta]}{n_r}},\\
    &\Bigg(\big(\sqrt{\lambda_{\max}(\bfY\bfY^\tp)} + \sqrt{\lambda_{\max}(\bfZ\bfZ^\tp)}\big) \sqrt{\frac{\ell c_N^2 \log [ 4(n+\ell)/\delta]}{n_r}} \Bigg)^2, \frac{\sqrt{\lambda_{\max}(\bfY\bfY^\tp)}\lambda_{\max}(\bfZ\bfZ^\tp)}{\lambda_{\min}^2 (\bfZ\bfZ^\tp)} \sqrt{\frac{\ell c_N^2 \log [ 8(n+\ell)/\delta]}{n_r}}, \\
    & \bigg(2 + \frac{\lambda_{\min}(\bfZ\bfZ^\tp)}{\lambda_{\max}(\bfZ\bfZ^\tp)} \bigg) \frac{\sqrt{\lambda_{\max}(\bfY\bfY^\tp)}\lambda_{\max}(\bfZ\bfZ^\tp)}{\lambda_{\min}(\bfZ\bfZ^\tp)} \sqrt{\frac{\ell c_N^2 \log \{8 (n+\ell) [9^{n+m} + \left( 16 \lambda_{\max} (\bfZ\bfZ^\tp)/\lambda_{\min} (\bfZ\bfZ^\tp) + 1 \right)^{n+m}]/\delta\}}{n_r}},\\
    &\Bigg(\frac{\sqrt{\lambda_{\max}(\bfY\bfY^\tp)}\lambda_{\max}(\bfZ\bfZ^\tp)}{\lambda_{\min}^2 (\bfZ\bfZ^\tp)} \sqrt{\frac{\ell c_N^2 \log [ 8(n+\ell)/\delta]}{n_r}} \Bigg)^2, \\
    & \Bigg( \bigg(2 + \frac{\lambda_{\min}(\bfZ\bfZ^\tp)}{\lambda_{\max}(\bfZ\bfZ^\tp)} \bigg) \frac{\sqrt{\lambda_{\max}(\bfY\bfY^\tp)}\lambda_{\max}(\bfZ\bfZ^\tp)}{\lambda_{\min}(\bfZ\bfZ^\tp)}\\
    &\sqrt{\frac{\ell c_N^2 \log \{8 (n+\ell) [9^{n+m} + \left( 16 \lambda_{\max} (\bfZ\bfZ^\tp)/\lambda_{\min} (\bfZ\bfZ^\tp) + 1 \right)^{n+m}]/\delta\}}{n_r}} \Bigg)^2 \Bigg\} \Bigg)\\
    &=
    \bigO \Bigg( \max\Bigg\{  \frac{\sqrt{\lambda_{\max}(\bfY\bfY^\tp)} + \sqrt{\lambda_{\max}(\bfZ\bfZ^\tp)}}{\lambda_{\min}(\bfZ\bfZ^\tp)} \sqrt{\frac{\ell c_N^2 \log [ 4(n+\ell)/\delta]}{n_r}}, \frac{\sqrt{\lambda_{\max}(\bfY\bfY^\tp)}\lambda_{\max}(\bfZ\bfZ^\tp)}{\lambda_{\min}^2 (\bfZ\bfZ^\tp)} \sqrt{\frac{\ell c_N^2 \log [ 8(n+\ell)/\delta]}{n_r}}, \\
    & \bigg(2 + \frac{\lambda_{\min}(\bfZ\bfZ^\tp)}{\lambda_{\max}(\bfZ\bfZ^\tp)} \bigg) \frac{\sqrt{\lambda_{\max}(\bfY\bfY^\tp)}\lambda_{\max}(\bfZ\bfZ^\tp)}{\lambda_{\min}(\bfZ\bfZ^\tp)} \sqrt{\frac{\ell c_N^2 \log \{8 (n+\ell) [9^{n+m} + \left( 16 \lambda_{\max} (\bfZ\bfZ^\tp)/\lambda_{\min} (\bfZ\bfZ^\tp) + 1 \right)^{n+m}]/\delta\}}{n_r}} \Bigg\} \Bigg)\\
    &=
    \bigO \Bigg( \max\Bigg\{  \frac{\sqrt{\lambda_{\max}(\bfY\bfY^\tp)} + \sqrt{\lambda_{\max}(\bfZ\bfZ^\tp)}}{\lambda_{\min}(\bfZ\bfZ^\tp)} \sqrt{\frac{\ell c_N^2 \log [ 4(n+\ell)/\delta]}{n_r}}, \frac{\sqrt{\lambda_{\max}(\bfY\bfY^\tp)}\lambda_{\max}(\bfZ\bfZ^\tp)}{\lambda_{\min}^2 (\bfZ\bfZ^\tp)} \sqrt{\frac{\ell c_N^2 \log [ 8(n+\ell)/\delta]}{n_r}}, \\
    &\sqrt{\lambda_{\max}(\bfY\bfY^\tp)} \bigg(1 + \frac{2\lambda_{\max}(\bfZ\bfZ^\tp)}{\lambda_{\min} (\bfZ\bfZ^\tp)} \bigg) \sqrt{\frac{\ell c_N^2 \log \{8 (n+\ell) [9^{n+m} + \left( 16 \lambda_{\max} (\bfZ\bfZ^\tp)/\lambda_{\min} (\bfZ\bfZ^\tp) + 1 \right)^{n+m}]/\delta\}}{n_r}} \Bigg\} \Bigg).
\end{align*}
The qualitative claim in Theorem \ref{thm:hatAB_bounded} follows by dropping dependence on quantities other than $\delta$, $n_r$, and $\ell$.
\end{pf*}

\begin{remark}
    It can be seen from the above bound that smaller $\lambda_{\max}(\bfY\bfY^\tp)$, $\lambda_{\max}(\bfZ\bfZ^\tp)$, $c_N$, and larger $\lambda_{\min}(\bfZ\bfZ^\tp)$, all of which depend on both system parameters and input design, yield faster convergence speed of Algorithm~\ref{alg:A}. The exponential term of $n+m$ is technical and could be tightened \citep{tropp2012user, wainwright2019high}. In addition, the estimation error has higher order terms, e.g., $\bigO(1/n_r)$, which may be relatively large when $n_r$ is small. This could explain the performance of Algorithm~\ref{alg:A} with small number of rollouts in simulation.
\end{remark}

\clearpage

\section{Proof of Theorem \ref{thm:hatSigmaAB_bounded}}\label{append:pf_thm:hatSigmaAB_bounded}
Now we derive bounds for $[\hat{\tilde{\Sigma}}_A'~\hat{\tilde{\Sigma}}_B'] - [\tilde{\Sigma}_A^\prime~\tilde{\Sigma}_B^\prime ]$. The proofs follow a similar structure to that of the proofs for bounds on $[\hat{A}~\hat{B}] - [A~B]$, but with more complicated expressions due to the greater complexity of the second-moment dynamic.

Throughout this section, small probability bounds are denoted by $\eta_{[\cdot]}$, where $[\cdot]$ are various subscripts, and each of these bounds decreases monotonically towards $0$ with increasing number of rollouts $n_r$.

Recalling notations in Section \ref{sec:parameterEstimation}, we have
\begin{align*}
    &\hat{\tilde{X}}_t = \frac{1}{n_r} P_1 \vect \left( \sum_{k = 1}^{n_r} x_t^{(k)} (x_t^{(k)})^\tp \right),~    \tilde{X}_t = P_1 \vect(\EE\{x_tx_t^\tp\}), \\
    &\tilde{U}_t = P_2 \vect(\bar{U}_t + \nu_t\nu_t^\tp),\\
    &\hat{W}_t = \frac{1}{n_r} \vect \left( \sum_{k = 1}^{n_r} x_t^{(k)} \nu_t^\tp \right),~ W_t = \vect(\EE\{x_t u_t^\tp\}),\\
    &\hat{W}_t' = \frac{1}{n_r} \vect \left( \sum_{k = 1}^{n_r} \nu_t ({x_t^{(k)}})^\tp \right), ~ W_t^\prime = \vect(\EE \{u_t^\tp x_t\}),\\
    &\hat{\tilde{A}} = P_1 (\hat{A} \otimes \hat{A}) Q_1, \quad \tilde{A} = P_1 (A \otimes A) Q_1,\\
    &\hat{\tilde{B}} = P_1 (\hat{B} \otimes \hat{B}) Q_2, \quad \tilde{B} = P_1 (B \otimes B) Q_2,\\
    &\hat{K}_{BA} = P_1 (\hat{B} \otimes \hat{A}), \quad K_{BA} = P_1 (B \otimes A),\\
    &\hat{K}_{AB} = P_1 (\hat{A} \otimes \hat{B}), \quad K_{AB} = P_1 (A \otimes B).
\end{align*}
Further denote
\begin{align*}
    \mathbf{M}_1
    \Let
    \big[\tilde{X}_{\ell-1} \cdots \tilde{X}_0 \big] , ~
    \mathbf{L}_1
    \Let
    \big[W_{\ell-1} \cdots W_0 \big] ,~
    \mathbf{U}
    \Let
    \big[\tilde{U}_{\ell-1} \cdots \tilde{U}_0 \big] .
\end{align*}

\begin{lemma}\label{lem:bounded_C_and_D}
Suppose Assumptions~\ref{asmp1} and~\ref{asmp:bounded_system} hold. Then for all $\eps > 0$,
\begin{align*}
    \PP \big\{\big\| \hat{\bfD} - \bfD \big\|_2 \ge \eps \big\} &\le \eta_D(\eps),
\end{align*}
where
\begin{align*}
    \eta_D(\eps)
    &\Let 
    \left( \frac{n(n+1)}{2} + \ell \right) \exp\left\{- \frac{3}{2} \cdot \frac{n_r \eps^2}{3 \ell c_F^2 + \eps \sqrt{\ell c_F^2}} \right\},
\end{align*}
and
\begin{align*}
    \PP \big\{\big\| \hat{\bfC} - \bfC \big\|_2 \ge \eps \big\} &\le \eta_C(\eps),
\end{align*}
where
\begin{align*}
    \eta_C(\eps)
    &\Let
    \eta_{D} (\eps/5)
    + \eta_{AM} (\eps/5)
    + 2 \eta_{KL} (\eps/5)
    + \eta_B \left( \frac{\eps}{5\|\mathbf{U}\|_2} \right) ,\\
    \eta_{AM} (\eps)
    &\Let
    \eta_A \left(\frac{\eps}{3 \| \mathbf{M}_1\|_2}\right)
    +
    \eta_D \left( \frac{\eps}{6 \| A \|_2^2 } \right) 
    +
    \eta_A( \sqrt{\eps/3} ) + \eta_D( \sqrt{\eps/3} ) ,\\
    \eta_{KL} (\eps)
    &\Let
    \eta_{AB} \left( \frac{\eps}{3 \| \mathbf{L}_1 \|_2 } \right)
    +
    \eta_{L} \left( \frac{\eps}{3 \| A \|_2 \| B \|_2} \right)
    +
    \eta_{AB} \left( \sqrt{\eps/3} \right) + \eta_{L} \left( \sqrt{\eps/3} \right) ,\\
    \eta_A(\eps)
    &\Let
    \delta_{AB} \left(\sqrt{\eps}/2\right) +  \delta_{AB} \left(\eps/(8\sqrt{\| A \|_2})\right) ,\\
    \eta_B(\eps)
    &\Let
    \delta_{AB} \left(\sqrt{\eps}/2\right) +  \delta_{AB} \left(\eps/(8\sqrt{\| B \|_2})\right) ,\\
    \eta_{AB}(\eps)
    &\Let
    2\delta_{AB} \left(\sqrt{\eps/3}\right) + \delta_{AB} \left(\eps/(3\sqrt{\| B \|_2})\right) + \delta_{AB} \left(\eps/(3\sqrt{\| A \|_2})\right) ,\\
    \eta_{L}(\eps)
    &\Let
    \left( nm + \ell \right) \exp\left\{- \frac{3}{2} \cdot \frac{n_r \eps^2}{3 \ell c_W^2 + \eps \sqrt{\ell c_W^2}} \right\}.
\end{align*}
Here, $\delta_{AB}(\eps) = \delta_{AB}(\eps,\eps_{\max})$, $\eps\in(0,\eps_{\max})$ and $\eps_{\max} \in (0,1)$, is defined in Theorem \ref{thm:hatAB_bounded_restate}, and we omit $\eps_{\max}$ for simplicity.
\end{lemma}

\begin{pf}
Denote
\begin{align*}
\begin{array}[t]{c@{\,} c@{\,} c@{\quad} c@{\,} c@{\,} c@{\,}}
    \mathbf{M}_1 &\Let &\big[\tilde{X}_{\ell-1} \cdots \tilde{X}_0 \big], &\hat{\mathbf{M}}_1 &\Let &\big[\hat{\tilde{X}}_{\ell-1} \cdots \hat{\tilde{X}}_0 \big], \\
    \mathbf{M}_2 &\Let &\big[\tilde{X}_{\ell} \cdots \tilde{X}_1 \big],   &\hat{\mathbf{M}}_2 &\Let &\big[\hat{\tilde{X}}_{\ell} \cdots \hat{\tilde{X}}_1 \big], \\
    \mathbf{L}_1 &\Let &[W_{\ell-1} \cdots W_0 ], &\hat{\mathbf{L}}_1 &\Let &\big[\hat{W}_{\ell-1} \cdots \hat{W}_0 \big], \\
    \mathbf{L}_2 &\Let &[W_{\ell-1}^\prime \cdots W_0^\prime ], &\hat{\mathbf{L}}_2 &\Let &\big[\hat{W}_{\ell-1}^\prime \cdots \hat{W}_0^\prime \big], \\
    \mathbf{U}   &\Let &\big[\tilde{U}_{\ell-1} \cdots \tilde{U}_0 \big], & \ & \ & \
\end{array}
\end{align*}
which will be used both for the development of the bound on $\big\| \hat{\bfC} - \bfC \big\|_2$ and on $\big\| \hat{\bfD} - \bfD \big\|_2$.

We begin by justifying the claim regarding a bound on $\big\| \hat{\bfD} - \bfD \big\|_2$.
We make the new definitions
\begin{align*}
    \mathbf{X}_k
    & \Let 
    \begin{bmatrix}
    \vect \Big( x_{\ell-1}^{(k)} ({x_{\ell-1}^{(k)}})^\tp - \EE \big\{ x_{\ell-1}^{(k)} ({x_{\ell-1}^{(k)}})^\tp \big\} \Big) & \cdots & \vect \Big( x_{0}^{(k)} ({x_{0}^{(k)}})^\tp - \EE \big\{ x_{0}^{(k)} ({x_{0}^{(k)}})^\tp \big\} \Big)
    \end{bmatrix}, \\
    \mathbf{\widetilde{X}}_k
    & \Let 
    P_1 \mathbf{X}_k,
\end{align*}
so that
\begin{align*}
    \hat{\mathbf{M}}_1 - \mathbf{M}_1 = \frac{1}{n_r} \sum_{k=1}^{n_r} \mathbf{\widetilde{X}}_k.
\end{align*}
Considering a single column of $\mathbf{X}_k$, we use the bound from Lemma \ref{lem:bounded_system} to obtain
\begin{align*}
    \big\| \mathbf{\widetilde{X}}_k \big\|_2
    &\leq
    \|P_1\|_2 \| \mathbf{X}_k \|_2 \tag{by submultiplicativity}\\
    &=
    \| \mathbf{X}_k \|_2 \tag{since $\|P_1\|_2=1$ by definition of $P_1$}\\
    &\leq
    \left\| \mathbf{X}_k  \right\|_F \tag{by ordering of $\| \cdot \|_2$ and $\| \cdot \|_F$}\\
    &= 
    \sqrt{ \sum_{t=0}^{\ell-1} \left\| \vect\left( x_{t}^{(k)} ({x_{t}^{(k)}})^\tp - \EE \big\{ x_{t}^{(k)} ({x_{t}^{(k)}})^\tp \big\} \right) \right\|^2 } \tag{by definition of $\mathbf{X}_k$, $\vect$, $\| \cdot \|_F$} \\
    &\leq
    \sqrt{\ell c_F^2}.
\end{align*}

Notice that
\begin{align}
    \big\| \hat{\bfD} - \bfD \big\|_2 
    &=
    \left\|  \begin{bmatrix}
    \hat{\tilde{X}}_{\ell-1} & \cdots & \hat{\tilde{X}}_0\\
    \tilde{U}_{\ell-1} & \cdots & \tilde{U}_0
    \end{bmatrix} - \begin{bmatrix}
    \tilde{X}_{\ell-1} & \cdots & \tilde{X}_0\\
    \tilde{U}_{\ell-1} & \cdots & \tilde{U}_0
    \end{bmatrix} \right\|_2   \nonumber \\
    &=
    \left\|  \begin{bmatrix}
    \hat{\tilde{X}}_{\ell-1} & \cdots & \hat{\tilde{X}}_0 \end{bmatrix} - \begin{bmatrix} \tilde{X}_{\ell-1} & \cdots & \tilde{X}_0 \end{bmatrix} \right\|_2 \nonumber\\
    &=
    \left\| \frac{1}{n_r} \sum_{k=1}^{n_r} \mathbf{\widetilde{X}}_k \right\|_2  
\end{align}
Thus we have the small probability bound
\begin{align*}
    \PP \big\{ \big\| \hat{\bfD} - \bfD \big\|_2 \geq \eps \big\}
    \leq 
    \eta_D (\eps),
\end{align*}
where
\begin{align*}
    \eta_D (\eps)
    \Let
    \left( \frac12 n(n+1) + \ell \right) \exp\left\{- \frac{3}{2} \cdot \frac{n_r \eps^2}{3 \ell c_F^2 + \eps \sqrt{\ell c_F^2}} \right\},
\end{align*}
which follows by applying Corollary \ref{cor:bernstein} with $Y_k = \mathbf{\widetilde{X}}_k$, $N = n_r$, and $M = \sqrt{ \ell c_F^2}$.

We now justify the claim regarding a bound on $\big\| \hat{\bfC} - \bfC \big\|_2$. The prior statement implies
\begin{align}
    \PP \big\{\big\| \hat{\mathbf{M}}_1 - \mathbf{M}_1 \big\|_2 \ge \eps \big\} 
    \leq
    \eta_D(\eps) . \label{eq:bound_of_M1}
\end{align}
An identical argument, but shifting the time indices of all terms by 1, leads to the bound
\begin{align}
    \PP \big\{\big\| \hat{\mathbf{M}}_2 - \mathbf{M}_2 \big\|_2 \ge \eps \big\} 
    \le 
    \eta_D(\eps) . \label{eq:bound_of_M2}
\end{align}
We will also need probabilistic bounds on the cross-terms $\hat{\mathbf{L}}_1 - \mathbf{L}_1$ and $\hat{\mathbf{L}}_2 - \mathbf{L}_2$.
To this end, make the new definition
\begin{align*}
    \mathbf{W}_k
    & \Let 
    \begin{bmatrix}
    \vect\left( x_{\ell-1}^{(k)} ({u_{\ell-1}^{(k)}})^\tp - \EE \big\{ x_{\ell-1}^{(k)} ({u_{\ell-1}^{(k)}})^\tp \big\} \right) & \cdots & \vect\left( x_{0}^{(k)} ({u_{0}^{(k)}})^\tp - \EE \big\{ x_{0}^{(k)} ({u_{0}^{(k)}})^\tp \big\} \right)
    \end{bmatrix},
\end{align*}
so that
\begin{align*}
    \hat{\mathbf{L}}_1 - \mathbf{L}_1 = \frac{1}{n_r} \sum_{k=1}^{n_r} \mathbf{W}_k .
\end{align*}
Considering a single column of $\mathbf{W}_k$, we use the bound from Lemma \ref{lem:bounded_system} to obtain
\begin{align*}
    \| \mathbf{W}_k \|_2 
    &\leq
    \left\| \mathbf{W}_k  \right\|_F \tag{by ordering of $\| \cdot \|_2$ and $\| \cdot \|_F$}\\
    &= 
    \sqrt{ \sum_{t=0}^{\ell-1} \left\| \vect\left( x_{t}^{(k)} ({u_{t}^{(k)}})^\tp - \EE \big\{ x_{t}^{(k)} ({u_{t}^{(k)}})^\tp \big\} \right) \right\|^2 } \tag{by definition of $\mathbf{W}_k$, $\vect$, $\| \cdot \|_F$} \\
    &\leq
    \sqrt{\ell c_W^2}.
\end{align*}
Thus we have the probability bound
\begin{align}
    \PP \big\{ \big\| \hat{\mathbf{L}}_1 - \mathbf{L}_1 \big\|_2 \geq \eps \big\}
    \leq 
    \eta_L(\eps), \label{eq:bound_of_L1}
\end{align}
where 
\begin{align*}
    \eta_L (\eps)
    \Let
    \left( nm + \ell \right) \exp\left\{- \frac{3}{2} \cdot \frac{n_r \eps^2}{3 \ell c_W^2 + \eps \sqrt{\ell c_W^2}} \right\} ,
\end{align*}
which follows by applying Corollary \ref{cor:bernstein} with $Y_k = \mathbf{W}_k$, $N = n_r$, and $M = \sqrt{ \ell c_W^2}$. 
An identical argument yields the same bound for $\hat{\mathbf{L}}_2 - \mathbf{L}_2$, i.e.
\begin{align}
    \PP \big\{ \big\| \hat{\mathbf{L}}_2 - \mathbf{L}_2 \big\|_2 \geq \eps \big\}
    \leq 
    \eta_L(\eps) . \label{eq:bound_of_L2}
\end{align}


Denote the optimal estimation error bounds on $A$ and $B$ as
\begin{align*}
    \delta_{A,*}(\eps) \Let \PP \big\{\|\hat{A} - A\|_2 \ge \eps \big\}, ~\delta_{B,*}(\eps) \Let \PP \big\{\|\hat{B} - B\|_2 \ge \eps \big\}.
\end{align*}
By Theorem \ref{thm:hatAB_bounded} we know $\delta_{A,*}(\eps) \leq \delta_{AB}(\eps)$ and $\delta_{B,*}(\eps) \leq \delta_{AB}(\eps)$, so we can use the computable bound $\delta_{AB}(\eps)$ in Theorem \ref{thm:hatAB_bounded} as a conservative approximation of $\delta_{A,*}(\eps)$ and $\delta_{B,*}(\eps)$.

From the assumption of Theorem \ref{thm:hatSigmaAB_bounded}, it holds that
\begin{align}
    &\PP \big\{ \big\| \hat{\tilde{A}} - \tilde{A} \big\|_2 \ge \eps \big\} \nonumber\\
    &=
    \PP \big\{ \big\| P_1 (\hat{A} \otimes \hat{A}) Q_1 - P_1 (A \otimes A) Q_1 \big\|_2 \ge \eps \big\} \tag{definition of $\hat{\tilde{A}}$, $\tilde{A}$} \nonumber \\
    &\le
    \PP \big\{ \|P_1\|_2\big \| \hat{A} \otimes \hat{A} - A \otimes A \big\|_2 \|Q_1\|_2 \ge \eps \big\} \tag{submultiplicativity} \nonumber\\
    &\le
    \PP \big\{2 \big\| \hat{A} \otimes \hat{A} - A \otimes A \big\|_2  \ge \eps \big\} \tag{by definition, $\| P_1 \|_2 = 1$, $\|Q_1\|_2 \leq 2$} \nonumber\\
    &=
    \PP \big\{ \big\|(\hat{A} - A) \otimes (\hat{A} - A) + (\hat{A} - A) \otimes A + A \otimes (\hat{A} - A) \big\|_2 \ge \eps/2 \big\}\nonumber\\
    &\le 
    \PP \big\{ \big\|(\hat{A} - A) \otimes (\hat{A} - A) \big\|_2 \ge \eps/4 \big\} + \PP \big\{ \big\| (\hat{A} - A) \otimes A + A \otimes (\hat{A} - A) \big\|_2 \ge \eps/4 \big\} \tag{by \ref{eq:prob_addition_bound}} \nonumber\\
    &= 
    \PP \big\{ \big\|\hat{A} - A \big\|_2 \ge \sqrt{\eps}/2 \big\} + \PP \big\{ \big\|\hat{A} - A \big\|_2 \ge \eps/(8\|A\|_2) \big\} \tag{$\|A\otimes B\|_2 = \|A\|_2 \|B\|_2$} \nonumber \\
    &= 
    \delta_{A,*}\left(\sqrt{\eps}/2\right) +  \delta_{A,*}\left(\eps/(8\sqrt{\| A \|})\right) \nonumber \\
    &\leq 
    \delta_{AB}\left(\sqrt{\eps}/2\right) +  \delta_{AB}\left(\eps/(8\sqrt{\| A \|})\right) \\
    & \teL
    \eta_A(\eps) \label{eq:bound_of_hat_tilde_A},
\end{align}
and by an identical argument
\begin{align}
    &\PP \big\{ \big\| \hat{\tilde{B}} - \tilde{B} \big\|_2 \ge \eps \big\} \nonumber\\
    &\leq \delta_{B,*}\left(\sqrt{\eps}/2\right) +  \delta_{B,*}\left(\eps/(8\sqrt{\| A \|})\right) \nonumber \\
    &\leq \delta_{AB}\left(\sqrt{\eps}/2\right) +  \delta_{AB}\left(\eps/(8\sqrt{\| B \|})\right) \teL \eta_B(\eps) . \label{eq:bound_of_hat_tilde_B}
\end{align}
Similarly,
\begin{align}\nonumber
    &\PP \big\{ \big\| \hat{K}_{AB} - K_{AB} \big\|_2 \ge \eps \big\}\\
    &=\PP \big\{ \big\| \hat{K}_{BA} - K_{BA} \big\|_2 \ge \eps \big\} \tag{by symmetry} \\\nonumber
    &=
    \PP \big\{ \big\| P_1 (\hat{A} \otimes \hat{B}) - P_1 (A \otimes B) \big\|_2 \ge \eps \big\}\\\nonumber
    &=
    \PP \big\{\| P_1 \|_2 \big\|(\hat{A} \otimes \hat{B}) - (A \otimes B) \big\|_2 \ge \eps \big\}\\\nonumber
    &\le
    \PP \big\{ \big\|(\hat{A} - A) \otimes (\hat{B} - B) + (\hat{A} - A) \otimes B + A \otimes (\hat{B} - B) \big\|_2 \ge \eps \big\}\\\nonumber
    &\le 
    \PP \big\{ \big\|(\hat{A} - A) \otimes (\hat{B} - B) \big\|_2 \ge \eps/3 \big\} + \PP \big\{ \big\|(\hat{A} - A) \otimes B \big\|_2 \ge \eps/3 \big\}  + \PP \big\{ \big\|A \otimes (\hat{B} - B) \big\|_2 \ge \eps/3 \big\} \\\nonumber
    &\le 
    \PP \big\{ \big\|\hat{A} - A \big\|_2 \ge \sqrt{\eps/3} \big\} + \PP \big\{ \big\|\hat{B} - B \big\|_2 \ge \sqrt{\eps/3} \big\} + \PP \big\{ \big\|\hat{A} - A \big\|_2 \ge \eps/(3\|B\|_2) \big\} + \PP \big\{ \big\|\hat{B} - B \big\|_2 \ge \eps/(3\|A\|_2) \big\}\\\nonumber
    &= 
    \delta_{A,*}\left(\sqrt{\eps/3}\right) + \delta_{B,*}\left(\sqrt{\eps/3}\right) + \delta_{A,*}\left(\eps/(3\sqrt{\| B \|})\right) + \delta_{B,*} \left(\eps/(3\sqrt{\| A \|})\right) \\
    &\leq
    2\delta_{AB}\left(\sqrt{\eps/3}\right) + \delta_{AB}\left(\eps/(3\sqrt{\| B \|})\right) + \delta_{AB} \left(\eps/(3\sqrt{\| A \|})\right) \\
    &\teL \eta_{AB}(\eps) \label{eq:bound_of_Kab}.
\end{align}

Consider the decomposition of $\hat{\bfC} - \bfC$ as
\begin{align}\nonumber
    &\hat{\bfC} - \bfC \\
    &=
    \big( \big[ \hat{\tilde{X}}_{\ell} \cdots \hat{\tilde{X}}_1 \big] - \big[\tilde{X}_{\ell} \cdots \tilde{X}_1 \big] \big)
    - \big(\hat{\tilde{A}} \big[\hat{\tilde{X}}_{\ell-1} \cdots \hat{\tilde{X}}_0 \big] - \tilde{A} \big[\tilde{X}_{\ell-1} \cdots \tilde{X}_0 \big]    \big) \nonumber \\
    &\quad - \big( \hat{K}_{BA} \big[\hat{W}_{\ell - 1} \cdots \hat{W}_0 \big] - K_{BA} \big[W_{\ell-1} \cdots W_0 \big]  \big)
    - \big( \hat{K}_{AB} \big[\hat{W}'_{\ell - 1} \cdots \hat{W}'_0 \big] - K_{AB} \big[W'_{\ell-1} \cdots W'_0 \big] \big) \nonumber \\
    &\quad - \big( \hat{\tilde{B}} \big[\tilde{U}_{\ell-1} \cdots \tilde{U}_0 \big] - \tilde{B} \big[\tilde{U}_{\ell-1} \cdots \tilde{U}_0 \big]  \big) \nonumber \\
    &=
    \big( \hat{\mathbf{M}}_2 - \mathbf{M}_2 \big) 
    - \big(\hat{\tilde{A}} \hat{\mathbf{M}}_1 - \tilde{A} \mathbf{M}_1 \big) 
    - \big( \hat{K}_{BA} \hat{\mathbf{L}}_1 - K_{BA} \mathbf{L}_1  \big) 
    - \big( \hat{K}_{AB} \hat{\mathbf{L}}_2 - K_{AB} \mathbf{L}_2 \big) 
    - \big( \hat{\tilde{B}} - \tilde{B}\big) \mathbf{U} \label{eq:form_of_hatC}.
\end{align}
We treat each of these five terms separately.

For the first term, $\hat{\mathbf{M}}_2 - \mathbf{M}_2$, we have the bound in \eqref{eq:bound_of_M2}.

For the second term, $\hat{\tilde{A}} \hat{\mathbf{M}}_1 - \tilde{A} \mathbf{M}_1$, we have the decomposition
\begin{align*}
    \hat{\tilde{A}} \hat{\mathbf{M}}_1 - \tilde{A} \mathbf{M}_1
    =
    \big(\hat{\tilde{A}} - \tilde{A}\big) \mathbf{M}_1 + \tilde{A} \big(\hat{\mathbf{M}}_1 - \mathbf{M}_1\big) + \big(\hat{\tilde{A}} - \tilde{A} \big) \big(\hat{\mathbf{M}}_1 - \mathbf{M}_1\big) .
\end{align*}
Considering a probability bound for each of these three subterms, we have
\begin{align*}
    &\PP \big\{ \big\| (\hat{\tilde{A}} - \tilde{A}) \mathbf{M}_1 \big\|_2 \geq \eps \big\} \\
    &\leq
    \PP \big\{ \big\| \hat{\tilde{A}} - \tilde{A} \big\|_2 \| \mathbf{M}_1 \|_2 \geq \eps \big\} \tag{by submultiplicativity} \\
    &=
    \PP \left\{ \big\| (\hat{\tilde{A}} - \tilde{A}) \big\|_2 \geq \frac{\eps}{\| \mathbf{M}_1 \|_2} \right\} \\
    &\leq
    \eta_A \left(\frac{\eps}{\left\| \mathbf{M}_1 \right\|_2}\right), \tag{ by \eqref{eq:bound_of_hat_tilde_A}}\\
    &\PP \big\{ \big\| \tilde{A} (\hat{\mathbf{M}}_1 - \mathbf{M}_1) \big\|_2 \geq \eps \big\} \\
    & \leq
    \PP \big\{ \big\| \tilde{A} \big\|_2  \big\| \hat{\mathbf{M}}_1 - \mathbf{M}_1 \big\|_2 \geq \eps \big\} \tag{by submultiplicativity} \\
    & \leq
    \PP \big\{ 2 \| A \|_2^2  \big\| \hat{\mathbf{M}}_1 - \mathbf{M}_1 \big\|_2 \geq \eps \big\} \tag{since $\big\| \tilde{A} \big\|_2 = \| P_1 (A \otimes A) Q_1 \|_2 \leq 2 \|A \otimes A\|_2 = 2 \|A\|_2^2$} \\
    & = 
    \PP \left\{ \big\| \hat{\mathbf{M}}_1 - \mathbf{M}_1 \big\|_2 \geq \frac{\eps}{2 \| A \|_2^2 } \right\} \\
    & \leq
    \eta_D \left( \frac{\eps}{2 \| A \|_2^2 } \right), \tag{by \eqref{eq:bound_of_M1}}
\end{align*}
and
\begin{align*}
    &\PP \big\{ \big\| (\hat{\tilde{A}} - \tilde{A}) (\hat{\mathbf{M}}_1 - \mathbf{M}_1)  \big\|_2 \geq \eps \big\} \\
    & \leq
    \PP \big\{ \big\| \hat{\tilde{A}} - \tilde{A} \big\|_2  \big\| \hat{\mathbf{M}}_1 - \mathbf{M}_1 \big\|_2 \geq \eps \big\} \tag{by submultiplicativity} \\
    & \leq
    \PP \big\{ \big\| \hat{\tilde{A}} - \tilde{A} \big\|_2 \geq \sqrt{\eps}  \big\} + \PP \big\{  \big\| \hat{\mathbf{M}}_1 - \mathbf{M}_1 \big\|_2 \geq \sqrt{\eps} \big\} \tag{by \eqref{eq:prob_times_bound}} \\
    & \leq
    \eta_A( \sqrt{\eps} ) + \eta_D( \sqrt{\eps} ). \tag{by \eqref{eq:bound_of_hat_tilde_A} and \eqref{eq:bound_of_M1} }
\end{align*}
Putting together the bounds for the three subterms,
\begin{align*}
    &\PP \big\{ \big\| \hat{\tilde{A}} \hat{\mathbf{M}}_1 - \tilde{A} \mathbf{M}_1 \big\|_2 \geq \eps \big\} \\
    & \leq
    \PP \big\{ \big\| (\hat{\tilde{A}} - \tilde{A}) \mathbf{M}_1 \big\|_2 \geq \eps/3 \big\}
    +
    \PP \big\{ \big\| \tilde{A} (\hat{\mathbf{M}}_1 - \mathbf{M}_1) \big\|_2 \geq \eps/3 \big\} +
    \PP \big\{ \big\| (\hat{\tilde{A}} - \tilde{A}) (\hat{\mathbf{M}}_1 - \mathbf{M}_1)  \big\|_2 \geq \eps/3 \big\} \tag{by \eqref{eq:prob_addition_bound}}\\
    & \leq
    \eta_A \left(\frac{\eps}{3 \left\| \mathbf{M}_1 \right\|_2}\right)
    +
    \eta_D \left( \frac{\eps}{6 \left\| A \right\|_2^2 } \right) 
    +
    \eta_A( \sqrt{\eps/3} ) + \eta_D( \sqrt{\eps/3} ) \\
    & \teL
    \eta_{AM} (\eps).
\end{align*}

For the third term, $\hat{K}_{BA} \hat{\mathbf{L}}_1 - K_{BA} \mathbf{L}_1$, we have the decomposition
\begin{align*}
    \hat{K}_{BA} \hat{\mathbf{L}}_1 - K_{BA} \mathbf{L}_1
    =
    (\hat{K}_{BA} - K_{BA}) \mathbf{L}_1 + K_{BA} (\hat{\mathbf{L}}_1 - \mathbf{L}_1 ) + (\hat{K}_{BA} - K_{BA}) (\hat{\mathbf{L}}_1 - \mathbf{L}_1 ).
\end{align*}
Considering a probability bound for each of these three subterms, we have
\begin{align*}
    &\PP \big\{ \big\| (\hat{K}_{BA} - K_{BA}) \mathbf{L}_1 \big\|_2 \geq \eps \big\}\\
    & \leq
    \PP \big\{ \big\| \hat{K}_{BA} - K_{BA} \big\|_2 \| \mathbf{L}_1 \|_2 \geq \eps \big\} \tag{by submultiplicativity} \\
    & =
    \PP \left\{ \big\| \hat{K}_{BA} - K_{BA} \big\|_2 \geq \frac{\eps}{\| \mathbf{L}_1 \|_2 } \right\} \\
    & \leq
    \eta_{AB} \left( \frac{\eps}{\left\| \mathbf{L}_1 \right\|_2 } \right), \tag{by \eqref{eq:bound_of_Kab}}
\end{align*}
\begin{align*}
    &\PP \big\{ \big\|  K_{BA} (\hat{\mathbf{L}}_1 - \mathbf{L}_1 ) \big\|_2 \geq \eps \big\}\\
    & \leq
    \PP \big\{ \| K_{BA} \|_2  \big\| \hat{\mathbf{L}}_1 - \mathbf{L}_1 \big\|_2 \geq \eps \big\} \tag{by submultiplicativity} \\
    & \leq
    \PP \big\{ \| A \| \| B \|  \big\| \hat{\mathbf{L}}_1 - \mathbf{L}_1 \big\|_2 \geq \eps \big\} \tag{since $\| K_{BA} \|_2 = \| P_1 ( B \otimes A ) \|_2 \leq \| P_1 \|_2 \| B \otimes A \|_2 = \|A \|_2 \|B\|_2$} \\
    & = 
    \PP \left\{ \big\| \hat{\mathbf{L}}_1 - \mathbf{L}_1 \big\|_2 \geq \frac{\eps}{\| A \|_2 \| B \|_2} \right\} \\
    &\leq
    \eta_{L} \left( \frac{\eps}{\| A \|_2 \| B \|_2} \right), \tag{by \eqref{eq:bound_of_L1}}
\end{align*}
and
\begin{align*}
    &\PP \big\{ \big\| (\hat{K}_{BA} - K_{BA}) (\hat{\mathbf{L}}_1 - \mathbf{L}_1 ) \big\|_2 \geq \eps \big\}\\
    & \leq 
    \PP \big\{ \big\| \hat{K}_{BA} - K_{BA} \big\|_2 \geq \sqrt{\eps} \big\} + \PP \big\{ \big\| \hat{\mathbf{L}}_1 - \mathbf{L}_1 \big\|_2 \geq \sqrt{\eps} \big\} \tag{by \eqref{eq:prob_times_bound}} \\
    & \leq
    \eta_{AB} \left( \sqrt{\eps} \right) + \eta_{L} \left( \sqrt{\eps} \right). \tag{by \eqref{eq:bound_of_Kab} and \eqref{eq:bound_of_L1}}
\end{align*}
Putting together the bounds for the three subterms,
\begin{align*}
    &\PP \big\{ \big\| \hat{K}_{BA} \hat{\mathbf{L}}_1 - K_{BA} \mathbf{L}_1 \big\|_2 \geq \eps \big\}\\
    & \leq
    \PP \big\{ \big\| (\hat{K}_{BA} - K_{BA}) \mathbf{L}_1 \big\|_2 \geq \eps/3 \big\}
    +
    \PP \big\{ \big\|  K_{BA} (\hat{\mathbf{L}}_1 - \mathbf{L}_1 ) \big\|_2 \geq \eps/3 \big\} +
    \PP \big\{ \big\| (\hat{K}_{BA} - K_{BA}) (\hat{\mathbf{L}}_1 - \mathbf{L}_1 ) \big\|_2 \geq \eps/3 \big\} \tag{by \eqref{eq:prob_addition_bound}}\\
    & \leq 
    \eta_{AB} \left( \frac{\eps}{3 \left\| \mathbf{L}_1 \right\|_2 } \right)
    +
    \eta_{L} \left( \frac{\eps}{3 \| A \|_2 \| B \|_2} \right)
    +
    \eta_{AB} \left( \sqrt{\eps/3} \right) + \eta_{L} \left( \sqrt{\eps/3} \right) \\
    &\teL 
    \eta_{KL} (\eps) .
\end{align*}

For the fourth term, $\hat{K}_{AB} \hat{\mathbf{L}}_2 - K_{AB} \mathbf{L}_2$, an identical argument to that for the third term using \eqref{eq:bound_of_Kab} and \eqref{eq:bound_of_L2} yields
\begin{align*}
    \PP \big\{ \big\| \hat{K}_{AB} \hat{\mathbf{L}}_2 - K_{AB} \mathbf{L}_2 \big\|_2 \geq \eps \big\}
    \leq
    \eta_{KL} (\eps) .
\end{align*}

For the fifth term, $\big( \hat{\tilde{B}} - \tilde{B}\big) \mathbf{U}$, we have
\begin{align*}
    &\PP \big\{ \big\| \big( \hat{\tilde{B}} - \tilde{B}\big) \mathbf{U} \big\|_2 \geq \eps \big\}\\
    & \leq
    \PP \big\{ \big\| \hat{\tilde{B}} - \tilde{B} \big\|_2 \| \mathbf{U} \|_2 \geq \eps \big\} \tag{by submultiplicativity} \\
    & =
    \PP \left\{ \big\| \hat{\tilde{B}} - \tilde{B} \big\|_2 \geq \frac{\eps}{ \| \mathbf{U} \|_2} \right\} \\
    & \leq
    \eta_B \left( \frac{\eps}{ \| \mathbf{U} \|_2} \right). \tag{by \eqref{eq:bound_of_hat_tilde_B}}
\end{align*}

Putting together the bounds for the five terms, we have
\begin{align*}
    &\PP \big\{\big\| \hat{\bfC} - \bfC \big\|_2 \ge \eps \big\}\\
    & \leq
    \PP \big\{\big\| \hat{\mathbf{M}}_2 - \mathbf{M}_2 \big\|_2 \ge \eps/5 \big\} 
    +
    \PP \big\{\big\| \hat{\tilde{A}} \hat{\mathbf{M}}_1 - \tilde{A} \mathbf{M}_1 \big\|_2 \ge \eps/5 \big\} +
    \PP \big\{\big\| \hat{K}_{BA} \hat{\mathbf{L}}_1 - K_{BA} \mathbf{L}_1 \big\|_2 \ge \eps/5 \big\} \\
    &\quad +
    \PP \big\{\big\| \hat{K}_{AB} \hat{\mathbf{L}}_2 - K_{AB} \mathbf{L}_2 \big\|_2 \ge \eps/5 \big\} 
    +
    \PP \big\{\big\| \big( \hat{\tilde{B}} - \tilde{B}\big) \mathbf{U} \big\|_2 \ge \eps/5 \big\} \tag{by \eqref{eq:prob_addition_bound}} \\
    & \leq
    \eta_{D} (\eps/5)
    + \eta_{AM} (\eps/5)
    + 2 \eta_{KL} (\eps/5)
    + \eta_B \left( \frac{\eps}{5\left\| \mathbf{U} \right\|} \right) \\
    & \teL
    \eta_{C}(\eps).
\end{align*}
\end{pf}


\begin{lemma}\label{lem:deltaCD}
Suppose Assumptions~\ref{asmp1} and~\ref{asmp:bounded_system} hold. Then for all $ \eps > 0$,
\begin{align*}
    \PP \big\{ \big\| \hat{\bfC}\hat{\bfD}^\tp - \bfC\bfD^\tp \big\|_2 \ge \eps \big\} \le \eta_{CD}(\eps),
\end{align*}
where
\begin{align*}
    \eta_{CD}(\eps)
    &\Let
    \eta_C \left( \sqrt{\frac{\eps}{3}} \right) 
    + \eta_D \left( \sqrt{\frac{\eps}{3}} \right) 
    + \eta_C \left( \frac{\eps}{3\left\| \bfD \right\|_2} \right) 
    + \eta_D \left( \frac{\eps}{3\left\| \bfC \right\|_2} \right).
\end{align*}
\end{lemma}

\begin{pf}
The proof follows from using the decomposition
\begin{align*}
    \hat{\bfC}\hat{\bfD}^\tp - \bfC\bfD^\tp 
    =
    \big(\hat{\bfC} - \bfC\big) \big(\hat{\bfD}-\bfD\big)^\tp 
    + 
    \big(\hat{\bfC} - \bfC\big)\bfD^\tp 
    + 
    \bfC \big(\hat{\bfD}- \bfD\big)^\tp 
\end{align*}
to provide conservative decompositions into terms of the form 
\begin{align*}
    \PP \big\{ \big\| \hat{\bfC} - \bfC \big\|_2 \geq \eps \big\} \leq \eta, ~\PP \big\{ \big\| \hat{\bfD} - \bfD \big\|_2 \geq \eps \big\} \leq \eta,
\end{align*}
which are suitable for the bounds of Lemma \ref{lem:bounded_C_and_D}.
\end{pf}

\begin{lemma}\label{lem:deltaDD}
Suppose Assumptions~\ref{asmp1} and~\ref{asmp:bounded_system} hold. Given a positive value $\eps_{\max}$, then for all $0 < \eps < \eps_{\max}$,
\begin{align*}
    \PP \big\{\big\|  (\hat{\bfD}\hat{\bfD}^\tp)^{\dagger} - (\bfD\bfD^\tp)^{-1} \big\|_2 \ge \eps \big\} \le \eta_{DD}(\eps,\eps_{\max}), 
\end{align*}
where
\begin{align*}
    \eta_{DD}(\eps, \eps_{\max}) 
    &\Let
    \eta_0 \left( \frac12 \lambda_{\min}^2 (\bfD\bfD^\tp)  \left(1 - \frac{\eps}{\eps_{\max}}\right) \eps  \right)
    +
    \eta_m \left( \frac{\eps \lambda_{\min}(\bfD\bfD^\tp)}{ \eps_{\max} ( 2 + \lambda_{\min}(\bfD\bfD^\tp)/\lambda_{\max}(\bfD\bfD^\tp) ) } \right),\\
    \eta_0(\eps)
    &\Let
    \eta_{D} \left(\sqrt{\lambda_{\max}(\bfD\bfD^\tp) + \eps} - \sqrt{\lambda_{\max}(\bfD\bfD^\tp)}  \right), \\
    \eta_m(\eps)
    &\Let
    \left(9^{[n(n+1)+m(m+1)]/2} + \left( \frac{ 16 \lambda_{\max} (\bfD\bfD^\tp)}{\lambda_{\min} (\bfD\bfD^\tp)} + 1 \right)^{[n(n+1)+m(m+1)]/2} \right) \eta_0(\eps).
\end{align*}
\end{lemma}

\begin{pf}
The proof follows an identical argument to Lemma \ref{lem:deltaZZ}:
\begin{enumerate}
    \item Replace $\bfZ$ by $\bfD$, $n$ by $n(n+1)/2$, and $m$ by $m(m+1)/2$.
    \item we apply \eqref{eq:difference_of_matrix_inverses} to obtain the decomposition
    \begin{align*}
        (\hat{\bfD}\hat{\bfD}^\tp)^{\dagger} - (\bfD\bfD^\tp)^{-1}
        &=
        (\bfD\bfD^\tp)^{-1}  (\hat{\bfD}\hat{\bfD}^\tp)^{\dagger} \big[ (\hat{\bfD}\hat{\bfD}^\tp) - (\bfD\bfD^\tp)  \big] .
    \end{align*}
    \item Apply Lemma \ref{lem:bounded_C_and_D} with the appropriate settings of $\eps$ to get the bound
    \begin{align}
        \PP \big\{\big\|\hat{\bfD}\hat{\bfD}^\tp - \bfD\bfD^\tp \big\|_2 \ge \eps \big\} 
        &\le 
        \eta_{D} \left(\sqrt{\lambda_{\max}(\bfD\bfD^\tp) + \eps} - \sqrt{\lambda_{\max}(\bfD\bfD^\tp)}  \right) \label{eq:bound_of_DD} 
        \teL \eta_0(\eps).
    \end{align}
    \item Apply a similar $\gamma$-net argument to obtain an upper bound of $ \lambda_{\max}(\hat{\bfD}\hat{\bfD}^\tp)$, then a lower bound of $\lambda_{\min} (\hat{\bfD}\hat{\bfD}^\tp)$, and finally the claimed bound.
\end{enumerate}
\end{pf}

\begin{theorem}[Theorem \ref{thm:hatSigmaAB_bounded} restated] \label{thm:hatSigmaAB_bounded_restate}
Suppose Assumptions~\ref{asmp1} and~\ref{asmp:bounded_system} hold. Given a positive value $\eps_{\max}$, then for all $0 < \eps < 3 \eps_{\max} \cdot \min\{\sqrt{\lambda_{\max} (\bfC\bfC^\tp)\lambda_{\max} (\bfD\bfD^\tp)}, \eps_{\max} \}$,
\begin{align*}
    \PP \Big\{\Big\|\Big[\hat{\tilde{\Sigma}}'_A~ \hat{\tilde{\Sigma}}'_B\Big] - \big[\tilde{\Sigma}'_A~ \tilde{\Sigma}'_B\big] \Big\|_2 \ge \eps \Big\} \le \eta(\eps),
\end{align*}
where
\begin{align*}
    \eta(\eps) 
    \Let
    \eta_{CD} \left( \frac13 \lambda_{\min} (\bfC\bfC^\tp) \eps \right)
    +
    \eta_{CD} \left( \sqrt{\frac{\eps}{3}} \right)
    +
    \eta_{DD} \left( \frac{\eps}{ 3 \sqrt{\lambda_{\max}(\bfC\bfC^\tp)\lambda_{\max}(\bfD\bfD^\tp)} }, \eps_{\max} \right)
    + \eta_{DD} \left( \sqrt{\frac{\eps}{3}}, \eps_{\max} \right) .
\end{align*}
\end{theorem}

\begin{pf}
The proof follows an identical argument to Theorem \ref{thm:hatAB_bounded_restate}:
\begin{enumerate}
    \item Replace $A$ and $B$ by $\tilde{\Sigma}'_A$ and $\tilde{\Sigma}'_B$, and replace $\bfY$ and $\bfZ$ by $\bfC$ and $\bfD$.
    \item Decompose the error matrix using the least-squares estimators as
    \begin{align*}
        &\Big[\hat{\tilde{\Sigma}}'_A~ \hat{\tilde{\Sigma}}'_B\Big]
        -
        \big[
        \tilde{\Sigma}'_A ~ \tilde{\Sigma}'_B
        \big]\\
        &= 
        \hat{\bfC}\hat{\bfD}^\tp (\hat{\bfD}\hat{\bfD}^\tp)^{-1} - \bfC\bfD^\tp (\bfD\bfD^\tp)^{-1}\\
        &= 
        \big[ \hat{\bfC}\hat{\bfD}^\tp - \bfC\bfD^\tp \big] (\bfD\bfD^\tp)^{-1} + \bfC\bfD^\tp \big[ (\hat{\bfD}\hat{\bfD}^\tp)^{-1} - (\bfD\bfD^\tp)^{-1} \big] + \big[ \hat{\bfC}\hat{\bfD}^\tp - \bfC\bfD^\tp \big] \big[ (\hat{\bfD}\hat{\bfD}^\tp)^{-1} - (\bfD\bfD^\tp)^{-1} \big].
    \end{align*}
    \item Consider a probability bound and use \eqref{eq:prob_addition_bound}.
    \item Apply Lemmas \ref{lem:deltaCD} and \ref{lem:deltaDD} with the appropriate settings of $\eps$ in each term.
    \item The conclusion follows by combining the probability bounds for each term.
\end{enumerate}
\end{pf}

\begin{pf*}{PROOF OF THEOREM \ref{thm:hatSigmaAB_bounded}.} \ \\
The qualitative claim in Theorem \ref{thm:hatSigmaAB_bounded} is found by inverting the bound of Theorem \ref{thm:hatSigmaAB_bounded_restate} and examining the behavior of the bound as $n_r \to \infty$. The argument is similar to the proof of Theorem~\ref{thm:hatAB_bounded}, so we just state the major steps.

From Lemma~\ref{lem:bounded_C_and_D}, it follows that for fixed $\delta \in (0,1)$ and $\eps_D(\delta) > 0$ such that $\PP \{\| \hat{\bfD} - \bfD \|_2 \ge \eps_D(\delta) \} \le \eta_D(\eps_D(\delta)) = \delta$,
\begin{align*}
    \eps_D(\delta) = \bigO \left( \sqrt{\frac{\ell c_F^2 \log \{ [n(n+1)/2+\ell]/\delta \}}{n_r}} \right).
\end{align*}
Write $\eta_C(\eps)$ in Lemma~\ref{lem:bounded_C_and_D} explicitly,
\begin{align*}
    \eta_C(\eps) &= 
    \eta_{D} (\eps/5) 
    + \eta_D \left( \frac{\eps}{30 \| A \|_2^2 } \right) 
    + \eta_D( \sqrt{\eps/15} ) 
    + 2 \eta_{L} \left( \frac{\eps}{15 \| A \|_2 \| B \|_2} \right) 
    + 2 \eta_{L} \left( \sqrt{\eps/15} \right) 
    + \delta_{AB} \left(\frac12 \sqrt{\frac{\eps}{15 \| \mathbf{M}_1\|_2}}\right)\\
    &\quad 
    + \delta_{AB} \left(\frac{\eps}{120 \| \mathbf{M}_1\|_2 \sqrt{\| A \|_2}}\right) 
    + \delta_{AB} \left(\frac12 \bigg(\frac{\eps}{15}\bigg)^{\frac14} \right) 
    + \delta_{AB} \left(\frac18 \sqrt{\frac{\eps}{15 \| A \|_2}} \right) 
    + \delta_{AB} \left(\frac12 \sqrt{\frac{\eps}{5 \| \mathbf{U}\|_2}}\right) \\
    &\quad
    + \delta_{AB} \left(\frac{\eps}{40 \| \mathbf{U}\|_2 \sqrt{\| B \|_2}}\right)
    + 4 \delta_{AB} \left(\sqrt{\frac{\eps}{45 \| \mathbf{L}_1\|_2}}\right) 
    + 2 \delta_{AB} \left(\frac{\eps}{45 \| \mathbf{L}_1\|_2 \sqrt{\| A \|_2}}\right) 
    + 2 \delta_{AB} \left(\frac{\eps}{45 \| \mathbf{L}_1\|_2 \sqrt{\| B \|_2}}\right) \\
    &\quad
    + 4 \delta_{AB} \left(\frac{1}{\sqrt{3}} \bigg(\frac{\eps}{15}\bigg)^{\frac14} \right)
    + 2 \delta_{AB} \left(\frac{1}{3 \sqrt{\|A\|_2}} \sqrt{\frac{\eps}{15}}\right) 
    + 2 \delta_{AB} \left(\frac{1}{3 \sqrt{\|B\|_2}} \sqrt{\frac{\eps}{15}}\right).
\end{align*}
Hence for fixed $\delta \in (0,1)$ we can find $\eps_C(\delta) > 0$ such that $\PP \{\| \hat{\bfC} - \bfC \|_2 \ge \eps_C(\delta) \} \le \eta_C(\eps_C(\delta)) \le \delta$ holds for large enough~$n_r$, $\eps_L(\delta)$ such that $\eta_L(\eps_L(\delta)) = \delta$, and $\eps_{AB}(\delta)$, given in~\eqref{eq_append_AB_epsAB}+. That is,
\begin{align*}
    &\eps_C(\delta) \\
    &= \bigO \Big( \max\Big\{ 
    5 \eps_D(\delta/17), 
    30 \|A\|_2^2 \eps_D(\delta/17), 
    15 \eps^2_D(\delta/17), 
    15 \|A\|_2 \|B\|_2 \eps_L(\delta/34),
    15 \eps_L^2(\delta/34),
    60 \|\mathbf{M_1}\|_2 \eps^2_{AB}(\delta/17),\\
    &\quad 120 \|\mathbf{M_1}\|_2 \sqrt{\|A\|_2} \eps_{AB}(\delta/17),
    240 \eps^4_{AB}(\delta/17),
    960 \|A\|_2 \eps^2_{AB}(\delta/17), 
    20 \|\mathbf{U}\|_2 \eps^2_{AB}(\delta/17),
    40 \|\mathbf{U}\|_2 \sqrt{\|B\|_2} \eps_{AB}(\delta/17),\\
    &\quad 
    45 \|\mathbf{L}_1\|_2 \eps^2_{AB}(\delta/68),
    45 \|\mathbf{L}_1\|_2 \sqrt{\|A\|_2} \eps_{AB}(\delta/34),
    45 \|\mathbf{L}_1\|_2 \sqrt{\|B\|_2} \eps_{AB}(\delta/34),
    135 \eps_{AB}^4(\delta/68), 
    135 \|A\|_2 \eps_{AB}^2(\delta/34),\\
    &\quad
    135 \|A\|_2 \eps_{AB}^2(\delta/34) \Big\} \Big)
    \\
    &=
    \bigO \big( \max\big\{ 
    5 \eps_D(\delta/17), 
    30 \|A\|_2^2 \eps_D(\delta/17), 
    15 \|A\|_2 \|B\|_2 \eps_L(\delta/34), 
    120 \|\mathbf{M_1}\|_2 \sqrt{\|A\|_2} \eps_{AB}(\delta/17),\\
    &\quad 
    40 \|\mathbf{U}\|_2 \sqrt{\|B\|_2} \eps_{AB}(\delta/17),
    45 \|\mathbf{L}_1\|_2 \sqrt{\|A\|_2} \eps_{AB}(\delta/34),
    45 \|\mathbf{L}_1\|_2 \sqrt{\|B\|_2} \eps_{AB}(\delta/34) \big\} \big)\\
    &= 
    \bigO \Bigg( \max\Bigg\{ \|A\|_2^2
    \sqrt{\frac{17 \ell c_F^2 \log \{ [n(n+1)/2+\ell]/\delta \}}{n_r}}, 
    \|A\|_2 \|B\|_2 \sqrt{\frac{34 \ell c_W^2 \log [ (nm+\ell)/\delta ]}{n_r}}, \\
    &\quad 
    \max\{ \|\mathbf{M_1}\|_2 \sqrt{\|A\|_2},  \|\mathbf{U}\|_2 \sqrt{\|B\|_2}, \|\mathbf{L}_1\|_2 \sqrt{\|A\|_2}, \|\mathbf{L}_1\|_2 \sqrt{\|B\|_2} \} \eps_{AB}(\delta/34) \Bigg\} \Bigg),
\end{align*}
where in the last equation we drop the constants and only show the dependence of the bound on system parameters, and $\eps_{AB}(\delta)$ is given in~\eqref{eq_append_AB_epsAB}.

Next, for $\delta\in (0,1)$, let 
\begin{align*}
    \eps_{CD}(\delta) 
    &= \max\{3 \eps_C^2(\delta/4), 3 \eps_D^2(\delta/4), 3\|\bfD\|_2 \eps_C(\delta/4), 3\|\bfC\|_2 \eps_D(\delta/4)\}\\
    &= \bigO \left( \max\{\|\bfD\|_2 \eps_C(\delta/4), \|\bfC\|_2 \eps_D(\delta/4) \} \right),
\end{align*}
and then from Lemma~\ref{lem:deltaCD} we know that
\begin{align*}
    \PP \big\{ \big\| \hat{\bfC}\hat{\bfD}^\tp - \bfC\bfD^\tp \big\|_2 \ge \eps_{CD}(\delta) \big\} \le \eta_{CD}(\eps_{CD}(\delta)) \le \delta.
\end{align*}

Under the condition of Lemma~\ref{lem:deltaDD}, for fixed $\delta \in (0,1)$, define $\eps_{\eta0}(\delta), \eps_{\eta m}(\delta) > 0$ as follows such that $\delta_0(\eps_{\eta0}(\delta)) = \delta$ and $\delta_m(\eps_{\eta m}(\delta)) = \delta$ 
\begin{align*}
    \eps_{\eta 0}(\delta) &:= \eps_D^2(\delta) + 2 \eps_D(\delta) \sqrt{\lambda_{\max}(\bfD\bfD^\tp)} = \bigO \left( \sqrt{\lambda_{\max}(\bfD\bfD^\tp)} \eps_D(\delta) \right),\\
    \eps_{\eta m}(\delta) &:= \eps_D^2(\delta/d(n,m)) + 2 \eps_D(\delta/d(n,m)) \sqrt{\lambda_{\max}(\bfD\bfD^\tp)} = \bigO \left( \sqrt{\lambda_{\max}(\bfD\bfD^\tp)} \eps_D(\delta/d(n,m)) \right),
\end{align*}
where 
\begin{align}\label{eq_append_sigmaAB_dnm}
    d(n,m) := 9^{[n(n+1)+m(m+1)]/2} + \left( \frac{ 16 \lambda_{\max} (\bfD\bfD^\tp)}{\lambda_{\min} (\bfD\bfD^\tp)} + 1 \right)^{[n(n+1)+m(m+1)]/2}.
\end{align}
For fixed $\delta,\eps_{\max} \in (0,1)$, let $\eps_{D1}(\delta) \in (0,\eps_{\max})$ such that
\begin{align*}
    \eps_{D1}(\delta) 
    = \frac{1}{2} \eps_{\max} \bigg(1 - \sqrt{1 - \frac{8 \eps_{\eta 0}(\delta)}{\lambda_{\min}^2 (\bfD\bfD^\tp)}} \bigg) 
    = \bigO \left( \frac{\sqrt{\lambda_{\max}(\bfD\bfD^\tp)}}{\lambda_{\min}^2 (\bfD\bfD^\tp)} \eps_D(\delta) \right),
\end{align*}
and set $\eps_{D2}(\delta) \in (0,\eps_{\max})$ such that
\begin{align*}
    \eps_{D2}(\delta) &= \frac{\eps_{\max}(2 + \lambda_{\min}(\bfD\bfD^\tp)/\lambda_{\max}(\bfD\bfD^\tp))}{\lambda_{\min}(\bfD\bfD^\tp)} \eps_{\eta m}(\delta)\\
    &=
    \bigO \left( \bigg(2 + \frac{\lambda_{\min}(\bfD\bfD^\tp)}{\lambda_{\max}(\bfD\bfD^\tp)} \bigg) \frac{\sqrt{\lambda_{\max}(\bfD\bfD^\tp)}}{\lambda_{\min}(\bfD\bfD^\tp)} \eps_D(\delta/d(n,m)) \right).
\end{align*}
Now define $\eps_{DD}(\delta) := \max\{\eps_{D1}(\delta/2),\eps_{D2}(\delta/2)\}$. Since for fixed $\delta, \eps_{\max} \in (0,1)$, when $n_r$ is large enough, $\eps_{D1}(\delta/2)$, $\eps_{D2}(\delta/2) < \eps_{\max}/2$, it holds that
\begin{align*}
    &\PP \big\{\big\| (\hat{\bfD}\hat{\bfD}^\tp)^{\dagger} - (\bfD\bfD^\tp)^{-1} \big\|_2 \ge \eps_{DD}(\delta) \big\} \\
    &\le \eta_{DD}(\eps_{DD}(\delta), \eps_{\max})\\
    &= 
    \eta_0 \left( \frac12 \lambda_{\min}^2 (\bfD\bfD^\tp)  \left(1 - \frac{\eps_{DD}(\delta)}{\eps_{\max}}\right) \eps_{DD}(\delta)  \right)
    +
    \eta_m \left( \frac{\eps_{DD}(\delta) \lambda_{\min}(\bfD\bfD^\tp)}{ \eps_{\max} ( 2 + \lambda_{\min}(\bfD\bfD^\tp)/\lambda_{\max}(\bfD\bfD^\tp) ) } \right)\\
    &\le
    \eta_0 \left( \frac12 \lambda_{\min}^2 (\bfD\bfD^\tp)  \left(1 - \frac{\eps_{D1}(\delta/2)}{\eps_{\max}}\right) \eps_{D1}(\delta/2)  \right)
    +
    \eta_m \left( \frac{\eps_{D2}(\delta/2) \lambda_{\min}(\bfD\bfD^\tp)}{ \eps_{\max} ( 2 + \lambda_{\min}(\bfD\bfD^\tp)/\lambda_{\max}(\bfD\bfD^\tp) ) } \right)
    =\delta.
\end{align*}
Finally, let
\begin{align*}
    \eps_{\Sigma}(\delta) := \max \bigg\{ \frac{3\eps_{CD}(\delta/4)}{\lambda_{\min}(\bfD\bfD^\tp)}, 3 \eps^2_{CD}(\delta/4), 3\sqrt{\lambda_{\max}(\bfC\bfC^\tp)\lambda_{\max}(\bfD\bfD^\tp)}\eps_{DD}(\delta/4), 3\eps^2_{DD}(\delta/4) \bigg\},
\end{align*}
and it can be observed that for fixed $\delta \in (0,1)$ and large enough $n_r$,
\begin{align*}
    &\PP \Big\{\Big\|\Big[\hat{\tilde{\Sigma}}'_A~ \hat{\tilde{\Sigma}}'_B\Big] - \big[\tilde{\Sigma}'_A~ \tilde{\Sigma}'_B\big] \Big\|_2 \ge \eps_{\Sigma}(\delta) \Big\} \\
    &\le \eta(\eps_{\Sigma}(\delta))\\
    &= 
    \eta_{CD} \left( \frac13 \lambda_{\min} (\bfD\bfD^\tp) \eps_{\Sigma}(\delta) \right)
    +
    \eta_{CD} \left( \sqrt{\frac{\eps_{\Sigma}(\delta)}{3}} \right)+
    \eta_{DD} \left( \frac{\eps_{\Sigma}(\delta)}{ 3 \sqrt{\lambda_{\max}(\bfC\bfC^\tp)\lambda_{\max}(\bfD\bfD^\tp)} }, \eps_{\max} \right)\\
    &\quad + \eta_{DD} \left( \sqrt{\frac{\eps_{\Sigma}(\delta)}{3}}, \eps_{\max} \right)\\
    &\le
    \eta_{CD} ( \eps_{CD}(\delta/4) )
    +
    \eta_{CD} ( \eps_{CD}(\delta/4) )+
    \eta_{DD} (\eps_{DD}(\delta/4), \eps_{\max}) + \eta_{DD} (\eps_{DD}(\delta/4), \eps_{\max}) = \delta.
\end{align*}
Moreover,
\begin{align*}
    &\eps_{\Sigma}(\delta) \\
    &=
    \bigO \Bigg( \max \bigg\{ \frac{\eps_{CD}(\delta/4)}{\lambda_{\min}(\bfD\bfD^\tp)}, \sqrt{\lambda_{\max}(\bfC\bfC^\tp)\lambda_{\max}(\bfD\bfD^\tp)}\eps_{DD}(\delta/4) \bigg\} \Bigg) \\
    &=
    \bigO \Bigg( \max \bigg\{
    \frac{\sqrt{\lambda_{\max}(\bfD\bfD^\tp)}}{\lambda_{\min}(\bfD\bfD^\tp)} \eps_{C}(\delta/4), \frac{\sqrt{\lambda_{\max}(\bfC\bfC^\tp)}}{\lambda_{\min}(\bfD\bfD^\tp)} \eps_{D}(\delta/4), \frac{\sqrt{\lambda_{\max}(\bfC\bfC^\tp)}\lambda_{\max}(\bfD\bfD^\tp)}{\lambda_{\min}^2 (\bfD\bfD^\tp)} \eps_D(\delta/8), \\
    &\quad
    \sqrt{\lambda_{\max}(\bfC\bfC^\tp)} \bigg(1 + \frac{2\lambda_{\max}(\bfD\bfD^\tp)}{\lambda_{\min} (\bfD\bfD^\tp)} \bigg) \eps_D(\delta/8d(n,m)) \bigg\} \Bigg)\\
    &=
    \bigO \Bigg( \max \bigg\{ \frac{\sqrt{\lambda_{\max}(\bfD\bfD^\tp)} \|A\|_2^2}{\lambda_{\min}(\bfD\bfD^\tp)} 
    \sqrt{\frac{17 \ell c_F^2 \log \{ 4[n(n+1)/2+\ell]/\delta \}}{n_r}},
    \frac{\sqrt{\lambda_{\max}(\bfC\bfC^\tp)}}{\lambda_{\min}(\bfD\bfD^\tp)} \sqrt{\frac{\ell c_F^2 \log \{ 4 [n(n+1)/2+\ell]/\delta \}}{n_r}},\\
    &\quad 
    \frac{\sqrt{\lambda_{\max}(\bfC\bfC^\tp)}\lambda_{\max}(\bfD\bfD^\tp)}{\lambda_{\min}^2 (\bfD\bfD^\tp)} \sqrt{\frac{\ell c_F^2 \log \{ 8 [n(n+1)/2+\ell]/\delta \}}{n_r}},\\
    &\quad
    \sqrt{\lambda_{\max}(\bfC\bfC^\tp)} \bigg(1 + \frac{2\lambda_{\max}(\bfD\bfD^\tp)}{\lambda_{\min} (\bfD\bfD^\tp)} \bigg) \sqrt{\frac{\ell c_F^2 \log \{ 8 [n(n+1)/2+\ell] d(n,m)/\delta \}}{n_r}} \\
    &\quad      \frac{\sqrt{\lambda_{\max}(\bfD\bfD^\tp)} \|A\|_2 \|B\|_2}{\lambda_{\min}(\bfD\bfD^\tp)}  \sqrt{\frac{34 \ell c_W^2 \log [ 4(nm+\ell)/\delta ]}{n_r}},\\
    &\quad
    \max\{ \|\mathbf{M_1}\|_2 \sqrt{\|A\|_2},  \|\mathbf{U}\|_2 \sqrt{\|B\|_2}, \|\mathbf{L}_1\|_2 \sqrt{\|A\|_2}, \|\mathbf{L}_1\|_2 \sqrt{\|B\|_2} \} \frac{\sqrt{\lambda_{\max}(\bfD\bfD^\tp)}}{\lambda_{\min}(\bfD\bfD^\tp)} \eps_{AB}(\delta/136) \bigg\} \Bigg),
\end{align*}
where $\eps_{AB}(\delta)$ is given in~\eqref{eq_append_AB_epsAB}, and $d(n,m)$ is given in~\eqref{eq_append_sigmaAB_dnm}. This completes the proof by noticing the bound of $\eps_{AB}(\delta)$ in the proof of Theorem~\ref{thm:hatAB_bounded}.
\end{pf*}

\bibliographystyle{apalike}        
\bibliography{bibliography.bib}           

\end{document}